\pdfoutput=1
\documentclass{article}

\usepackage{arxiv}
\usepackage{amsmath,amssymb,amsfonts,amsthm,mathtools}
\usepackage[utf8]{inputenc}
\usepackage[T1]{fontenc}
\usepackage[round]{natbib}
\usepackage{url}
\usepackage{booktabs}
\usepackage{array}
\usepackage{tabularx}
\usepackage{microtype}
\usepackage{xcolor}
\usepackage{graphicx}
\usepackage{caption}
\usepackage{algorithm}
\usepackage{algorithmic}
\usepackage{enumitem}
\usepackage{needspace}
\usepackage[section]{placeins}
\usepackage{hyperref}
\hypersetup{colorlinks=true,
  linkcolor={black!45!blue}, citecolor={black!45!blue}, urlcolor={black!45!blue},
  pdftitle={Pair-Dependent Judge Errors Change the Minimax Rate of Semantic Entropy},
  pdfauthor={Jiamiao Liu; Dewen Qiao; Yu Zhang; Xuetao Chen},
  pdfkeywords={entropy estimation, large language models, method of moments, minimax rates, uncertainty quantification}}
\usepackage[capitalize,noabbrev]{cleveref}
\newcolumntype{L}[1]{>{\raggedright\arraybackslash}p{#1}}

\theoremstyle{plain}
\newtheorem{theorem}{Theorem}
\newtheorem{proposition}[theorem]{Proposition}
\newtheorem{lemma}[theorem]{Lemma}
\newtheorem{corollary}[theorem]{Corollary}
\theoremstyle{definition}
\newtheorem{definition}[theorem]{Definition}
\theoremstyle{remark}

\newcommand{\Real}{\mathbb{R}}
\newcommand{\Cpx}{\mathbb{C}}
\newcommand{\Exp}{\mathbb{E}}
\newcommand{\Prob}{\mathbb{P}}
\newcommand{\Var}{\mathrm{Var}}
\newcommand{\Cov}{\mathrm{Cov}}
\newcommand{\Bern}{\mathrm{Bern}}
\newcommand{\TV}{\mathrm{TV}}
\newcommand{\Ind}[1]{\mathbf{1}\{#1\}}
\newcommand{\Ent}{H}
\newcommand{\hH}{\widehat{H}}
\newcommand{\Simplex}{\Theta_3}
\newcommand{\Int}{\Theta_3^\circ}
\newcommand{\htheta}{\widehat{\theta}}
\newcommand{\eps}{\varepsilon}
\newcommand{\one}{\mathbf{1}}
\newcommand{\tM}{\widetilde{M}}
\newcommand{\cU}{\mathcal{U}}
\newcommand{\cD}{\mathcal{D}}
\newcommand{\cT}{\mathcal{T}}
\newcommand{\cS}{\mathcal{S}}
\newcommand{\cJ}{\mathcal{J}}
\newcommand{\Sym}{\mathrm{Sym}_3}
\newcommand{\tFthree}{\widetilde{F}^{(3)}}
\newcommand{\Sw}{\mathrm{Sw}}
\newcommand{\Pu}{\mathsf{P}}
\newcommand{\DS}{\mathsf{D}}
\newcommand{\SG}{S_{\mathrm{G}}}
\newcommand{\Sobs}{S_{\mathrm{obs}}}
\newcommand{\Rem}{\mathrm{Rem}}
\newcommand{\hF}{\widehat{F}}

\DeclareMathOperator{\rank}{rank}
\DeclareMathOperator{\diag}{diag}
\DeclareMathOperator{\aff}{aff}

\graphicspath{{figures/}}

\renewcommand{\shorttitle}{Pair-Dependent Judge Errors and Semantic Entropy Rates}

\title{Pair-Dependent Judge Errors Change the\\Minimax Rate of Semantic Entropy}

\author{%
  Jiamiao Liu \quad Dewen Qiao \quad Yu Zhang \quad Xuetao Chen\thanks{Corresponding author.} \\
  Department of Information, Xinqiao Hospital \\
  Army Medical University (Third Military Medical University) \\
  Chongqing 400037, China \\
  \texttt{xuetao@tmmu.edu.cn}
}

\date{}

\begin{document}

\maketitle

\begin{abstract}
Semantic entropy, the entropy of the meanings among a language model's sampled answers, is estimated from the
verdicts of a judge that declares pairs of answers equivalent or not, and automatic judges err in ways that depend
on the answers they compare. The existing minimax theory covers only judges whose false-equivalence probability is
the same for all pairs. We model a known judge that declares two answers with meanings $a$ and $b$ equivalent with
probability $q+\Delta M_{ab}$, for a symmetric matrix $M$ and $q$ bounded away from $0$ and $1$, and show
that for three meanings the rate is set by tree statistics on four answers, while on an open set of channels the
consistency threshold is set by trees on five. For every channel in an explicit open class, the minimax squared
risk is of order $N^{-1/3}$ with $N=n^4\Delta^6$ on polynomial weak windows on which $n^5\Delta^8$ exceeds a multiple
of $\log n$, and of order $\min\{1,N^{-1/3}\}$ on every polynomial weak window when the tree functions on at most four
vertices are injective. The exponent comes from a cubic degeneracy at the equitable point. On an open set of channels in the
class, the four-vertex tree functions take equal values at two distributions with different entropies. For such a
channel, on every polynomial weak window, the minimax risk is of the order of the sum of $\min\{1,N^{-1/3}\}$ and a
term that decays exponentially in $n^5\Delta^8$, so no estimator is
consistent while $n^5\Delta^8$ stays bounded, even as $N\to\infty$. For $\Delta=n^{-\beta}$ with $\beta>1/2$, channels of the class with injective
four-vertex tree functions allow consistent estimation exactly for $\beta<2/3$, and the channels of this open set
exactly for $\beta<5/8$, at which the risk exponent drops from $1/12$ to $0$. Computer-assisted certificates show that
both kinds of channels occur on open sets.

\end{abstract}


\section{Introduction}
\label{sec:intro}

Semantic entropy, the entropy of the meanings among the sampled answers of a language model, has to be estimated
from the verdicts of a judge that compares pairs of answers and declares them equivalent or not, because the
meanings are not observed \citep{kuhn2023semantic,farquhar2024detecting}. Automatic answer judges err in ways
that depend on the answers they compare \citep{wang2023evaluating,kamalloo2023evaluating}, so the probability of a
false equivalence can depend on the pair of meanings, for instance being larger for close meanings than for
distant ones. We ask how such pair-dependent errors change the
difficulty of estimating semantic entropy when the error probabilities themselves are known.

We model the judge as a known channel. Given the meanings $Z_1,\dots,Z_n$ of $n$ answers, drawn independently
from a distribution $\theta$ on three meanings, answers $i$ and $j$ are declared equivalent independently with
probability $q+\Delta M_{Z_iZ_j}$, where $q$ is bounded away from $0$ and $1$, and $M$ is a known symmetric matrix of scaled offsets. Two answers of meaning
$c$ are declared equivalent with probability $q+\Delta M_{cc}$, the sensitivity of the judge inside $c$, and two
answers of meanings $c\ne d$ with probability $q+\Delta M_{cd}$, its false-equivalence probability for the pair
$\{c,d\}$. The existing minimax theory for this model covers two special channels, both in the weak regime
$n\Delta^2\to0$, in which the meanings of the answers cannot be recovered uniformly over the simplex. For a homogeneous judge, $M=I$, the
minimax squared risk for three meanings is of order $\min\{1,N_3^{-1/2}\}$, with $N_v:=n^v\Delta^{2v-2}$, on
every fixed polynomial window, and the least-favorable distributions sit near a vertex of the simplex
\citep{yu2026rarity}. For a judge whose sensitivity depends on the meaning but whose false-equivalence
probability is the same for all pairs, $M=\mathrm{diag}(w)$, the rate becomes $\min\{1,N_5^{-1/5}\}$ for almost
every $w\in(0,\infty)^3$, and the difficulty moves to the interior point $\theta\propto1/w$ \citep{meaningdependent2026}. Both
channels are degenerate inside the family of symmetric matrices, and both analyses assume one
false-equivalence probability for all pairs of meanings, which the meaning-dependent analysis lists among its limitations.

\begin{figure}[t]
  \centering
  \includegraphics{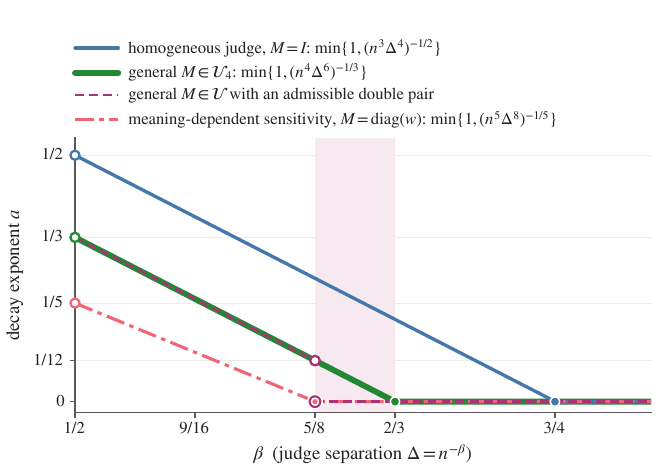}
  \caption{Polynomial decay exponent $a$ of the minimax squared risk $R^*_n=n^{-a+o(1)}$ of semantic entropy for three meanings and a known judge,
  when $\Delta=n^{-\beta}$ and $q\in[q_0,1-q_0]$ is fixed. In the shaded band $5/8<\beta<2/3$, channels in $\cU_4$ allow
  consistent estimation and channels with an admissible double pair do not, although $N_4=n^4\Delta^6\to\infty$
  for both. Blue: a homogeneous judge
  \citep{yu2026rarity}. Red, dash-dotted: a judge with meaning-dependent sensitivities and one common
  false-equivalence probability, for almost every positive sensitivity profile \citep{meaningdependent2026}. Green:
  general channels in $\cU_4$, whose tree functions on at most four vertices are injective. Purple, dashed:
  general channels in $\cU$ with an admissible double pair, two interior distributions with equal tree functions
  on at most four vertices and different entropies; their exponent jumps at
  $\beta=5/8$ from the left limit $1/12$, open circle, to the value $0$, large purple dot. Small
  filled dots mark the thresholds of the other three curves, beyond which each exponent is $0$ and the curves
  coincide; open circles at $\beta=1/2$ mark the excluded endpoint $n\Delta^2=1$. The rates hold on fixed polynomial
  windows, which cover every fixed $\beta>1/2$.}
  \label{fig:hero}
\end{figure}

For a general channel the rate has an exponent of its own. We define an explicit open class $\cU$ of symmetric
matrices and prove that for every $M$ in it the minimax squared risk is of order
\[
  N_4^{-1/3}=\big(n^4\Delta^6\big)^{-1/3}
\]
on every fixed polynomial window on which $N_5\ge C_{\log}\log n$ for a constant $C_{\log}$, and is of order
$\min\{1,N_4^{-1/3}\}$ on every fixed polynomial window when the tree functions on at most four vertices are injective,
a condition that defines an open subclass $\cU_4$ (\Cref{thm:main}). The rate is attained by an estimator that inverts
averages of products of centered verdicts along trees on at most five answers. It is slower than the rate of a
homogeneous judge and faster than that of a judge with meaning-dependent sensitivities alone (\Cref{fig:hero},
\Cref{tab:comparison}). The estimator uses only the verdicts and the known channel, and its grid version takes
$O(n^5)$ arithmetic operations. For $q$ bounded away from $0$ and $1$, the lower bound of order
$\min\{1,N_4^{-1/3}\}$ holds on every fixed polynomial window for every channel that satisfies the nondegeneracy
conditions (N), whether or not it lies in $\cU$. A further theorem
extends both parts of \Cref{thm:main}, with the same estimator, to relatively open classes of channels with equal
diagonal entries. Among them are judges with one sensitivity for every meaning, as for the homogeneous judge, and with
false-equivalence probabilities that depend on the pair, such as the certified channels $M_6,\dots,M_9$ of
\Cref{tab:certificates}.

On an open set of channels, which contains an explicit channel $M_0$ and every channel within $5\times10^{-7}$ of it
entrywise, the four tree functions on at most four vertices take the same values at two distributions with
different entropies. For such a channel in $\cU$ the minimax risk on every fixed polynomial window is of order
\[
  \min\{1,N_4^{-1/3}\}+e^{-\Theta(N_5)},
\]
that is, between $c(\min\{1,N_4^{-1/3}\}+e^{-CN_5})$ and $C(\min\{1,N_4^{-1/3}\}+e^{-cN_5})$.
The second term is the cost of choosing between the two distributions: the smallest trees whose statistics tell them
apart have five vertices, and statistics of graphs with cycles carry information of order at most $(n\Delta^2)^3$. No estimator is consistent while $N_5$ stays bounded, although $N_4\to\infty$, and the rate $N_4^{-1/3}$ returns
once $N_5$ exceeds a multiple of $\log N_4$, so the condition on $N_5$ in \Cref{thm:main} can be weakened only in its
constant. For $\Delta=n^{-\beta}$ with $\beta>1/2$ the two parts of the class therefore behave differently: a channel
in $\cU_4$ allows consistent estimation exactly for $\beta<2/3$, while for a channel with such a double pair the risk is
of order $n^{-(4/3-2\beta)}$ for $1/2<\beta<5/8$ and of constant order for $\beta\ge5/8$. The rate is set by trees on
four answers and the consistency threshold by trees on five, and the risk exponent drops from $1/12$ to $0$ at
$\beta=5/8$. Diagonal channels $\mathrm{diag}(w)$ for almost every $w\in(0,\infty)^3$ have the same threshold, at
which the exponent $1-8\beta/5$ of their rate reaches $0$ continuously \citep{meaningdependent2026}. For a channel
in $\cU$ with an admissible double pair the exponent jumps there instead, because the threshold comes from the cost
of telling the two points of the pair apart, while the exponent $4/3-2\beta$ of the four-vertex rate is still $1/12$. In a random sample of channels in $\cU$, about a quarter lie in $\cU_4$ and the others have
entropy-distinct double pairs.

\begin{table}[t]
  \centering
  \caption{Three meanings: how the structure of a known judge sets the minimax squared risk of semantic
  entropy and the consistency threshold $\beta_c$ for $\Delta=n^{-\beta}$, with $q\in[q_0,1-q_0]$ fixed and
  $N_v:=n^v\Delta^{2v-2}$. Every rate with a judge holds on fixed polynomial windows $n\Delta^2\le n^{-\kappa}$;
  in the last row, $e^{-\Theta(N_5)}$ is the cost of the double pair, with constants in the exponent that depend on
  the channel. The fourth column says where a least-favorable pair of distributions sits, and the fifth which
  trees first tell its two members apart.}
  \label{tab:comparison}
  \small
  \setlength{\tabcolsep}{4pt}
  \begin{tabular}{@{}L{4.61cm}L{3.43cm}L{1.06cm}L{2.84cm}L{2.95cm}@{}}
    \toprule
    Observation and channel & minimax squared risk & $\beta_c$ & least-favorable pair & separated by \\
    \midrule
    meanings observed & $1/n$ & none & interior & the labels \\
    homogeneous judge, $M=I$ \citep{yu2026rarity} & $\min\{1,N_3^{-1/2}\}$ & $3/4$ & near a vertex & trees on $3$ vertices \\
    meaning-dependent sensitivity, $M=\mathrm{diag}(w)$, almost every $w\in(0,\infty)^3$ \citep{meaningdependent2026} &
      $\min\{1,N_5^{-1/5}\}$ & $5/8$ & near $\theta\propto1/w$ & trees on $5$ vertices \\
    general $M\in\mathcal U_4$ & $\min\{1,N_4^{-1/3}\}$ & $2/3$ & near $\theta^*$ & trees on $4$ vertices \\
    general $M\in\mathcal U$ with an admissible double pair &
      $\min\{1,N_4^{-1/3}\}+e^{-\Theta(N_5)}$ & $5/8$ &
      near $\theta^*$ for the rate; the double pair for the threshold & trees on $4$ vertices near $\theta^*$, on $5$ at the double pair \\
    \bottomrule
  \end{tabular}
\end{table}

\paragraph{Why four vertices.}
Write $d:=M\theta$. Every tree function is stationary at the equitable point $\theta^*$ at which $d$ is
constant, $\theta^*=M^{-1}\one/\one^\top M^{-1}\one$. With $u=\sqrt{\theta^*}$ and the normalized channel
$\tM:=\diag(u)M\diag(u)$, the second-order part of every tree function at $\theta^*$ depends on a perturbation only
through the squares $x_1^2,x_2^2$ of its coordinates, taken in an eigenbasis of $\tM$ on the plane orthogonal to $u$. To second order, therefore, no tree function tells a perturbation from its reflections
in the two eigen-axes, while the entropy changes at first order because $\theta^*$ is not the uniform
distribution. The edge and wedge functions, those of the trees on two and three vertices, even take exactly the same values
at a perturbation and at a reflection of it corrected at second order, and the trees on four vertices tell the two
apart at third order. Near $\theta^*$ the entropy therefore satisfies a cubic modulus, $|\Delta H|^3\lesssim\cD_4$ for the entropy difference $\Delta H$ and the largest difference $\cD_4$ of the four-vertex tree functions between two points,
and the Hölder exponent $1/3$ cannot be improved. Unbiased tree statistics estimate the
four-vertex tree functions with variance of order $1/N_4$, which yields the rate $N_4^{-1/3}$. For a diagonal
channel, $\tM$ has a single eigenvalue on that plane, the second-order part sees only $x_1^2+x_2^2$, and more moments
can be matched \citep{meaningdependent2026}; for a homogeneous judge, $\theta^*$ is uniform and the entropy is
stationary there.

\paragraph{Techniques.}
The upper bound rests on a cubic entropy modulus that must hold across the double pairs and at three kinds of
degenerate points: the equitable point, the three equitable points of the faces and the boundary of the simplex.
The first two are handled by a scaling limit and a weighted blow-up. The double pairs are handled by giving
the five-vertex statistics the weight $(N_5/N_4)^{1/2}$ in the inversion, together with a moderate-deviation bound for
tree statistics that follows from a concentration inequality for polynomials in independent variables
\citep{adamczak2015concentration}. For the lower bound, with $M$ fixed, the edge and wedge functions determine the
distribution near the two least-favorable points, so the channel has to move along any path that keeps them constant.
We rotate the eigenbasis of $\tM$ by $\pi$, which brings the channel back to itself. The five-vertex threshold needs a
path between two points at distance of order one; it comes from the connectivity of fibers of a three-point moment
problem, crossed through collisions of atoms at two Gauss nodes. Both paths feed one information inequality for
moving general channels (\Cref{prop:info}), which extends the argument for diagonal channels
\citep{meaningdependent2026}. Along the second path the expansion behind that inequality gives the log-likelihood ratio itself,
up to an error that a stopping time controls. While $N_5$ is at most a multiple of $\log n$, under the law at one end of the path, the log-likelihood ratio of the other end is at least $-C(N_5+1)$ with probability at least $5/8$. Exact elimination and interval
arithmetic certify the conditions for explicit channels.

\paragraph{Contributions.}
\begin{enumerate}[leftmargin=*,itemsep=1pt,topsep=2pt]
  \item The minimax rate $N_4^{-1/3}$ of semantic entropy for three meanings under a general known channel, on an
    explicit open class, attained by a weighted tree-statistic estimator on every polynomial window on which $N_5\ge C_{\log}\log n$, and the rate
    $\min\{1,N_4^{-1/3}\}$ on the subclass whose four-vertex tree map is injective, attained by the same estimator on every
    polynomial window.
  \item For channels with an admissible double pair, inconsistency of every estimator while
    $N_5$ stays bounded, and for those in $\cU$ the minimax risk $\min\{1,N_4^{-1/3}\}+e^{-\Theta(N_5)}$, a five-vertex
    consistency threshold, and the resulting phase diagram for
    $\Delta=n^{-\beta}$, in which the rate is set by trees on at most four vertices and the threshold by trees on five.
  \item A cubic entropy modulus with the optimal Hölder exponent $1/3$ on the whole simplex, after
    localization with five-vertex tree functions, and without it when the four-vertex tree map is injective; an extension to channels with equal diagonal entries, among them judges with one sensitivity
    for every meaning and pair-dependent false-equivalence probabilities.
  \item Two lower-bound paths along which the channel moves, an eigenbasis rotation and a path through
    collisions of atoms at the Gauss nodes, under one information inequality, with an expansion of the
    log-likelihood ratio along the second path; a moderate-deviation bound for tree statistics; and
    computer-assisted certificates showing that both parts of the class have nonempty interior.
\end{enumerate}

The proofs are in the appendix, except for the parts of the proof of \Cref{prop:info} that are unchanged from
the argument for diagonal channels. These parts are proved as in \citet{meaningdependent2026}. The appendix also reports sampled
numerical checks of steps of the proofs.

\section{Setup}
\label{sec:setup}

\subsection{Model}

A language model is sampled $n$ times on one prompt. Answer $i$ has a meaning $Z_i\in[3]:=\{1,2,3\}$, and
$Z_1,\dots,Z_n$ are independent with distribution $\theta\in\Simplex:=\{\theta\in[0,1]^3:\sum_c\theta_c=1\}$; zero
coordinates are allowed, and $\Int$ denotes the interior of $\Simplex$. The target is the entropy
$\Ent(\theta):=-\sum_c\theta_c\log\theta_c$, with $0\log0:=0$. A judge compares every pair of answers and reports
$A_{ij}\in\{0,1\}$; given the meanings, the verdicts are independent, with
\begin{equation}
  \label{eq:model}
  A_{ij}\mid Z\ \sim\ \Bern\big(q+\Delta M_{Z_iZ_j}\big),\qquad1\le i<j\le n .
\end{equation}
The channel $(q,\Delta,M)$ is known: $M\in\Sym$, the real symmetric $3\times3$ matrices, $\Delta>0$, and
\begin{equation*}
  \tag{P}
  \label{eq:P}
  q\in[0,1]\quad\text{and}\quad q+\Delta M_{ab}\in[0,1]\quad\text{for all }a,b .
\end{equation*}
Two answers of meaning $c$ are declared equivalent with probability $q+\Delta M_{cc}$, the sensitivity of the
judge inside meaning $c$, and two answers of meanings $c\ne d$ with probability $q+\Delta M_{cd}$, its
false-equivalence probability for the pair $\{c,d\}$. The entries of $M$ are scaled offsets from $q$ and may be
negative. The homogeneous
judge is $M=I$, and the judge with meaning-dependent sensitivities is $M=\diag(w)$. We write $U:=\max_{ab}|M_{ab}|$
and $\bar U:=\max\{1,U\}$, and for a set $\mathcal K$ of channels, $\bar U_{\mathcal K}:=\max\{1,\sup_{M\in\mathcal K}\max_{ab}|M_{ab}|\}$; for one
channel, $\mathcal K=\{M\}$ and $\bar U_{\mathcal K}=\bar U$.

Only the matrix $q\one\one^\top+\Delta M$ enters \eqref{eq:model}. Adding $a\one\one^\top$ to $M$ and subtracting
$\Delta a$ from $q$ leaves every success probability, and hence the law of the verdicts, unchanged, so a rate
proved for $(q,\Delta,M)$ holds for $(q-\Delta a,\Delta,M+a\one\one^\top)$ whenever $q-\Delta a$ satisfies the
condition imposed on $q$ in the statement. For instance, $q\in[q_0,1-q_0]$ and $\Delta|a|\le q_0/2$ give
$q-\Delta a\in[q_0/2,1-q_0/2]$. The conditions below are stated for one representative $M$ and
are not invariant under such shifts: for $a=-c(M)$, with $c(M)$ defined in \Cref{sec:equitable}, the shifted matrix
annihilates the equitable point and violates (N0). The rescaling $(\Delta,M)\mapsto(\Delta/\lambda,\lambda M)$,
$\lambda>0$, also leaves \eqref{eq:model} unchanged, and every condition below is invariant under it, because each
tree function is homogeneous in $M$.

We write $\delta:=n\Delta^2$ and $N_v:=n\delta^{v-1}=n^v\Delta^{2v-2}$. The weak regime is $\delta\to0$, and a
polynomial window is $\{\delta\le n^{-\kappa}\}$ for a fixed $\kappa\in(0,1)$; for $\Delta=n^{-\beta}$ these
windows cover every fixed $\beta>1/2$. The minimax risk is
\[
  R^*_n(M):=\inf_{\hH}\sup_{\theta\in\Simplex}\Exp_\theta\big(\hH-\Ent(\theta)\big)^2,
\]
over all estimators $\hH$ computed from the verdicts $(A_{ij})_{i<j}$, the known channel and auxiliary randomness.
For positive $a,b$ we write $a\lesssim b$ if $a\le Cb$ and $a\asymp b$ if $a\lesssim b\lesssim a$, where $C$ does not
depend on $n$, $q$, $\Delta$, $\theta$ or the amplitude $s$ used in the lower bounds; the dependence of $C$ on the
channel and on fixed parameters is stated in each result.

\subsection{Tree functions}
\label{sec:trees}

\begin{figure}[t]
  \centering
  \includegraphics[width=\linewidth]{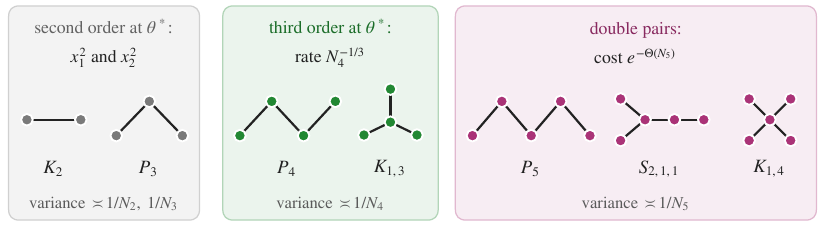}
  \caption{The seven trees on at most five vertices, grouped by the role of their tree functions. For fixed $q\in(0,1)$
  and $M$, in the weak regime, the statistic of a tree on $v$ vertices has variance of order $1/N_v$,
  and $N_{v+1}=\delta N_v$, so the statistics on more vertices are noisier. For a channel in $\cU$, the edge and wedge
  functions determine $x_1^2$ and $x_2^2$ at the equitable point to leading order, and the two trees on four vertices
  separate a perturbation from its reflections in the eigen-axes at third order,
  which is why the four-vertex statistics set the rate $N_4^{-1/3}$. The trees on five vertices separate the two points
  of an admissible double pair, at which the tree functions on at most four vertices agree. The five-vertex trees enter the risk through
  the term $e^{-\Theta(N_5)}$.}
  \label{fig:trees}
\end{figure}

Let $X_{ij}=X_{ji}:=(A_{ij}-q)/\Delta$, so that $\Exp[X_{ij}\mid Z]=M_{Z_iZ_j}$. For a tree $T$ with vertex set
$V(T)$ and edge set $E(T)$, the \emph{tree function} is
\[
  F_T(\theta;M):=\sum_{z\in[3]^{V(T)}}\ \prod_{x\in V(T)}\theta_{z_x}\prod_{xy\in E(T)}M_{z_xz_y}.
\]
For distinct answers $\varphi(x)$, $x\in V(T)$, the product of the centered verdicts $X_{\varphi(x)\varphi(y)}$ over
the edges $xy$ of $T$ has mean $F_T(\theta;M)$, so tree functions are what tree statistics
estimate. We use the seven trees with at most five vertices (\Cref{fig:trees}): the edge $K_2$, the wedge $P_3$, the path $P_4$, the
star $K_{1,3}$, the path $P_5$, the spider $S_{2,1,1}$ with legs of lengths $2,1,1$, and the star $K_{1,4}$. With
$d:=M\theta$ and $g:=\theta\odot d$ (entrywise product),
\begin{equation}
  \label{eq:closed}
  \begin{gathered}
    F_{K_2}=\sum_b\theta_bd_b,\qquad F_{P_3}=\sum_b\theta_bd_b^2,\qquad F_{P_4}=g^\top Mg,\qquad
    F_{K_{1,3}}=\sum_b\theta_bd_b^3,\\
    F_{P_5}=\sum_b\theta_b(Mg)_b^2,\qquad F_{S_{2,1,1}}=\sum_b\theta_b(Mg)_bd_b^2,\qquad F_{K_{1,4}}=\sum_b\theta_bd_b^4 .
  \end{gathered}
\end{equation}
Each follows by summing out the label of every leaf, which contributes $d_b$ when its neighbor has label $b$, and then
the label of a vertex whose remaining neighbor has label $b$, which contributes $(Mg)_b$. As polynomials in $\theta$, the
tree functions are defined on all of $\Real^3$. Let $\cT_4:=\{K_2,P_3,P_4,K_{1,3}\}$ and $\cT_5:=\cT_4\cup\{P_5,S_{2,1,1},K_{1,4}\}$, let
$F_{\le4}:=(F_T)_{T\in\cT_4}:\Simplex\to\Real^4$ and $F_{\le5}:=(F_T)_{T\in\cT_5}:\Simplex\to\Real^7$, and
\[
  \cD_v(\theta,\theta'):=\max_{T\in\cT_v}\big|F_T(\theta)-F_T(\theta')\big|,\qquad v\in\{4,5\}.
\]
We write $F_T^M$ and $\cD_v^M$ when the channel matters. Three of the four functions in $F_{\le4}$ are moments:
$F_{K_2},F_{P_3},F_{K_{1,3}}$ are the first three moments of $d_Z$ for $Z\sim\theta$, and $F_{P_4}$ is linear in $M$
for fixed $\theta$ and $d$.

\subsection{The equitable point}
\label{sec:equitable}

When $M$ is invertible and $\one^\top M^{-1}\one\ne0$, let
\[
  \theta^*:=\frac{M^{-1}\one}{\one^\top M^{-1}\one},\qquad c:=\frac1{\one^\top M^{-1}\one},\qquad\text{so that }M\theta^*=c\one .
\]
At $\theta^*$ the vector $d$ is constant; we call $\theta^*$ the \emph{equitable point}. If $\theta^*>0$ entrywise,
let $u:=\sqrt{\theta^*}$ (entrywise, so $|u|=1$), $D_u:=\diag(u)$ and $\tM:=D_uMD_u$. Then $\tM u=cu$, the plane
$u^\perp$ is invariant under $\tM$, and $\tM$ restricted to $u^\perp$ has eigenvalues $\lambda_1,\lambda_2$ with
orthonormal eigenvectors $e^\perp_1,e^\perp_2$. The map $x\mapsto\theta^*+D_ux$ is an affine isomorphism from $u^\perp$
onto the affine hull of $\Simplex$. Writing $x=x_1e^\perp_1+x_2e^\perp_2$, every tree $T$ on $v$ vertices satisfies
(\Cref{lem:a:expansion})
\begin{equation}
  \label{eq:expansion}
  \begin{gathered}
  F_T(\theta^*+D_ux)=c^{v-1}+\kappa_{T,1}x_1^2+\kappa_{T,2}x_2^2+F^{(3)}_T(x)+F^{(\ge4)}_T(x),\\
  \kappa_{T,i}:=\sum_{\{a,b\}}c^{v-1-d_T(a,b)}\lambda_i^{d_T(a,b)},
  \end{gathered}
\end{equation}
where the sum runs over unordered pairs of distinct vertices, $d_T$ is the distance in $T$, $F^{(3)}_T$ is a cubic
form and $F^{(\ge4)}_T$ collects the terms of degree at least four. Every tree function is stationary at
$\theta^*$, and its second-order part depends on $x$ only through $x_1^2$ and $x_2^2$. For the edge,
$\kappa_{K_2,i}=\lambda_i$, and for the wedge, $\kappa_{P_3,i}=\lambda_i^2+2c\lambda_i$; the edge function has no
higher terms and the wedge function has no terms of degree four or more.

\subsection{Conditions}
\label{sec:conditions}

\begin{description}[leftmargin=2.2em,style=nextline,itemsep=2pt]
  \item[(N)] (N0) $M$ is invertible and $\one^\top M^{-1}\one\ne0$; (N1) $\theta^*_c>0$ for $c=1,2,3$; (N2) $\theta^*$ is
    not the uniform distribution; (N3) $\lambda_1\ne\lambda_2$; (N4) $\lambda_1\lambda_2\ne0$.
  \item[(F0)] For each $k\in[3]$, with $\{i,j\}=[3]\setminus\{k\}$: $M_{ii}-2M_{ij}+M_{jj}\ne0$ and
    $p_k:=(M_{jj}-M_{ij})/(M_{ii}-2M_{ij}+M_{jj})\in(0,1)$. Then $f_k:=p_ke_i+(1-p_k)e_j$ lies in the open face
    $\{\theta_k=0\}$ and $(Mf_k)_i=(Mf_k)_j$; we call $f_k$ the \emph{equitable point of the face}.
  \item[(U1)] The coefficients $\kappa_{K_2,i},\kappa_{P_3,i}$, $i=1,2$, form a matrix with determinant $\lambda_1\lambda_2(\lambda_2-\lambda_1)$,
    so under (N3) and (N4) there are unique $\alpha_T,\beta_T$ with $\kappa_{T,i}=\alpha_T\kappa_{K_2,i}+\beta_T\kappa_{P_3,i}$, $i=1,2$. For $T\in\{P_4,K_{1,3}\}$ let
    $\tFthree_T:=F^{(3)}_T-\beta_TF^{(3)}_{P_3}=\sum_{j=0}^3\gamma_{T,j}x_1^{3-j}x_2^j$. Then (i) $\tFthree_{P_4}$ and
    $\tFthree_{K_{1,3}}$ have no common real zero $x\ne0$; (ii) no $(X,Y)\in[0,\infty)^2\setminus\{0\}$ satisfies
    $\gamma_{T,0}X+\gamma_{T,2}Y=0$ for both $T$; (iii) no $(X,Y)\in[0,\infty)^2\setminus\{0\}$ satisfies
    $\gamma_{T,3}X+\gamma_{T,1}Y=0$ for both $T$.
  \item[(U2)] For each face $k$, with the face coordinates $\theta_k(t,y):=f_k+t(e_i-e_j)+y\big(e_k-\tfrac12(e_i+e_j)\big)$,
    write $F_T(\theta_k(t,y))=\sum_{a,b\ge0}\phi^T_{ab}t^ay^b$ and
    $J_k:=\big(\begin{smallmatrix}\phi^{K_2}_{01}&\phi^{K_2}_{20}\\ \phi^{P_3}_{01}&\phi^{P_3}_{20}\end{smallmatrix}\big)$.
    Then (i) $\det J_k\ne0$; (ii) with $(\alpha'_T,\beta'_T)$ defined by
    $(\phi^T_{01},\phi^T_{20})=\alpha'_T(\phi^{K_2}_{01},\phi^{K_2}_{20})+\beta'_T(\phi^{P_3}_{01},\phi^{P_3}_{20})$ and
    $\psi^T_{ab}:=\phi^T_{ab}-\alpha'_T\phi^{K_2}_{ab}-\beta'_T\phi^{P_3}_{ab}$, no $(X,Y)\in[0,\infty)^2\setminus\{0\}$
    satisfies $\psi^T_{11}X+\psi^T_{30}Y=0$ for both $T\in\{P_4,K_{1,3}\}$.
  \item[(U3)] For every $\theta\in\Simplex\setminus\{\theta^*,f_1,f_2,f_3\}$, the differential of $F_{\le4}$ at $\theta$,
    restricted to the tangent plane $\one^\perp$, has rank $2$.
  \item[(U4)] $F_{\le5}$ is injective on $\Simplex$.
\end{description}

Let $\Omega$ be the set of $M\in\Sym$ satisfying (N), let $\cU$ be the set of $M\in\Sym$ satisfying (N), (F0) and
(U1) to (U4), and let $\cU_4:=\{M\in\cU:F^M_{\le4}\text{ is injective on }\Simplex\}$.

The conditions describe the geometry of $F_{\le4}$. By (U3) its only degenerate points in the closed simplex
are $\theta^*$ and the three face equitable points. At $\theta^*$, \eqref{eq:expansion} shows that the edge and
wedge functions determine $(x_1^2,x_2^2)$ to leading order; (U1) says that the reduced cubic forms of the two
four-vertex trees separate every $x\ne0$ from its three reflections $(-x_1,x_2)$, $(x_1,-x_2)$ and $-x$, to first
order on the eigen-axes. At a face equitable point every tree function has zero derivative along the face
(\Cref{lem:a:face}), the edge and wedge functions determine $(y,t^2)$ to leading order, and (U2) asks the
four-vertex trees to separate $t$ from $-t$ at the weighted cubic order $ty$, $t^3$. Finally (U4) says that the
tree functions on at most five vertices identify $\theta$. The conditions do not depend on the signs chosen for
$e^\perp_1,e^\perp_2$ (\Cref{app:modulus}). Conditions (N2) to (N4) exclude the homogeneous judge, for which $\theta^*$ is
uniform, and diagonal channels, for which $\tM=\mu I$ on $u^\perp$. Two equal diagonal entries make (U2) and (U4)
fail exactly; a later theorem treats these channels separately.

\needspace{6\baselineskip}
\section{Main results}
\label{sec:results}

\begin{theorem}[minimax rate]
  \label{thm:main}
  Let $q_0\in(0,1/2]$ and $\kappa\in(0,1)$.
  \begin{enumerate}[label=(\alph*),leftmargin=*]
    \item Let $\mathcal K\subset\cU$ be compact. There are $c_A>0$, $C_A$, $C_{\log}$ and $n_A$ depending only on
      $(\mathcal K,q_0,\kappa)$ such that for every $M\in\mathcal K$, $n\ge n_A$, $q\in[q_0,1-q_0]$ and $\Delta>0$ with
      \[
        n\Delta^2\le n^{-\kappa}\qquad\text{and}\qquad N_5\ge C_{\log}\log n,
      \]
      the estimator $\hH_n$ of \Cref{alg:estimator} satisfies
      \[
        c_AN_4^{-1/3}\le R^*_n(M)\le\sup_{\theta\in\Simplex}\Exp_\theta\big(\hH_n-\Ent(\theta)\big)^2\le C_AN_4^{-1/3}.
      \]
      For large $n$, such $\Delta$ exist if and only if $\kappa<1/4$.
    \item Let $\mathcal K\subset\cU_4$ be compact. There are $c'_A>0$, $C'_A$ and $n'_A$ depending only on
      $(\mathcal K,q_0,\kappa)$ such that for every $M\in\mathcal K$, $n\ge n'_A$, $q\in[q_0,1-q_0]$ and $\Delta>0$ with
      $n\Delta^2\le n^{-\kappa}$, the same estimator satisfies
      \[
        c'_A\min\{1,N_4^{-1/3}\}\le R^*_n(M)\le\sup_{\theta\in\Simplex}\Exp_\theta\big(\hH_n-\Ent(\theta)\big)^2\le
        C'_A\min\{1,N_4^{-1/3}\}.
      \]
  \end{enumerate}
\end{theorem}

The upper and lower bounds come from different results. The upper bounds hold under
\eqref{eq:P} alone, for every $n\ge8$, with no window and no $q_0$: on compact subsets of $\cU$ the risk of $\hH_n$ is at most $C_U(N_4^{-1/3}+e^{-c_UN_5})$ when
$\delta\le1$, and on compact subsets of $\cU_4$ it is at most a constant times $\max\{1/n,1/N_4\}^{1/3}$
(\Cref{thm:upper}). The lower bound $R^*_n(M)\ge c_L\min\{1,N_4^{-1/3}\}$ holds uniformly over every compact set of
channels satisfying (N), whether or not they lie in $\cU$ (\Cref{thm:lower}). The constants come from compactness, and
$C_{\log}=1/(3c_U)$. The window of part (a) is the set $\{(C_{\log}\log n/n)^{1/4}\le\delta\le n^{-\kappa}\}$, on which
$N_4^{-1/3}$ ranges from $n^{-(1-3\kappa)/3}$ to $n^{-1/12}(C_{\log}\log n)^{-1/4}$. The bounds of part (a) also hold, with
the same constants, when $N_5\ge C_{\log}\log n$ is replaced by $N_4\ge1$ and $N_5\ge C_{\log}\log N_4$, since then
$e^{-c_UN_5}\le N_4^{-1/3}$ and $\min\{1,N_4^{-1/3}\}=N_4^{-1/3}$.

The condition on $N_5$ in part (a) is needed, up to the value of $C_{\log}$, for the channels in $\cU\setminus\cU_4$
that the next definition describes.

\begin{definition}[admissible double pair]
  \label{def:admissible}
  An \emph{admissible double pair} of $M\in\Sym$ is a pair of points $\theta^0\ne\theta^1$ of $\Int$ with
  $F^M_{\le4}(\theta^0)=F^M_{\le4}(\theta^1)$ and $\Ent(\theta^0)\ne\Ent(\theta^1)$, such that the entries of $M\theta^0$
  are pairwise distinct and the entries of $M\theta^1$ are pairwise distinct. Its entropy gap is
  $\omega:=|\Ent(\theta^0)-\Ent(\theta^1)|$.
\end{definition}

For such a pair the variance $v:=F_{P_3}(\theta^0)-F_{K_2}(\theta^0)^2=\sum_a\theta^0_a(d^0_a-F_{K_2}(\theta^0))^2$ of
$d^0_Z=(M\theta^0)_Z$ under $Z\sim\theta^0$ is positive, because $\theta^0$ is interior and $d^0$ is not constant; it
takes the same value at $\theta^1$.

\begin{theorem}[five-vertex threshold]
  \label{thm:threshold}
  Let $(\theta^0,\theta^1)$ be an admissible double pair of $M\in\Sym$ and $q_0\in(0,1/2]$.
  \begin{enumerate}[label=(\alph*),leftmargin=*]
    \item There are $c_1>0$ and $C_1$, depending on $M$, the pair and $q_0$, and for every $c'>0$ there is $n_1$,
      depending also on $c'$, such that
      \[
        R^*_n(M)\ge c_1e^{-C_1N_5}\quad\text{whenever}\quad n\ge n_1,\quad q\in[q_0,1-q_0]\quad\text{and}\quad
        N_5=n^5\Delta^8\le c'\log n .
      \]
    \item If $M\in\cU$ and $\kappa\in(0,1)$, let $c_U$ be the constant of \Cref{thm:upper}(a) for $\mathcal K=\{M\}$.
      There are $c_T>0$, $C_T$, $C_E$ and $n_T$, depending on $M$, the pair, $q_0$ and $\kappa$, such that for $n\ge n_T$,
      $q\in[q_0,1-q_0]$ and $n\Delta^2\le n^{-\kappa}$,
      \[
        c_T\Big(\min\{1,N_4^{-1/3}\}+e^{-C_EN_5}\Big)\le R^*_n(M)\le C_T\Big(\min\{1,N_4^{-1/3}\}+e^{-c_UN_5}\Big).
      \]
  \end{enumerate}
\end{theorem}

The channel need not belong to $\cU$ in part (a), and the lower bound of (b) holds for every $M\in\Omega$ with an
admissible double pair, because its proof uses only part (a) and \Cref{thm:lower}. Since $N_4=N_5/\delta$, the quantity $N_4$ can
tend to infinity while $N_5$ stays bounded: with $N_5=1$, $\delta=n^{-1/4}$ and $N_4=n^{1/4}$. By part (a) no estimator
is consistent along such a sequence, and by part (b) the risk of a channel in $\cU$ with an admissible double pair tends to zero if and only if
$N_5\to\infty$. The second term in (b) is the cost of telling the two points of the pair apart. The smallest trees whose functions
separate them have five vertices; graphs with cycles may separate them as well, but the information that they carry is
of order at most $\delta^3$, which tends to zero on every polynomial window. For $N_4\ge1$ the rate $N_4^{-1/3}$ holds once $N_5\ge\log N_4/(3c_U)$, and the risk exceeds it by a
power of $N_4$ once $N_5\le a\log N_4$ with $a<1/(3C_E)$. The constants $c_U$ and $C_E$ are not identified and need
not agree, so the crossover between the two terms is located at $N_5\asymp\log N_4$ only up to a constant factor.

\begin{figure}[t]
  \centering
  \includegraphics[width=\linewidth]{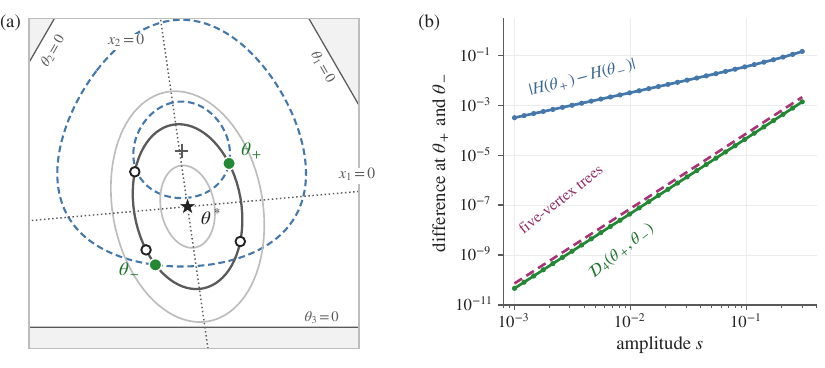}
  \caption{The two points of the lower bound at the equitable point of the channel $M_0$. (a) Near $\theta^*$ (star), the
  level curves of the edge function (gray), $F_{K_2}(\theta^*+D_ux)=c+\lambda_1x_1^2+\lambda_2x_2^2$, are ellipses
  centered at $\theta^*$; the dotted lines are the eigen-axes $x_1=0$ and $x_2=0$ of $\tM$, the dashed curves the level
  curves of the entropy through $\theta_+$ and $\theta_-$, the plus sign marks the uniform distribution, and the shaded
  regions lie outside the simplex. A perturbation $x=s\zeta_0$, here with $s=1/4$, and its three reflections in the
  eigen-axes (circles) lie on one ellipse, the darker one, while the entropy separates $x$ from its reflection $-x$ at
  first order. The filled points are $\theta_+=\theta^*+sD_u\zeta_0$ and $\theta_-$, the reflection corrected at second
  order, at which the edge and wedge functions take their values at $\theta_+$. (b) The same
  construction for $31$ amplitudes $s$ from $10^{-3}$ to $0.3$: the entropy gap grows like $s$, and
  $\cD_4(\theta_+,\theta_-)$ and the largest difference of the five-vertex tree functions grow like $s^3$, with fitted
  log-log slopes $1.00$, $3.00$ and $3.00$ for $s\le0.01$; the edge and wedge functions agree to within $10^{-45}$. So
  $|\Ent(\theta_+)-\Ent(\theta_-)|^3\asymp\cD_4(\theta_+,\theta_-)$, as in the cubic modulus, and
  at $s\asymp N_4^{-1/6}$ the difference $\cD_4$ is of the same order $N_4^{-1/2}$ as the errors of the four-vertex
  statistics, while the squared entropy gap is of order $N_4^{-1/3}$.}
  \label{fig:equitable}
\end{figure}

\begin{corollary}[phase diagram for power laws]
  \label{cor:phase}
  Let $q_0\in(0,1/2]$, $q_n\in[q_0,1-q_0]$ and $\Delta_n=n^{-\beta}$ with $\beta>1/2$.
  \begin{enumerate}[label=(\roman*),leftmargin=*]
    \item If $M\in\cU_4$, then $R^*_n(M)\asymp n^{-(4/3-2\beta)}$ for $\beta<2/3$ and $R^*_n(M)\asymp1$ for $\beta\ge2/3$.
    \item If $M\in\cU$ has an admissible double pair, then $R^*_n(M)\asymp n^{-(4/3-2\beta)}$ for $\beta<5/8$ and
      $R^*_n(M)\asymp1$ for $\beta\ge5/8$, although $N_4=n^{4-6\beta}\to\infty$ for $\beta<2/3$.
  \end{enumerate}
  Hence the exponent $a(\beta):=\lim_n-\log R^*_n(M)/\log n$ exists for every $\beta>1/2$ in both cases. In case (ii),
  $a(\beta)\to1/12$ as $\beta\uparrow5/8$ and $a(\beta)=0$ for $\beta\ge5/8$.
\end{corollary}

It follows from \Cref{thm:main,thm:threshold}.
So the two parts of $\cU$ have different consistency thresholds, $2/3$ and $5/8$, and on the second part the rate
below the threshold is set by trees on four vertices, while the threshold is set by trees on five. At the threshold
itself the risk is of constant order. Diagonal channels $\diag(w)$ for almost every $w\in(0,\infty)^3$ have the threshold $5/8$ as
well, with the slower rate $n^{-(1-8\beta/5)}$ below it \citep{meaningdependent2026}.

Both parts of the class have nonempty interior, with explicit channels in each. Let
\[
  M_0:=\frac1{20}\begin{pmatrix}20&-6&1\\-6&17&4\\1&4&15\end{pmatrix},\quad
  M_1:=\frac1{20}\begin{pmatrix}21&-4&-2\\-4&25&6\\-2&6&26\end{pmatrix},\quad
  M_2:=\frac1{20}\begin{pmatrix}27&6&1\\6&23&-4\\1&-4&14\end{pmatrix}.
\]

\begin{proposition}[the condition class]
  \label{prop:class}
  \begin{enumerate}[label=(\alph*),leftmargin=*]
    \item $\cU$ and $\cU_4$ are open subsets of $\Sym$.
    \item $M_0\in\cU\setminus\cU_4$. The map $F^{M_0}_{\le4}$ takes equal values at exactly one pair of distinct points of
      $\Simplex$; this pair is an admissible double pair with entropy gap $0.1709\ldots$, and $\cD_5\ge1.98472\times10^{-3}$
      on it.
    \item $M_1,M_2\in\cU_4$.
    \item Let $P_r:=\{M\in\Sym:\max_{ab}|M_{ab}-(M_0)_{ab}|\le r\}$ with $r=5\times10^{-7}$. Every $M\in P_r$ has an
      admissible double pair with entropy gap at least $0.16651$. Hence $P_r\cap\cU_4=\emptyset$, the interior of
      $\cU\setminus\cU_4$ contains a neighborhood of $M_0$, and \Cref{cor:phase}(ii) holds for every channel in the open
      set $\operatorname{int}(P_r)\cap\cU$.
    \item The box $\{M\in\Sym:\max_{ab}|M_{ab}-(M_0)_{ab}|\le0.01\}$ is contained in $\Omega$.
  \end{enumerate}
\end{proposition}

Part (a) is proved in the appendix, and parts (b) to (e) are computer-assisted. The three neighborhoods of $M_0$ in the
proposition serve different purposes: the box of radius $0.01$ is certified to satisfy (N), so the lower bound of
\Cref{thm:lower} holds uniformly on it; the box $P_r$ is certified to carry admissible double pairs but is not
certified to lie in $\cU$; and openness gives an unspecified neighborhood of $M_0$ inside $\cU$, so that $P_r\cap\cU$
contains a neighborhood of $M_0$.

Channels with two equal diagonal entries are not in $\cU$: if $M_{ii}=M_{jj}$, the restriction of the model to the
face $\{\theta_k=0\}$ is symmetric under exchanging $i$ and $j$, every pair of points of that face exchanged by the
symmetry is an exact double point of every tree function, and (U2) and (U4) fail. The two points of such a pair
have equal entropy, so the entropy can still be estimated at the same rate. For
$\cJ\in\{\{1\},\{2\},\{3\},\{1,2,3\}\}$ let $\cS_\cJ$ be the linear subspace of $M\in\Sym$ with $M_{ii}=M_{jj}$ for every
$k\in\cJ$, where $\{i,j\}=[3]\setminus\{k\}$; two equalities imply the third, so these are all the patterns.

\begin{theorem}[equal diagonal entries]
  \label{thm:symmetric}
  For each $\cJ$ let $\cU^\sigma_\cJ\supset\cU^\sigma_{\cJ,4}$ be the classes of \Cref{app:symmetric}: the channels $M\in\cS_\cJ$
  whose diagonal entries satisfy no equality beyond those of $\cJ$ and that satisfy (N), (F0) and (U1); (U2) on the faces
  $k\notin\cJ$; a modified local condition and a transversality condition on the faces $k\in\cJ$; (U3); and injectivity
  of $F_{\le5}$, respectively $F_{\le4}$, up to exchanging symmetric pairs. Both classes are relatively open in $\cS_\cJ$.
  \Cref{thm:modulus}(a) holds with $\cU$ replaced by $\cU^\sigma_\cJ$, and \Cref{thm:modulus}(b) with $\cU_4$ replaced by
  $\cU^\sigma_{\cJ,4}$. \Cref{thm:main}(a) and (b) hold with the same replacements and the same estimator. The channels
  $M_3,\dots,M_9$ of \Cref{tab:certificates} belong to the classes listed there; in particular the submodel with a common within-meaning sensitivity,
  $M_{11}=M_{22}=M_{33}$, contains channels in $\cU^\sigma_{[3],4}$ and channels in $\cU^\sigma_{[3]}\setminus\cU^\sigma_{[3],4}$.
\end{theorem}

\paragraph{Where the difficulty sits.}
The rate comes from pairs near $\theta^*$, at which every meaning has probability bounded away from zero. The
lower bound uses $\theta_\pm\approx\theta^*\pm sD_u\zeta_0$ with $|\Ent(\theta_+)-\Ent(\theta_-)|\asymp s$ and
$\cD_4(\theta_+,\theta_-)=O(s^3)$ (\Cref{fig:equitable}), and the amplitude $s\asymp\min\{1,N_4^{-1/6}\}$ makes the two laws
indistinguishable. The threshold comes from a double pair, two points at
distance of order one that the tree functions on at most four vertices cannot tell apart at all.

\section{The upper bound}
\label{sec:upper}

\subsection{Tree statistics}

For a tree $T$ on the vertex set $[v]$ and a set $S$ of $m\ge v$ answers, let $\mathrm{Inj}(v,S)$ be the set of
injective maps $\varphi:[v]\to S$, of size $(m)_v:=m(m-1)\cdots(m-v+1)$, and
\[
  \widehat U_T[S]:=\frac1{(m)_v}\sum_{\varphi\in\mathrm{Inj}(v,S)}\ \prod_{xy\in E(T)}X_{\varphi(x)\varphi(y)},\qquad
  \hF_T:=\widehat U_T[[n]] .
\]
Let $N_v(m):=m^v\Delta^{2v-2}$, so that $N_v=N_v(n)$.

\begin{lemma}[tree statistics]
  \label{lem:ustat}
  Assume \eqref{eq:P}, let $T$ be a tree on $v\ge2$ vertices and $|S|=m\ge v$. Then
  $\Exp\,\widehat U_T[S]=F_T(\theta;M)$. If $m\ge2v-2$,
  \[
    \Var\big(\widehat U_T[S]\big)\le B_v\max\Big\{\frac1m,\frac1{N_v(m)}\Big\},\qquad
    B_v:=2^{2v+1}\,v!\,(5/4)^{v-1}\,\bar U^{2v-2}.
  \]
  For fixed $q\in(0,1)$ and $M$, with $\Delta\to0$ and $m\Delta^2\to0$, uniformly in $\theta\in\Simplex$, the variance is asymptotic to
  $|\mathrm{Aut}(T)|\,(q(1-q))^{v-1}/((m)_v\Delta^{2v-2})$, so the order $1/N_v(m)$ is attained.
\end{lemma}

The proof needs only $|\Exp[X_e\mid Z]|\le U$, conditional independence of the verdicts, and
the fact that two embedded copies of a tree share a forest of edges; it holds for any tree and any symmetric $M$,
including negative entries.

\begin{lemma}[moderate deviations]
  \label{lem:md}
  Assume \eqref{eq:P}, $\delta\le1$ and $n\ge8$, and let $T\in\cT_5$ have $v$ vertices. There is $c_C>0$, depending
  only on $\bar U$ and nonincreasing in it, such that for all $\theta\in\Simplex$ and $x>0$,
  \[
    \Prob_\theta\Big(|\hF_T-F_T(\theta)|\ge\frac x{\sqrt{N_v}}\Big)\le4\exp\big(-c_C\min\{x^2,n^{1/3}\}\big).
  \]
\end{lemma}

So the standardized deviations have Gaussian tails up to $x$ of order $n^{1/6}$. Given the labels, $\hF_T$ is a
multilinear polynomial of degree $v-1$ in the independent centered variables $A_{ij}-q-\Delta M_{Z_iZ_j}$, and the
concentration inequality of \citet{adamczak2015concentration} for polynomials in independent sub-Gaussian variables
bounds its tails, through the norms of the expected derivative tensors with respect to partitions of their indices.
The Hilbert--Schmidt norms give the Gaussian term. Every partition into two or more blocks gains a factor $n^{-1/2}$
over it, by the Schur test: two blocks of edges of a subtree share a vertex, and joining the components of a
disconnected forest costs at least one factor $\Delta^{-2}\ge n$. The labels enter through a U-statistic with bounded
differences \citep{mcdiarmid1989method}. The proof is in the appendix.

\subsection{The estimator}

\begin{algorithm}[t]
  \caption{Tree-statistic estimator of semantic entropy}
  \label{alg:estimator}
  \begin{algorithmic}[1]
    \REQUIRE verdicts $(A_{ij})_{i<j}$ on $n\ge5$ answers and the channel $(q,\Delta,M)$
    \STATE $X_{ij}\leftarrow(A_{ij}-q)/\Delta$; $\hF_T\leftarrow\widehat U_T[[n]]$ for $T\in\cT_5$;
      $\lambda\leftarrow\min\{1,\sqrt n\,\Delta\}$
    \STATE $\Phi_w(\vartheta):=\max\big\{\max_{T\in\cT_4}|F_T(\vartheta;M)-\hF_T|,\
      \lambda\max_{T\in\cT_5\setminus\cT_4}|F_T(\vartheta;M)-\hF_T|\big\}$
    \STATE \textbf{return} $\hH_n:=\Ent(\htheta)$ for a minimizer $\htheta$ of $\Phi_w$ over $\Simplex$
  \end{algorithmic}
\end{algorithm}

The estimator uses only the observations and the known channel; it uses neither the constants of the
analysis nor the location of $\theta$. Minimizers exist because $\Phi_w$ is continuous on the compact simplex, and
every selection rule gives an estimator because the observation space is finite. The grid version minimizes $\Phi_w$
over the grid $\Simplex\cap(\mathbb Z/n)^3$ and takes the lexicographically first minimizer; it can be computed with
$O(n^5)$ arithmetic operations, enumerating injective maps for the statistics and evaluating the closed forms
\eqref{eq:closed} on the grid.

The two groups of trees play different roles. The statistics on at most four vertices carry the rate; their errors
are of order $N_4^{-1/2}$ or smaller. The five-vertex statistics rule out the far member of a double pair, which the
four-vertex statistics alone cannot exclude. In the weak regime their weight is $\lambda=\sqrt\delta=(N_5/N_4)^{1/2}$,
which puts their contribution on the same scale $N_4^{-1/2}$ as the four-vertex errors. A minimizer then lies within a
fixed distance of $\theta$ in all seven tree functions, unless some statistic deviates by a multiple of $\sqrt{N_5}$
standard deviations, which by \Cref{lem:md} has probability at most of order $e^{-cN_5}+e^{-cn^{1/3}}$. The second term is smaller than $N_4^{-1/3}$, since $N_4\le n$ when $\delta\le1$. When $F_{\le4}$ is injective there
is nothing to rule out, and the five-vertex terms cost only a constant factor.

\subsection{The cubic entropy modulus}

\begin{theorem}[cubic entropy modulus]
  \label{thm:modulus}
  \begin{enumerate}[label=(\alph*),leftmargin=*]
    \item For every compact $\mathcal K\subset\cU$ there are $\eta>0$ and $C<\infty$ such that for all $M\in\mathcal K$ and
      $\theta,\theta'\in\Simplex$,
      \[
        \cD^M_5(\theta,\theta')\le\eta\quad\Longrightarrow\quad\big|\Ent(\theta)-\Ent(\theta')\big|^3\le C\,\cD^M_4(\theta,\theta').
      \]
    \item For every compact $\mathcal K\subset\cU_4$ there is $C_4<\infty$ such that
      $|\Ent(\theta)-\Ent(\theta')|^3\le C_4\,\cD^M_4(\theta,\theta')$ for all $M\in\mathcal K$ and $\theta,\theta'\in\Simplex$.
    \item If $F^M_{\le4}$ takes equal values at two points with different entropies, as for an admissible double pair,
      the inequality of (b) fails for $M$. For every $M\in\Omega$, every $\eta>0$ and every $p\in(0,3)$, the supremum of
      $|\Ent(\theta)-\Ent(\theta')|^p/\cD^M_4(\theta,\theta')$ over pairs with $\cD^M_5(\theta,\theta')\le\eta$ is infinite, so the
      Hölder exponent $1/3$ in (a) and (b) is optimal.
  \end{enumerate}
\end{theorem}

The constants $\eta$, $C$ and $C_4$ come from compactness, and
part (c) follows from (b) and from the two points of the lower bound, which satisfy $|\Delta\Ent|\asymp s$ and
$\cD_4=O(s^3)$.

\paragraph{Proof idea.}
Near pairs, $|\theta-\theta'|<\delta_0$, are handled by a covering argument and three local lemmas, each of which
gives a lower bound on $\cD_4$ strong enough to control the entropy; far pairs are handled by (U4) and
compactness. (1) At points where $F_{\le4}$ has rank two, which by (U3) is everywhere except at $\theta^*$ and the
$f_k$, an inverse-function bound gives $\cD_4\gtrsim|\theta-\theta'|$, and the logarithmic modulus of continuity of
the entropy at the boundary is absorbed by the cube: $|\Delta\Ent|^3\lesssim|\theta-\theta'|^3\log^3(1/|\theta-\theta'|)
\lesssim|\theta-\theta'|$. (2) At $\theta^*$, write $x=s\zeta$ with $|\zeta|\le1$ and transform $F_{\le4}$ linearly
into $(X,G)$, where $X$ is the pair of edge and wedge functions solved for $(x_1^2,x_2^2)$ and $G$ consists of the
four-vertex functions with their second-order parts removed. By homogeneity the rescaled map
$\Phi_s(\zeta):=(s^{-2}X(s\zeta),s^{-3}G(s\zeta))$ equals $\Phi_0+s\Psi$ exactly, with
\[
  \Phi_0(\zeta)=\big(\zeta_1^2,\ \zeta_2^2,\ \tFthree_{P_4}(\zeta),\ \tFthree_{K_{1,3}}(\zeta)\big)
\]
and $\Psi$ a fixed polynomial map. Condition (U1) makes $\Phi_0$ injective on pairs with $\max\{|\zeta|,|\zeta'|\}=1$
and an immersion on the unit circle, so $|\Phi_0(\zeta)-\Phi_0(\zeta')|\ge c_0|\zeta-\zeta'|$ there, and for small $s$
the same holds for $\Phi_s$. Scaling back gives $\cD_4\gtrsim s^2|\theta-\theta'|$ with $s\ge\max\{|\theta-\theta^*|,
|\theta'-\theta^*|\}$, and the entropy is Lipschitz near $\theta^*$, so $|\Delta\Ent|^3\lesssim s^2|\theta-\theta'|
\lesssim\cD_4$. (3) At a face equitable point every tree function is stationary along the face, so the tangential
coordinate $t$ enters at second order and the normal coordinate $y$ at first order. The weighted rescaling
$t=s\tau$, $y=s^2\upsilon$ produces a limit map $(\upsilon,\tau^2,\psi^{P_4}_{11}\tau\upsilon+\psi^{P_4}_{30}\tau^3,
\psi^{K_{1,3}}_{11}\tau\upsilon+\psi^{K_{1,3}}_{30}\tau^3)$ that (U2) makes injective and immersive, whence
$\cD_4\gtrsim\max\{s^2|t-t'|,s|y-y'|\}$. The entropy has a term $-y\log y$ in the normal direction; its logarithm
is absorbed by the extra power of $s$, because $|y-y'|\le s^2$. A pure Hölder bound on the parameters would not
suffice here. The constants of all three lemmas are uniform for channels near a given one, which also gives
openness of $\cU$.

For pairs at distance at least $\delta_0$, the set of $(M,\theta,\theta')$ with $\cD_4=0$ is compact, and by (U4)
$\cD_5$ has a positive minimum $2\eta$ on it; on the compact set where $\cD_5\le\eta$, the quantity $\cD_4$ is bounded below and the
entropy difference is at most $\log3$. On $\cU_4$ the set with $\cD_4=0$ is empty, which gives (b).

\subsection{Upper bounds}

\begin{theorem}[upper bounds]
  \label{thm:upper}
  Assume \eqref{eq:P} and $n\ge8$.
  \begin{enumerate}[label=(\alph*),leftmargin=*]
    \item For every compact $\mathcal K\subset\cU$ there are $c_U>0$ and $C_U$ depending only on $\mathcal K$ such that for all
      $M\in\mathcal K$, $\Delta$ with $\delta\le1$, and $\theta\in\Simplex$,
      $\Exp_\theta(\hH_n-\Ent(\theta))^2\le C_U\big(N_4^{-1/3}+e^{-c_UN_5}\big)$.
    \item For every compact $\mathcal K\subset\cU_4$, with $C_4$ from \Cref{thm:modulus}(b) and $B_4,B_5$ from
      \Cref{lem:ustat} for $\bar U=\bar U_{\mathcal K}$, for all $M\in\mathcal K$, $\Delta>0$ and $\theta\in\Simplex$,
      \[
        \Exp_\theta\big(\hH_n-\Ent(\theta)\big)^2\le\min\Big\{(\log3)^2,\ (2C_4)^{2/3}(4B_4+3B_5)^{1/3}\max\{1/n,1/N_4\}^{1/3}\Big\}.
      \]
  \end{enumerate}
  The grid version satisfies the same bounds with other constants depending only on $\mathcal K$.
\end{theorem}

In the weak regime $\delta\le1$ we have $N_4\le n$, and the bound of (b) is a constant times $N_4^{-1/3}$. The proof is
short given \Cref{thm:modulus,lem:ustat,lem:md}. With $\hat\eps_4:=\max_{T\in\cT_4}|\hF_T-F_T(\theta)|$ and $\hat\eps_5$
the same maximum over the five-vertex trees, $\Phi_w(\theta)=\max\{\hat\eps_4,\lambda\hat\eps_5\}$; minimality gives
$\cD_4(\htheta,\theta)\le2\Phi_w(\theta)$ and $\cD_5(\htheta,\theta)\le2\Phi_w(\theta)/\lambda$, and \Cref{lem:ustat} gives
$\Exp\Phi_w(\theta)^2\le(4B_4+3B_5)\max\{1/n,1/N_4\}$. In part (a), on the event $\{\Phi_w(\theta)\le\lambda\eta/2\}$, the
modulus applies and gives $|\hH_n-\Ent(\theta)|^3\le2C\Phi_w(\theta)$. The event fails only if the standardized deviation
$\sqrt{N_v}|\hF_T-F_T(\theta)|$ of some statistic is at least $\eta\sqrt{N_5}/2$, which by \Cref{lem:md} has probability
at most $28\exp(-c_C\min\{\eta^2N_5/4,n^{1/3}\})$. Part (b) needs no event. The full proof, including the grid
version, is in the appendix.

\section{The lower bound}
\label{sec:lower}

For a graph $G$ and $(\theta,M)$ let $\hom(G;\theta,M):=\sum_z\prod_{x\in V(G)}\theta_{z_x}\prod_{xy\in E(G)}M_{z_xz_y}$, so
that $F_T=\hom(T;\cdot,\cdot)$ for a tree $T$.

\paragraph{Two points.}
Let $M\in\Omega$. The entropy has gradient $-(\log\theta+\one)$, so along
$s\mapsto\theta^*+sD_u\zeta$ with $\zeta\in u^\perp$ its derivative at $s=0$ is
$\ell(\zeta):=-\sum_cu_c\zeta_c\log\theta^*_c$. This linear form does not vanish on $u^\perp$: otherwise
$u\odot\log\theta^*$ would be parallel to $u$, $\log\theta^*_c$ would not depend on $c$, and $\theta^*$ would be uniform,
against (N2). Take $\zeta_0:=(e^\perp_1+\sigma e^\perp_2)/\sqrt2$ with $\sigma\in\{\pm1\}$ such that
$|\ell(\zeta_0)|\ge\|\ell|_{u^\perp}\|/\sqrt2>0$; such a $\sigma$ exists because
$\max_\sigma|\ell(e^\perp_1)+\sigma\ell(e^\perp_2)|=|\ell(e^\perp_1)|+|\ell(e^\perp_2)|\ge\|\ell|_{u^\perp}\|$. The two points of the lower bound are
\[
  \theta_+:=\theta^*+sD_u\zeta_0,\qquad\theta_-:=\theta^*-sD_u\big(\zeta_0+s\rho(1,s)\big),
\]
with a correction $\rho(1,s)=O(1)$ in $u^\perp$ fixed below; their entropies differ by $2s\ell(\zeta_0)+O(s^2)$. By
\eqref{eq:expansion}, a perturbation $x$ and its reflection $-x$ have the same second-order parts of every tree
function, and the correction makes the edge and wedge functions agree exactly. Under the fixed channel $M$ the
two laws are therefore close, but the path argument needs more.

\paragraph{Why the channel moves.}
To bound the distance between the two laws, we use a path of laws from one to the other along which the Fisher
information is small. The path keeps the edge and wedge functions exactly constant, which removes the leading terms
of the information; with $M$ fixed, no such path leaves $\theta_+$.

\begin{lemma}[the channel must move]
  \label{lem:move}
  Let $M\in\Omega$. In the eigencoordinates of $u^\perp$, the differential of $x\mapsto(F^M_{K_2},F^M_{P_3})(\theta^*+D_ux)$
  at $x=s\zeta_0$ has determinant $4s^2\lambda_1\lambda_2(\lambda_2-\lambda_1)\zeta_{0,1}\zeta_{0,2}+O(s^3)$, which is
  nonzero for small $s>0$. Hence no nonconstant continuous path $t\mapsto\theta_t\in\Simplex$ with $\theta_0=\theta_+$
  keeps $F^M_{K_2}(\theta_t)$ and $F^M_{P_3}(\theta_t)$ constant.
\end{lemma}

Only the two endpoints enter the two-point bound, so the intermediate
laws may leave the experiment with channel $M$; the path below moves the channel and returns to it.

\paragraph{The rotation path.}
Let $R_t:=uu^\top+\cos(\pi t)(e^\perp_1e^{\perp\top}_1+e^\perp_2e^{\perp\top}_2)+\sin(\pi t)(e^\perp_2e^{\perp\top}_1-e^\perp_1e^{\perp\top}_2)$
be the rotation by the angle $\pi t$ in $u^\perp$ that fixes $u$, and
\[
  \tM_t:=R_t\tM R_t^\top,\qquad M_t:=D_u^{-1}\tM_tD_u^{-1},\qquad\theta_t:=D_u\big(u+sR_t(\zeta_0+s\rho(t,s))\big).
\]
Then $M_t\theta^*=c\one$ for all $t$, and at $t=1$ the rotation is $-I$ on $u^\perp$, so the path returns to $\tM$ and to $M$. The
two vertices of a tree that carry the perturbation contribute $c^{v-1-d}\zeta^\top\tM^d\zeta$, which is invariant
under rotating $\zeta$ and $\tM$ together (\Cref{lem:a:expansion}); the correction $\rho$ solves two equations
that keep the edge and wedge functions constant.

\begin{lemma}[rotation path]
  \label{lem:rotation}
  Let $\mathcal K\subset\Omega$ be compact and $r\ge4$. There are $s_P>0$ and $L\ge1$ such that for every $M\in\mathcal K$ and
  $0<s\le s_P$ there is $\rho(t,s)\in u^\perp$, real-analytic in $(t,s)$ with $\rho(0,s)=0$, for which the path
  $(\theta_t,M_t)$, $t\in[0,1]$, starts at $(\theta_+,M)$, ends at $(\theta_-,M)$, satisfies $\theta_t\ge\theta^*/2$ entrywise,
  and
  \begin{enumerate}[label=(\roman*),leftmargin=*]
    \item $F_{K_2}(\theta_t;M_t)$ and $F_{P_3}(\theta_t;M_t)$ do not depend on $t$;
    \item $|\partial_tF_T(\theta_t;M_t)|\le Ls^3$ for every tree $T$ on $4\le v\le r$ vertices;
    \item $|\partial_t\hom(G;\theta_t,M_t)|\le Ls$ for every connected unicyclic graph $G$ on at most $r$ vertices;
    \item $|\partial_t\hom(G;\theta_t,M_t)|\le L$ for every connected graph $G$ on at most $r$ vertices;
    \item $\|\dot\theta_t\|_1+\|\dot M_t\|_\infty\le L$ and $\max_c|\dot\theta_{t,c}|/\theta_{t,c}\le Ls$.
  \end{enumerate}
  Moreover $|\Ent(\theta_+)-\Ent(\theta_-)|\ge s|\ell(\zeta_0)|$ for $0<s\le s_P$.
\end{lemma}

The channel travels a distance of order one along the path.
The information inequality below needs bounded derivatives, not small ones. The two endpoints differ by $-I$ on $u^\perp$ and need no relabeling of
the meanings.

\paragraph{Information along a moving general channel.}
Let $P_t$ be the law of the verdicts under $(\theta_t,M_t)$ and $I(t):=\sum_AP_t(A)(\partial_t\log P_t(A))^2$ its Fisher
information. For diagonal channels $M_t=\diag(w(t))$, the information $I(t)$ is bounded with a sweep \citep{meaningdependent2026}: the
locally averaged score is expanded in the edge variables $\eps_{ij}:=\Delta(A_{ij}-q)/(q(1-q))$, the Taylor
coefficients of small motifs are derivatives along the path of polynomials in motif functions, and overlapping
motifs are paired by their cycle structure. The argument extends to general channels: the likelihood ratio of a
verdict is $1+M_t[Z_i,Z_j]\eps_{ij}$ for every pair rather than only for pairs inside one meaning, the Glauber factors
take a product over all answers, the channel term of the score loses its indicator, and the motif coefficients
become polynomials in the possibly negative functions $\hom(Q;\theta_t,M_t)$ of the connected subgraphs $Q$ of a
motif. We state the result in the generality that both of our paths need.

\begin{proposition}[information inequality]
  \label{prop:info}
  Let $(\theta_t,M_t)$, $t\in[t_-,t_+]$, be real-analytic in $t$ on a neighborhood of $[t_-,t_+]$, with
  $\theta_{t,c}\ge\theta_{\min}>0$, $\max_{t,a,b}|M_t[a,b]|\le M_{\rm hi}$, $\|\dot\theta_t\|_1+\|\dot M_t\|_\infty\le L$ and
  $\max_c|\dot\theta_{t,c}|/\theta_{t,c}\le L$, where $L\ge1$. Let $m\ge2$, $r\ge1$ and $D\ge r+2$ be integers, $q_0\in(0,1/2]$, and
  $0\le\eps_{\rm tree}\le\eps_{\rm uni}\le1$. Assume that $F_T(\theta_t;M_t)$ is constant in $t$ for every tree $T$ on at
  most $m-1$ vertices, that $|\partial_tF_T(\theta_t;M_t)|\le L\eps_{\rm tree}$ for every tree on $m\le v\le r$ vertices,
  and that $|\partial_t\hom(G;\theta_t,M_t)|\le L\eps_{\rm uni}$ for every connected unicyclic graph $G$ on at most $r$
  vertices. There are $C_I$, $n_I$ and $\varsigma_0>0$, depending only on $(M_{\rm hi},L,q_0,r,D)$, such that for
  $n\ge n_I$, $q\in[q_0,1-q_0]$, $\delta\le1$ and $\sqrt\delta\log n\le\varsigma_0$, for all $t\in[t_-,t_+]$,
  \[
    I(t)\le C_I\Big[N_m\eps_{\rm tree}^2+\delta^3\eps_{\rm uni}^2+\frac{\delta^5}n+n\delta^r
      +n^{r+3}\delta^{D+1}(\log n)^{2(D-r)}+n^{4D+1}e^{-q_0^2\log^2n/36}\Big].
  \]
\end{proposition}

The four terms in front come from pairing motifs that are trees on at least $m$ vertices, unicyclic, multicyclic,
or on more than $r$ vertices. The last two are the remainder of the Taylor expansion of degree $D$ on an event
where all fields are small, and off it. The constants do not depend on $\eps_{\rm tree}$ and $\eps_{\rm uni}$, which
matters because the rotation path has $(m,\eps_{\rm tree},\eps_{\rm uni})=(4,s^3,s)$ with $s\to0$; the path for
the five-vertex threshold has $(5,1,1)$. \Cref{app:lower} states each step of the argument for diagonal channels in the form used here, with its
hypotheses and constants, and
proves each changed step.

\begin{theorem}[lower bound]
  \label{thm:lower}
  Let $\mathcal K\subset\Omega$ be compact, $q_0\in(0,1/2]$ and $\kappa\in(0,1)$. There are $c_L>0$ and $n_L$ depending only
  on $(\mathcal K,q_0,\kappa)$ such that $R^*_n(M)\ge c_L\min\{1,N_4^{-1/3}\}$ for every $M\in\mathcal K$, $n\ge n_L$,
  $q\in[q_0,1-q_0]$ and $\Delta>0$ with $n\Delta^2\le n^{-\kappa}$.
\end{theorem}

\begin{proof}[Proof sketch]
  Take $r:=\max\{4,\lfloor1/\kappa\rfloor+1\}$, so that $n\delta^r\le n^{1-\kappa r}\to0$, and $D:=\max\{r+2,\lceil(r+4)/\kappa\rceil\}$, so
  that the first remainder vanishes; the last term decays faster than any power of $n$, and $\delta^3s^2$ and
  $\delta^5/n$ vanish on the window. With $s:=\min\{s_1,a_0N_4^{-1/6}\}$ for constants $s_1,a_0>0$, the tree term
  $C_IN_4s^6$ is at most $1/8$, so $\sup_tI(t)\le1/4$ for $n\ge n_L$. The Hellinger path bound
  (\Cref{lem:a:hellinger}) gives $\TV(P_0,P_1)\le\frac12\int_0^1\sqrt{I(t)}\,\mathrm dt\le1/4$, and the two-point bound
  (\Cref{lem:a:lecam}) with the entropy gap $s|\ell(\zeta_0)|$ gives $R^*_n(M)\ge s^2\ell(\zeta_0)^2/16\ge
  c_L\min\{1,N_4^{-1/3}\}$. The full proof, with constants uniform on $\mathcal K$, is in the appendix.
\end{proof}

\section{The five-vertex threshold}
\label{sec:threshold}

\Cref{thm:threshold} needs a path between the two points of an admissible double pair along which all tree
functions on at most four vertices stay constant. The expansion behind the information inequality then sees no tree on
fewer than five vertices, and along such a path it gives the log-likelihood ratio between the two points. The two points
are at distance of order one, so the local rotation cannot be used. The path comes from the
structure of $F_{\le4}$.

\paragraph{Moment reduction.}
For $\theta\in\Int$ and $d\in\Real^3$ let $m_k(\theta,d):=\sum_a\theta_ad_a^k$ be the moments of the measure
$\mu_{\theta,d}:=\sum_a\theta_a\delta_{d_a}$, and $g:=\theta\odot d$. If $d=M\theta$, then
$(F_{K_2},F_{P_3},F_{K_{1,3}})=(m_1,m_2,m_3)$ and $F_{P_4}=g^\top Mg$. So a path keeps all four tree functions constant
if, along it, the configuration $(\theta_t,d_t)$ keeps its first three moments and the channel $M_t$ satisfies
$M_t\theta_t=d_t$ and $g_t^\top M_tg_t=F^M_{P_4}(\theta^0)$. The channel can be chosen given the
configuration.

\begin{lemma}[channel projection]
  \label{lem:projection}
  Let $\theta\in\Int$, $d\in\Real^3$ with $d$ not constant, $c_P\in\Real$ and $M\in\Sym$. The linear map
  $X\mapsto(X\theta,g^\top Xg)$ from $\Sym$ to $\Real^4$ is onto, so $\{X\in\Sym:X\theta=d,\ g^\top Xg=c_P\}$ is an affine
  plane of dimension two, and its point closest to $M$ in the Frobenius norm is a real-analytic function of
  $(\theta,d)$.
\end{lemma}

At an endpoint of a double pair, $M$ itself satisfies the
constraints and is therefore the closest point, so the projected channel returns to $M$ at both ends. It remains to connect the two configurations $(\theta^0,M\theta^0)$ and
$(\theta^1,M\theta^1)$ inside the set of positive three-point configurations with given first three moments.

\paragraph{Three-point moment fibers.}
Fix $(m_1,m_2,m_3)$ with $v:=m_2-m_1^2>0$, and let $\mathsf L$ be the linear functional on polynomials of degree at most
three with $\mathsf L(z^k)=m_k$, $m_0:=1$. The monic quadratic $\pi_2(z)=z^2-\sigma_1z+\sigma_2$ with
$\mathsf L(\pi_2)=\mathsf L(z\pi_2)=0$ exists, satisfies $\pi_2(m_1)=-v<0$, and has two real roots $\xi_1<m_1<\xi_2$, the
Gauss nodes; the two-point measure on them with positive weights has moments $1,m_1,m_2,m_3$. Let
$\pi_1(z):=z-m_1$ and, for $(a,b)\in\Real^2$,
\[
  p_{a,b}(z):=(z+a)\pi_2(z)+b\,\pi_1(z),\qquad w_{a,b}(x):=\frac{(x+a)(x+m_1-\sigma_1)+b}{p'_{a,b}(x)} .
\]

\begin{lemma}[three-point fibers]
  \label{lem:fibers}
  \begin{enumerate}[label=(\roman*),leftmargin=*]
    \item A configuration $(\theta,d)\in\Int\times\Real^3$ with pairwise distinct atoms $d_1,d_2,d_3$ has moments
      $m_1,m_2,m_3$ if and only if $\prod_c(z-d_c)=p_{a,b}(z)$ for some $(a,b)$ and $\theta_c=w_{a,b}(d_c)$ for every $c$.
    \item Every such configuration has $b<0$. For every $(a,b)$ with $b<0$, the polynomial $p_{a,b}$ has three simple real roots
      $x_1<\xi_1<x_2<\xi_2<x_3$, analytic in $(a,b)$, and none of the three weights vanishes; hence if one point of the
      half-plane $\{b<0\}$ carries positive weights, all points do.
    \item Let $\xi\in\{\xi_1,\xi_2\}$ and $c>0$. Along $(a,b)=(-\xi+\alpha,-c\alpha^2)$, $\alpha\in\Real$, the roots and
      the weights follow real-analytic branches with positive weights, two roots meet at $\xi$ when $\alpha=0$, and their
      order is reversed between $\alpha<0$ and $\alpha>0$: the atoms in positions $2,3$ exchange at $\xi_2$, and those in
      positions $1,2$ at $\xi_1$.
  \end{enumerate}
\end{lemma}

So the positive configurations with distinct atoms and given moments are parametrized by the half-plane
$\{b<0\}$, together with an assignment of the three meanings to the three positions. The collisions sit on the
boundary $b=0$, at the two points $a=-\xi_1$ and $a=-\xi_2$, and crossing them exchanges two adjacent positions. The
two adjacent exchanges generate all permutations of three positions.

\begin{lemma}[fiber path]
  \label{lem:fiber-path}
  Let $(\theta^0,\theta^1)$ be an admissible double pair of $M\in\Sym$. There are $0=t_0<t_1<\dots<t_J=1$ with $J\le7$
  and a continuous path $t\mapsto(\theta_t,M_t)\in\Int\times\Sym$, real-analytic on each $[t_{j-1},t_j]$, such that
  the path starts at $(\theta^0,M)$, ends at $(\theta^1,M)$, and $F^{M_t}_{\le4}(\theta_t)=F^M_{\le4}(\theta^0)$ for every $t$.
\end{lemma}

\begin{proof}[Proof sketch]
  The two configurations have the same moments, so by \Cref{lem:fibers} they correspond to points of the half-plane
  with two assignments of meanings to positions that differ by a permutation, a product of at most three adjacent
  exchanges. Move along straight segments in the convex half-plane to a point near the boundary, cross through the
  collision that realizes the next exchange, and repeat; finish with a straight segment to the second endpoint.
  Along the way the configuration is analytic, the atoms stay distinct off the collisions, the weights stay
  positive, and the variance $v$ is fixed. Project the channel by \Cref{lem:projection}. The full proof is in
  the appendix.
\end{proof}

\begin{figure}[t]
  \centering
  \includegraphics[width=\textwidth]{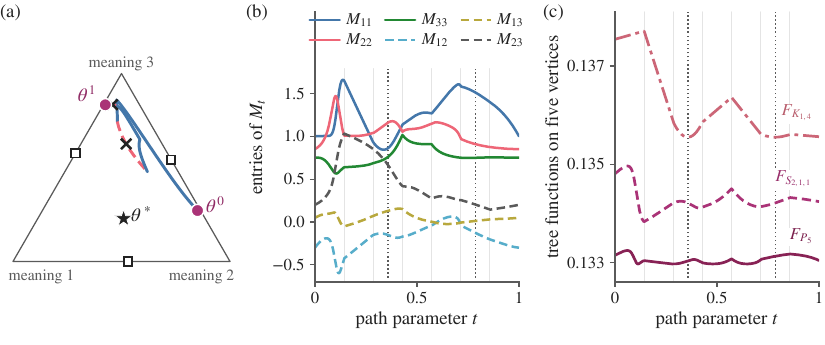}
  \caption{The fiber path for the channel $M_0$, between the two points $\theta^0$ and
  $\theta^1$ of its admissible double pair, whose entropies are $0.654$ and $0.483$. (a) The distributions
  $\theta_t$ in the simplex: straight pieces in the parameter half-plane (blue, solid) and the two crossings through
  collisions of atoms at the Gauss nodes (red, dashed), with the collision points marked by crosses, the second of which lies next to $\theta^1$; the star is
  the equitable point $\theta^*$ and the squares are the three face equitable points. (b) The entries of the
  channel $M_t$, equal to $M_0$ at both ends. (c) The three tree functions on five vertices. The
  four tree functions on at most four vertices are constant along the path; at the
  $1081$ evaluated points, in $50$-digit arithmetic, they differ from their values at the endpoints by at most
  $9.2\times10^{-49}$. Of the seven tree functions, only the three on five vertices move. Light vertical lines separate
  the seven pieces and dotted lines mark the collisions.}
  \label{fig:fiber}
\end{figure}

\Cref{fig:fiber} shows the path for $M_0$, computed in $50$-digit arithmetic from the construction. The information inequality
constrains the size of the motion only through bounds on the channel, on the derivatives and on the smallest
coordinate, so the channel may move far: by about $0.83$ entrywise at its farthest, while the smallest coordinate of
the distribution is $0.00836$.

\paragraph{The likelihood ratio along the path.}
On each piece of the fiber path the tree functions on at most four vertices are constant and every derivative is
bounded, so the path satisfies the hypotheses of \Cref{prop:info} with $(m,\eps_{\rm tree},\eps_{\rm uni})=(5,1,1)$. With
the Hellinger path bound (\Cref{lem:a:hellinger}) this gives $\TV(P_{\theta^0,M},P_{\theta^1,M})\lesssim\sqrt{N_5}+o(1)$, which
is informative only while $N_5$ is small. Part (a) of \Cref{thm:threshold} needs more: that $\log(P_1/P_0)\ge-CN_5-C$ with
probability at least $1/2$ under $P_0$, where $P_0,P_1$ are the laws at the two points. Then $P_1\ge e^{-CN_5-C}P_0$ on
that event, and every estimator misses one of the two entropies by $\omega/2$ with probability at least of order
$e^{-CN_5}$ under one of the two laws. The expansion in the proof of \Cref{prop:info} gives the log-likelihood ratio itself.
The score $\partial_t\log P_t(A)$ along the path is the conditional expectation, given the verdicts, of a swept
complete-data score, whose Taylor coefficients in the edge variables are, for motifs on at most $r$ vertices,
derivatives $\partial_tC_\Gamma$ of polynomials in the motif functions. Integrating along the path,
\[
  \log\frac{P_t(A)}{P_0(A)}=\Lambda_t(A)-\Lambda_0(A)+\int_0^t\Exp_s[X_s\mid A]\,\mathrm ds,
\]
where $\Lambda_t$ is an explicit sum of motif statistics and $X_s$, whose second moment under the law at $s$ is
polynomially small in $n$, collects the larger motifs and the Taylor remainder (\Cref{lem:e:decomp,lem:e:error}). Under $P_0$ the
motifs whose underlying graph is a tree on at most four vertices drop out of $\Lambda_t-\Lambda_0$, because their
coefficients are polynomials in tree functions that stay constant along the path, and the five-vertex trees
contribute $O(N_5+\sqrt{N_5})$ in mean (\Cref{lem:e:main}). The error term is controlled under the law at $s$, not under $P_0$. As long as
$\log(P_s/P_0)>-K\log n$, however, $P_0<n^KP_s$, so up to the first time at which the ratio reaches $-K\log n$ the
error is small under $P_0$ as well, and on that event the expansion shows that this time is never reached
(\Cref{lem:e:stop}). The constants are chosen in this order: the five-vertex constant, then $K$, then the sweep order and
the Taylor degree, then $n$; the first does not depend on the later ones. Part (b) combines (a) with \Cref{thm:upper,thm:lower}.
The details are in the appendix.

\begin{proof}[Proof sketch of \Cref{cor:phase}]
With $\Delta=n^{-\beta}$ we have $\delta=n^{1-2\beta}$, $N_4=n^{4-6\beta}$ and $N_5=n^{5-8\beta}$; take $\kappa:=\min\{2\beta-1,1/2\}$, so
that $n\Delta^2\le n^{-\kappa}$, and note that \eqref{eq:P} holds for large $n$. In case (i), \Cref{thm:main}(b) gives
$R^*_n\asymp\min\{1,n^{-(4/3-2\beta)}\}$. In case (ii) with $\beta<5/8$, the quantity $N_5$ grows polynomially and exceeds
$C_{\log}\log n$ for large $n$, and $\kappa<1/4$, so \Cref{thm:main}(a) gives $R^*_n\asymp N_4^{-1/3}$. For $\beta\ge5/8$ we have
$N_5\le1$, and \Cref{thm:threshold}(a) gives a positive lower bound that does not depend on $n$. As $\beta\uparrow5/8$ the
exponent $4/3-2\beta$ tends to $1/12$. The full proof is in the appendix.
\end{proof}

\section{Certificates and the size of the class}
\label{sec:certificates}

The conditions defining $\cU$ are explicit, but (U3) and (U4) are global statements about the closed simplex. For
each channel reported here we decide all of them exactly, by elimination in exact rational arithmetic with
SymPy \citep{meurer2017sympy}, and enclose the irrational quantities that remain by interval arithmetic with
outward rounding in mpmath \citep{mpmath,moore2009introduction,rump2010verification}. \Cref{app:certificates} gives
every quantity for $M_0$ and the matrices of the named channels $M_0,\dots,M_9$, and the repository given in \Cref{sec:conclusion}
holds those of all other certified channels. A number that bounds a quantity, as a lower or
upper bound or as an enclosing interval, is rounded outward, here and in the appendices.

\begin{table}[t]
  \centering
  \caption{Certified channels (entries in \Cref{app:certificates}) and two random samples. Double pairs are
  pairs of distinct points of $\Theta_3$ at which $F_{\le4}$ agrees; for the strata $\mathcal U^\sigma$, only pairs
  outside the symmetric pairs are counted. A channel with a counted double pair lies outside the corresponding class with subscript 4. The separation, given in the fourth column, is the interval lower bound of $\mathcal D_5$ over the
  double pairs, rounded down; the entropy gap is $|H(\theta)-H(\theta')|$ for each double pair. The grid channels
  have diagonal entries in $\{12,\dots,28\}/20$ and off-diagonal entries in $\{-6,\dots,6\}/20$, and were drawn
  conditioned on $\min_c\theta^*_c\ge1/20$; the random
  channels have independent entries, uniform on the grid of step $1/200$, diagonal entries in $[1/2,3/2]$ and
  off-diagonal entries in $[-1/2,1/2]$, the second sample with one diagonal draw shared by the three diagonal
  entries. All classifications are exact rational computations in SymPy with outward-rounded interval
  arithmetic in mpmath. The proportions are empirical for these distributions.}
  \label{tab:certificates}
  \small
  \begin{tabularx}{0.97\linewidth}{@{}ll*{3}{>{\centering\arraybackslash}X}@{}}
    \toprule
    Channel & class & double pairs & smallest $\mathcal D_5$ & entropy gap \\
    \midrule
    $M_0$ & $\mathcal U\setminus\mathcal U_4$ & 1 & $1.98\times10^{-3}$ & 0.171 \\
    $M_1$ & $\mathcal U_4$ & 0 & none & none \\
    $M_2$ & $\mathcal U_4$ & 0 & none & none \\
    $M_3$ & $\mathcal U^\sigma_{\{3\}}$ & 1 & $6.02\times10^{-4}$ & 0.114 \\
    $M_4$ & $\mathcal U^\sigma_{\{2\}}$ & 1 & $1.33\times10^{-3}$ & 0.207 \\
    $M_5$ & $\mathcal U^\sigma_{\{1\}}$ & 1 & $5.70\times10^{-4}$ & 0.0784 \\
    $M_6$ & $\mathcal U^\sigma_{\{1,2,3\},4}$ & 0 & none & none \\
    $M_7$ & $\mathcal U^\sigma_{\{1,2,3\}}$ & 1 & $2.22\times10^{-4}$ & 0.00312 \\
    $M_8$ & $\mathcal U^\sigma_{\{1,2,3\}}$ & 1 & $9.07\times10^{-4}$ & 0.00189 \\
    $M_9$ & $\mathcal U^\sigma_{\{1,2,3\},4}$ & 0 & none & none \\
    \midrule
    \multicolumn{5}{@{}L{0.97\linewidth}@{}}{Twenty grid channels: 13 in $\mathcal U$, of which 2 in $\mathcal U_4$; the other 7 have an exact coincidence of entries. Random sample of 200: 172 have $\theta^*>0$, 4 of these have an exact coincidence of entries, and all other 168 lie in $\mathcal U$, 43 of them in $\mathcal U_4$; the remaining 125 have 396 double pairs in total, with smallest entropy gap $1.50\times10^{-3}$ and smallest separation $1.62\times10^{-8}$. Random sample of 100 with a common diagonal entry: 84 have $\theta^*>0$, 1 has an exact coincidence of off-diagonal entries, and all other 83 lie in $\mathcal U^\sigma_{\{1,2,3\}}$, 25 of them in $\mathcal U^\sigma_{\{1,2,3\},4}$.} \\
    \bottomrule
  \end{tabularx}
\end{table}

\paragraph{Local conditions.}
For rational $M$, the equitable point $\theta^*$, the constant $c$ and the characteristic polynomial of $\tM$ on $u^\perp$
are rational; for $M_0$ we have $\theta^*=(73/194,\,307/776,\,177/776)$, $c=835/3104$ and $\lambda_1\approx0.13829$,
$\lambda_2\approx0.47633$. The reduced cubic forms of (U1) are computed in rational coordinates of the tangent plane,
and they have no common real zero because their resultant is a nonzero rational number. Conditions (U1)(ii) and (iii) are decided
in the eigenbasis, whose coordinates lie in a quadratic field, where a $2\times2$ determinant of the form
$a+b\sqrt{\mathrm{disc}}$ with rational $a,b\ne0$ cannot vanish because $\mathrm{disc}$ is not the square of a rational. The face equitable points of $M_0$ are
$(0,\tfrac{11}{24},\tfrac{13}{24})$, $(\tfrac{14}{33},0,\tfrac{19}{33})$ and $(\tfrac{23}{49},\tfrac{26}{49},0)$, and the
determinants of (U2) are nonzero rationals.

\paragraph{Degenerate points and double pairs.}
For (U3), a point $\theta\ne\theta^*$ at which $F_{\le4}$ has rank at most one has a kernel direction $h(m)$ in one of
two affine charts. The edge function forces $\theta$ onto a line $\theta^*+\lambda r(m)$, and the three remaining
conditions become polynomials in $(\lambda,m)$. The greatest common divisor of their resultants in $\lambda$ has
finitely many roots, all found exactly; for $M_0$ they give exactly the three face equitable points and $\theta^*$
itself. For (U4), a double pair $\theta\ne\theta'$ is written in terms of its midpoint $u$ and half-difference $t\,h(m)$.
Equality of the edge function puts $u$ on the same lines, and the parts of the other three tree functions that
are odd in $t$ give three polynomial equations in $(\lambda,m,\tau)$ with $\tau=t^2$. Eliminating $\tau$ and $\lambda$
leaves a univariate resultant $R(m)$, of degree $23$ for $M_0$ in each chart, which is factored over $\mathbb Q$. Each real
root is isolated exactly and classified: it gives no real double pair ($\tau\le0$, or a pair outside the simplex),
or a double pair in the simplex, whose five-vertex separation is then bounded below by interval arithmetic. For
$M_0$ exactly one pair lies in the simplex, $\theta\approx(0.0142202,0.7148095,0.2709703)$ and
$\theta'\approx(0.1580727,0.0083592,0.8335681)$, with $|\Ent(\theta)-\Ent(\theta')|\approx0.170955$ and
$|F_{K_{1,4}}(\theta)-F_{K_{1,4}}(\theta')|\ge1.98472\times10^{-3}$. A second implementation builds the tree
polynomials by enumerating labelings, computes a Gröbner basis for the rank conditions and searches for double
pairs with Powell's hybrid method in SciPy \citep{virtanen2020scipy} from $800$ random starts, refining each solution by Newton's method. It finds the same
degenerate points and the same pair.

\paragraph{Persistence.}
Let $G(z;M):=F^M_{\le4}(\theta)-F^M_{\le4}(\theta')$ with $z=(\theta_1,\theta_2,\theta'_1,\theta'_2)$. Around a rational
approximation $z_0$ of the double pair of $M_0$ we take the box $X:=z_0+[-1/5000,1/5000]^4$ and the parameter box
$P_r$ with $r=1/2000000$, both built from rational endpoints with outward rounding. The Krawczyk test
with a preconditioner $Y$,
\[
  K(X,P_r):=z_0-Y\,G(z_0;P_r)+\big(I-Y\,DG(X;P_r)\big)(X-z_0)\subset\operatorname{int}X,
\]
certifies that for every $M\in P_r$ the map $G(\cdot;M)$ has exactly one zero in $X$ \citep{moore2009introduction,rump2010verification}. Here
$K$ is the operator of \citet{krawczyk1969newton}, and $G(z_0;P_r)$ is enclosed by a centered form in the six entries of $M$, computed by
forward differentiation in interval arithmetic. Interval
evaluation over $X\times P_r$ then shows that both points are interior, with coordinates at least $0.01402$ and
$0.00815$, that the entries of $M\theta$ are separated by at least $0.30995$ at the first point and $0.07025$ at the
second, and that $\Ent(\theta)-\Ent(\theta')\in[0.16651,0.17540]$. This proves \Cref{prop:class}(d), and with it that $\cU\setminus\cU_4$ has interior
points, which any positive certified radius would show. The radius is set by the conditioning of the double pair of
$M_0$, at which the Jacobian of $G$ has smallest singular value about $0.016$ and one coordinate of $\theta'$ is about
$0.0084$, and by the overestimation of interval evaluation.

\paragraph{Equal diagonal entries.}
The certificates of \Cref{thm:symmetric} use the same elimination with three changes. On a symmetric face the
local condition is checked at the midpoint; the transversality condition holds because the four $3\times3$ minors of
the relevant $4\times3$ matrix, polynomials in the face coordinate $t$, are divisible by $t^2$ and their quotients
have no common complex root; and the direction of the face is a root of the elimination along which every double
pair is verified exactly to be a symmetric pair. \Cref{tab:certificates} lists the seven channels $M_3,\dots,M_9$.

\paragraph{How large the class is.}
We applied the certificate to two random samples whose distribution and selection rule were fixed before any
channel was examined (\Cref{tab:certificates}). Of $200$ channels with independent grid entries, $172$ have
$\theta^*>0$; $4$ of these have an exact coincidence of entries, and all other $168$ lie in $\cU$, $43$ of them in
$\cU_4$. The remaining $125$ have between $1$ and $12$ double pairs, $396$ in total, all with different entropies and
all separated by five-vertex trees. All of these pairs are certified to be interior; \Cref{thm:threshold} applies to them if,
in addition, the entries of $M\theta$ are pairwise distinct at both points of a pair, which we checked in floating point but did
not certify. Of $100$ channels with a common diagonal entry, $84$ have $\theta^*>0$, $1$ has an
exact coincidence of off-diagonal entries, and the other $83$ lie in the stratum class of \Cref{thm:symmetric}.
These proportions describe the two sampling distributions. Among channels with $\theta^*>0$, every failure of the conditions that we observed comes from an exact
coincidence of entries: $M_{ki}=M_{kj}$ makes the derivative of
every tree function along $e_i-e_j$ vanish on the segment from $f_k$ to $e_k$, which contains $\theta^*$, so (U3)
fails; $M_{ii}=M_{jj}$ is the symmetric case. In the first sample the smallest five-vertex separation
is $1.6\times10^{-8}$; the separations enter the constant $\eta$ of \Cref{thm:modulus}, not the rate.

\section{Related work}
\label{sec:related}

\paragraph{Semantic entropy and its minimax theory.}
Semantic entropy groups sampled answers into meanings with a bidirectional entailment judge
\citep{kuhn2023semantic}, and its discrete variant takes the entropy of the cluster proportions
\citep{farquhar2024detecting}. Other uncertainty measures are built directly from pairwise semantic similarities
\citep{nikitin2024kernel,nguyen2025beyond}, whereas we estimate the Shannon entropy of the latent meanings through a
noisy binary channel. With few samples the discrete variant underestimates semantic entropy, and an estimate of the
semantic alphabet size corrects it for sample coverage \citep{mccabe2026estimating}; that bias comes from meanings not
yet sampled rather than from errors of the judge. For a homogeneous pairwise model, \citet{liu2026statistically} proposes a spectral-clustering
plug-in. Whenever spectral clustering misclassifies a vanishing fraction of the answers, which
\citet{lei2015consistency} prove for stochastic block models under an eigengap condition, this plug-in is consistent
by continuity of the entropy. With observed labels, entropy estimation on a fixed alphabet has rate $1/n$, and its
large-alphabet theory rests on polynomial approximation \citep{jiao2015minimax,wu2016minimax}. Minimax rates for the weak regime are known for a
homogeneous judge, graded by how many meanings can be rare, with the global rate $N_K^{-1/(K-1)}$ for $K$ meanings
\citep{yu2026rarity}, and for a judge with meaning-dependent sensitivities and one common false-equivalence
probability, with rate $N_{2K-1}^{-1/(2K-1)}$ for almost every positive sensitivity profile
\citep{meaningdependent2026}. We use the same kind of tree statistics and inversion, and the information
inequality of the meaning-dependent analysis for channels that move along a path, which we extend from diagonal to general channels. What is new for a general
channel is the geometry: the least-favorable pair sits at the equitable point and is resolved by a cubic
degeneracy instead of a high-order moment match, the four-vertex tree map can have entropy-distinct double points
far apart, and both the estimator and the lower bounds have to deal with them.

\paragraph{Mixtures, moments and singular information.}
For mixing distributions with $K$ atoms, moments of order up to $2K-1$ control the Wasserstein distance with
exponent $1/(2K-1)$, and projecting noisy moments onto the moment space gives optimal estimators of Gaussian
mixtures \citep{wu2020optimal}. \citet{moitra2010settling} compare moments through sign-preserving polynomials,
and moment matching drives the lower bounds for mixtures of two Gaussians
\citep{hardt2015tight}. In multi-reference alignment, a Gaussian mixture whose centers lie on the orbit of a group, the
moment tensors control the Kullback--Leibler divergence, and the order of the first tensor that separates two signals
sets the power of the noise level in the minimax rate \citep{bandeira2019optimal}. In orbit recovery the sample complexity is set by the
degree of the invariant polynomials that determine the orbit \citep{bandeira2023estimation}. When the number of components of a finite mixture is not known,
\citet{chen1995optimal} gave $n^{-1/4}$ as the best possible rate for the mixing distribution, and
\citet{heinrich2018strong} showed that the optimal local minimax rate around a mixing distribution with $m_0$
components, among those with $m$ components, is $n^{-1/(4(m-m_0)+2)}$, correcting that claim. The local structure at $\theta^*$ follows a pattern known for likelihoods
with a singular information matrix \citep{rotnitzky2000likelihood}: when the first nonvanishing derivative in a
direction has even order, the magnitude of the parameter is identified at that order and its sign only at the
next, and the uniform rate in shrinking neighborhoods is $n^{-1/(2s+2)}$ for order $s$. Here the squares of the two
eigencoordinates are identified at second order and their signs at third, with $N_4$ in place of $n$ and $s=2$,
which gives the amplitude $N_4^{-1/6}$ and the squared risk $N_4^{-1/3}$. For a functional of an i.i.d.\ sample, the
modulus of continuity with respect to the Hellinger distance bounds the minimax rate from below, and the bound is
attained when the functional is linear and the class convex \citep{donoho1991geometrizing}. The cubic modulus
plays the corresponding role with respect to the tree functions, whose estimation error is known,
while our lower bounds come from the information inequality along moving channels. Our observations are not
independent samples: they are a graph of verdicts that share latent labels. The target is the entropy of the proportions
rather than a mixing distribution, and the degeneracy comes from the tree functions at the equitable point.

\paragraph{Truncated moment problems.}
Positive measures with finitely many prescribed moments and their Gauss-quadrature representations are classical
\citep{curto1991recursiveness,schmudgen2017moment}. The meaning-dependent analysis interpolates between two
reflected points through Gauss-quadrature measures with a fixed number of atoms \citep{meaningdependent2026}. Our
fiber is the set of positive three-point measures with three prescribed moments, a half-plane with an assignment of
labels to atoms, and the path must change the assignment; it does so through the two collisions of atoms, which
occur exactly at the Gauss nodes.

\paragraph{Block models and subgraph counts.}
The verdict graph is a stochastic block model with the known connectivity matrix $q\one\one^\top+\Delta M$. Work on
such models usually targets the partition, as in exact recovery \citep{abbe2016exact}, or the connectivity
parameters: in the sparse two-community planted partition model the parameters cannot be estimated consistently
below the detection threshold \citep{mossel2015reconstruction}, the likelihood estimators are asymptotically normal when the expected degree grows faster than logarithmically \citep{bickel2013asymptotic}, and uniformly over models with nearly equal blocks the
connectivity estimates converge slowly \citep{castillo2022uniform}. We fix the connectivity and estimate a
functional of the unknown proportions as the signal vanishes. \citet{rush2026easier} ask whether global properties of
the communities, such as their number and sizes, can be inferred where finding the communities is computationally
hard, and show in the low-degree polynomial framework that for certain pairs of community structures testing between
them is as hard as finding the communities. These results are computational lower bounds for testing, while ours are minimax
bounds over all estimators for a functional of the proportions. Our statistics are noise-corrected subgraph averages
\citep{chang2022estimation}, in the spirit of unbiased moment estimates along graph structures
\citep{kong2017spectrum}. Bounds on their deviations are needed only up to a power of $n$ standard deviations, and conditionally on
the labels they follow from the concentration inequality of \citet{adamczak2015concentration} for polynomials in
independent sub-Gaussian variables, which those authors also apply to subgraph counts in Erd\H{o}s--R\'enyi graphs.

\paragraph{Moment estimators for block models.}
When the connectivity is unknown, counts of trees identify and estimate the parameters of a block model.
\citet{bickel2011method} prove that normalized counts of acyclic patterns are consistent and asymptotically normal for
every range of the average degree, that the counts of finitely many trees identify a $K$-block model under a
linear-independence condition, and that, for average degrees bounded below and of order $o(n^{1/2})$, inverting them
gives $\sqrt n$-consistent estimates at parameters where the moment map has a differential of full rank. The graph pencil method \citep{gunderson2023graph} recovers the expected
degrees and the sizes of the blocks from the densities of stars, which are the moments of the degree distribution, by a
matrix pencil, and the connection probabilities from the densities of bistars. It requires the expected degrees of the
blocks to be distinct. Both nondegeneracy conditions fail at the point that sets our rate: $M\theta^*=c\one$, so at
$\theta^*$ every answer has the same expected degree, and every tree function is stationary. Knowing the channel does
not remove this singularity. For a channel in $\cU$, near $\theta^*$, the entropy is only Hölder continuous with
exponent $1/3$ in the four-vertex tree functions. The lower bound $c_L\min\{1,N_4^{-1/3}\}$,
which holds for every channel satisfying (N) on every fixed polynomial window with $q$ bounded away
from $0$ and $1$, comes from two points near $\theta^*$, and the tree-statistic estimator attains this order on $\cU_4$ and, once
$N_5\ge C_{\log}\log n$, on $\cU$. The regimes differ as well: those analyses keep the block
structure fixed relative to the edge density as the graph grows, while here the offsets $\Delta M$ vanish against a
dense background, $n\Delta^2\le n^{-\kappa}$.

\paragraph{Tree densities and identification.}
The tree functions are the homomorphism densities of trees in the step kernel with step widths $\theta$ and
values $M$ \citep{lovasz2012large}. The numbers of homomorphisms from all trees determine a finite graph up to
fractional isomorphism \citep{dell2018lovasz}, and the tree densities determine a graphon up to the analogous
notion for graphons \citep{grebik2022fractional}. Here the kernel is known and only the step widths vary, and the
question is which finite set of trees identifies the entropy of the widths and at which rate: trees on at most
four vertices set the rate, and on part of $\cU$, trees on five vertices are needed to identify $\theta$.
Gaussian latent-variable models at low signal-to-noise ratio show a similar split \citep{balanov2026generalized}: the
moment order that locally identifies the parameter sets how the smallest eigenvalue of the Fisher information scales
with the signal-to-noise ratio, and it can be smaller than the order that globally identifies the parameter, which a
globally consistent moment estimator needs. There the observations are independent, and the sample size grows at a
fixed signal strength before the signal vanishes. In our model the verdicts share latent labels, $n$ and $\Delta$ move
jointly through $N_4$ and $N_5$, and the risk is the worst case over the simplex. For a channel in $\cU$ with an
admissible double pair and $q$ bounded away from $0$ and $1$, the trees on five vertices enter the minimax risk on
every fixed polynomial window through the term $e^{-\Theta(N_5)}$, the cost of telling the two
points of the pair apart.

\paragraph{Computer-assisted proofs.}
Interval Newton methods and the Krawczyk operator \citep{krawczyk1969newton} certify existence and uniqueness of zeros of systems with interval
parameters \citep{moore2009introduction,rump2010verification}, and elimination by resultants with
exact real root isolation decides polynomial systems over the reals \citep{cox2015ideals,basu2006algorithms}. We use
both to verify the conditions of a statistical theorem for explicit channels, and to show that the two parts of
the condition class have interior points. Computer-assisted procedures also certify the algebraic properties that set
the sample complexity of orbit recovery \citep{bandeira2023estimation}.

\section{Conclusion}
\label{sec:conclusion}

When the errors of a known judge depend on the pair of meanings and $q$ is bounded away from $0$ and $1$, the minimax
rate of semantic entropy for three meanings is $N_4^{-1/3}=(n^4\Delta^6)^{-1/3}$ on an explicit open class of channels, on
every polynomial window on which $N_5\ge C_{\log}\log n$, and $\min\{1,N_4^{-1/3}\}$ on every polynomial window for the subclass whose four-vertex tree map is injective. Tree
statistics on at most four answers carry the rate, because at the equitable point the tree functions see only squared eigencoordinates to
second order and the four-vertex trees resolve their signs at third order. Both rates hold, with the same
estimator, on relatively open classes of channels with equal diagonal entries, which include judges with one sensitivity for
every meaning and pair-dependent false-equivalence probabilities. On channels in $\cU$ with an admissible double pair,
two distributions with different entropies at which the four-vertex tree map takes equal values, five-vertex
statistics are needed as well: the risk is $\min\{1,N_4^{-1/3}\}+e^{-\Theta(N_5)}$. For $\Delta=n^{-\beta}$ with
$\beta>1/2$ the consistency threshold falls from $2/3$ to $5/8$ while the rate below it stays $N_4^{-1/3}$, so at $\beta=5/8$ the risk exponent drops from $1/12$ to $0$.
For almost every $w\in(0,\infty)^3$, the exponent of the diagonal channel $\mathrm{diag}(w)$ instead reaches $0$ continuously. The upper bound inverts tree statistics with a five-vertex weight and
uses a cubic entropy modulus, and the lower bounds follow two paths along which the channel moves.

\paragraph{Limitations.}
The number of meanings is fixed at three; the conditions and certificates for general $K$, where the depth of the
trees that resolve the equitable point may grow with $K$, are not worked out. The channel is assumed known, and the
verdicts are conditionally independent given the meanings. The estimator takes the channel as input, and how an
error in the channel changes its risk is not analyzed. Conditional independence enters the variance bound for
tree statistics and the likelihood of the verdicts, so for a judge whose verdicts are dependent given the meanings,
neither the upper nor the lower bound is established. The matching lower bound is proved on every fixed polynomial window
$n\Delta^2\le n^{-\kappa}$, with constants depending on $\kappa$, but not on the whole weak window; for $\Delta=n^{-\beta}$
these windows cover every fixed $\beta>1/2$, and what they leave out is $n\Delta^2$ tending to zero more slowly than
every power of $n$. For channels in $\cU$ with an admissible double pair the exponential term is determined only up to
its constant: the
upper and lower bounds decay like $e^{-c_UN_5}$ and $e^{-C_EN_5}$ with constants that are not identified and need not
agree, so the crossover
between the four-vertex rate and the cost of the double pair is located at $N_5$ of order $\log N_4$, only up to a
constant factor; at $N_5=\log n$, for example, the bounds do not decide which term dominates. The constants of the
lower bound for the double pair depend on the channel and the pair, and are not shown to be uniform over compact sets
of channels, so \Cref{thm:threshold} and \Cref{cor:phase}(ii) are statements about one channel at a time. The constants
$\eta$, $C$, $c_U$, $C_U$, $C_4$ and $c_L$ come from compactness, and $c_U$ also from the constant
of the concentration inequality; none are evaluated, and each is uniform on compact subsets of the class on which it is proved,
$\Omega$ for $c_L$, $\cU_4$ for $C_4$, $\cU$ for the others and each stratum of \Cref{thm:symmetric} for their analogues
there. How each constant behaves as a channel approaches the boundary of its class, for instance as two diagonal
entries approach each other, is not studied. That $\cU$ has full measure in the open set of channels whose equitable
points $\theta^*$ and $f_1,f_2,f_3$ are interior is not proved; the evidence is the random samples,
in which every failure inside that set came from an exact coincidence of entries. The
certificates rely on computer algebra and interval arithmetic and cover the channels of \Cref{tab:certificates}, the
twenty grid channels, the $300$ channels of the two random samples, the box of radius $5\times10^{-7}$ around $M_0$ on which
every channel has an admissible double pair, and the box of radius $0.01$ inside $\Omega$; \Cref{prop:class}(b) to (e)
and the classes of $M_3,\dots,M_9$ in \Cref{thm:symmetric} rest on them. They do not show that the box of radius
$5\times10^{-7}$ lies in $\cU$, which openness gives only on an unspecified neighborhood of $M_0$, so this radius is not a
certified radius for \Cref{cor:phase}(ii). For the $396$ double pairs of the first random sample, the pairwise
distinctness of the entries of $M\theta$ that \Cref{thm:threshold} needs was checked in floating point and is not
certified.

\paragraph{Outlook.}
For general $K$ the equitable point remains a point where every tree function is stationary; whether the depth
at which trees resolve its reflections determines the local rate is open. A proof that $\cU$ has full measure in that open set
would extend the rate to almost every channel whose equitable points are interior. The exact exponent of the cost of a
double pair, a Chernoff-type constant of the test between its two points, would locate the crossover in $N_5$
precisely. With an unknown channel the cyclic motifs may carry information that tree statistics ignore.

\paragraph{Code availability.}
The scripts that produce every certificate of this paper and the numerical checks of \Cref{app:numerics}, together with their
outputs and the matrices of every certified channel, are in the repository \url{https://github.com/yahiko-l/pair-dependent-semantic-entropy}; everything runs on a CPU.

\section*{Acknowledgment}
This work was supported by the Young Doctoral Talent Incubation Program of the Second Affiliated Hospital, Army Medical
University, under Grant 2025YQB034.

\bibliographystyle{plainnat}
\bibliography{references}

\clearpage
\appendix
\counterwithin{theorem}{section}
\counterwithin{equation}{section}
\phantomsection\addcontentsline{toc}{section}{Appendix}%
\noindent{\Large\bfseries Appendix}\par
\begin{figure}[h]
  \centering
  \includegraphics[width=\linewidth]{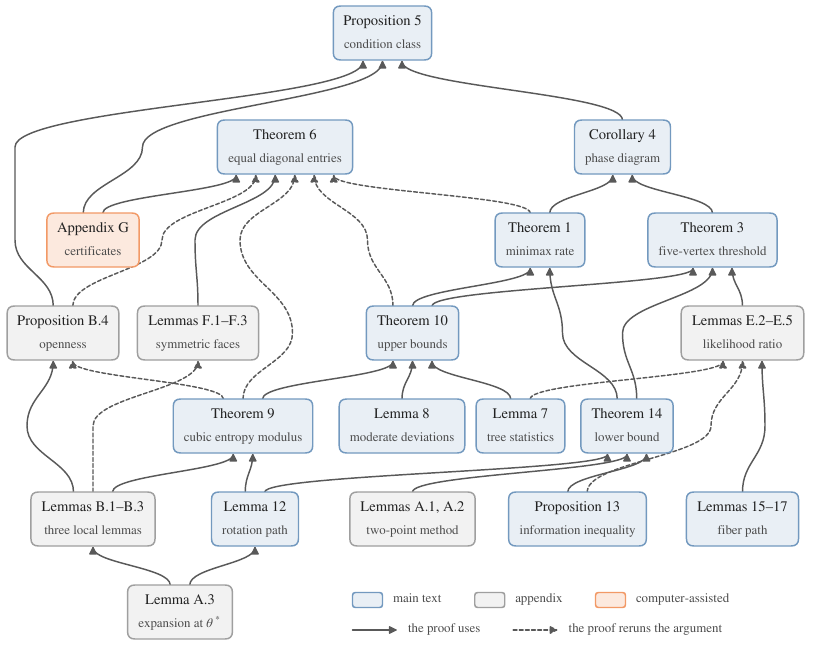}
  \caption{How the proofs depend on one another. An arrow from one result to another means that the proof of the
  second uses the first, and a dashed arrow that it reruns an argument from the proof of the first, with the results
  that argument uses: \Cref{thm:symmetric} reruns the proofs of \Cref{thm:modulus,thm:upper,thm:main} and of
  \Cref{prop:b:open} on each stratum, adding \Cref{lem:f:sym,lem:f:mid,lem:f:pairs}, and takes the classes of
  $M_3,\dots,M_9$ from the certificates of \Cref{app:certificates}; \Cref{prop:b:open} reruns the covering argument of
  the proof of \Cref{thm:modulus}(a); \Cref{lem:f:mid} reruns, for another map, the lower bound in the proof of
  \Cref{lem:b:rank}; and \Cref{lem:e:decomp,lem:e:error,lem:e:main,lem:e:stop} use the objects and steps of the proof of
  \Cref{prop:info} and, in \Cref{lem:e:main}, the factorization in the proof of \Cref{lem:ustat}. Not shown are
  \Cref{lem:move}, which shows why the path of the lower bound must move the channel and on which no
  proof depends, \Cref{lem:a:face,lem:a:entropy,lem:a:lip}, used in the proofs of
  \Cref{lem:b:rank,lem:b:face,lem:f:mid,lem:f:pairs} and of \Cref{thm:upper}, and \Cref{lem:e:gauss} on the Gauss
  nodes, used in the proof of \Cref{lem:fibers}.}
  \label{fig:roadmap}
\end{figure}

\Cref{fig:roadmap} summarizes how the proofs depend on one another.

\section{Notation, tools and tree-function identities}
\label{app:tools}

\paragraph{Notation.}
$\Simplex$ is the closed simplex, $\aff\Simplex=\{\theta:\sum_c\theta_c=1\}$ its affine hull and $\one^\perp$ its tangent
plane. $|\cdot|$ is the Euclidean norm, $|\cdot|_\infty$ the maximum norm on $\Real^4$ and $\Real^7$, and for matrices,
$\|M\|=\|M\|_\infty:=\max_{ab}|M_{ab}|$, $\|\cdot\|_{\rm op}$ is the operator norm and $\|\cdot\|_F$ the Frobenius norm. $\bar B(\theta,r)$ is a closed ball in $\aff\Simplex$ and $\sigma_{\min}$ the smallest
singular value of a linear map. The entropy uses the natural logarithm; $\varphi(x):=-x\log x$ with $\varphi(0):=0$,
$\omega(\delta):=\delta(2+\log(1/\delta))$ for $0<\delta\le1$ with $\omega(0):=0$, and
$\Xi_a:=\sup_{0<\delta\le1}\delta^a(2+\log(1/\delta))^3$, which is finite for every $a>0$. For a graph $G$,
$\hom(G;\theta,M)=\sum_z\prod_{x}\theta_{z_x}\prod_{xy\in E(G)}M_{z_xz_y}$. Estimators may use auxiliary randomness
independent of the data.

\needspace{5\baselineskip}
\subsection{Two general tools}

The lower bounds use two facts about families of distributions on a finite set; they do not involve the
channel.

\begin{lemma}[Hellinger path bound]
  \label{lem:a:hellinger}
  Let $\{P_t\}_{t\in[t_-,t_+]}$ be positive probability distributions on a finite set $\mathcal A$, continuously
  differentiable in $t$, and $I(t):=\sum_{A\in\mathcal A}P_t(A)(\partial_t\log P_t(A))^2$. Then
  $\TV(P_{t_-},P_{t_+})\le\mathcal H(P_{t_-},P_{t_+})\le\frac12\int_{t_-}^{t_+}\sqrt{I(t)}\,\mathrm dt$, where
  $\mathcal H(P,Q):=(\sum_A(\sqrt{P(A)}-\sqrt{Q(A)})^2)^{1/2}$. If $[0,1]$ is split into finitely many pieces on each of
  which the family is continuously differentiable, then
  $\TV(P_0,P_1)\le\frac12\sum_j\int_{t_{j-1}}^{t_j}\sqrt{I(t)}\,\mathrm dt$.
\end{lemma}

\begin{proof}
  By Cauchy--Schwarz, $\sum_A|P(A)-Q(A)|\le\mathcal H(P,Q)(\sum_A(\sqrt{P(A)}+\sqrt{Q(A)})^2)^{1/2}\le2\mathcal H(P,Q)$, so
  $\TV(P,Q)=\frac12\sum_A|P(A)-Q(A)|\le\mathcal H(P,Q)$. The map $t\mapsto(\sqrt{P_t(A)})_A\in\Real^{\mathcal A}$ is continuously
  differentiable with $\partial_t\sqrt{P_t(A)}=\frac12\sqrt{P_t(A)}\,\partial_t\log P_t(A)$, so the length of the curve bounds
  the distance between its ends:
  $\mathcal H(P_{t_-},P_{t_+})\le\int_{t_-}^{t_+}(\sum_A(\partial_t\sqrt{P_t(A)})^2)^{1/2}\mathrm dt=\frac12\int_{t_-}^{t_+}\sqrt{I(t)}\,\mathrm dt$.
  For several pieces, $\mathcal H$ is the Euclidean distance between the vectors $(\sqrt{P(A)})_A$ and satisfies the
  triangle inequality.
\end{proof}

\begin{lemma}[two-point bound]
  \label{lem:a:lecam}
  Let $\omega\ge0$. If $\theta^{(0)},\theta^{(1)}\in\Simplex$ satisfy $|\Ent(\theta^{(0)})-\Ent(\theta^{(1)})|\ge\omega$ and the laws of
  the observations under the same channel satisfy $\TV(P_{\theta^{(0)}},P_{\theta^{(1)}})\le1/2$, then $R^*_n(M)\ge\omega^2/16$.
\end{lemma}

\begin{proof}
  For an estimator $\hH$ let $\mathcal E_i:=\{|\hH-\Ent(\theta^{(i)})|\ge\omega/2\}$. Since the two entropies are at least $\omega$
  apart, $\mathcal E_0^c\subseteq\mathcal E_1$, so
  $P_{\theta^{(0)}}(\mathcal E_0)+P_{\theta^{(1)}}(\mathcal E_1)\ge P_{\theta^{(0)}}(\mathcal E_0)+P_{\theta^{(1)}}(\mathcal E_0^c)\ge1-\TV\ge1/2$; auxiliary
  randomness enters both laws in the same way. One of the two probabilities is at least $1/4$, and by Markov's
  inequality the corresponding risk is at least $(\omega/2)^2/4=\omega^2/16$
  \citep[Section~2.3]{tsybakov2009introduction}.
\end{proof}

\subsection{Tree functions at the equitable point and on the faces}

\begin{lemma}[expansion at the equitable point]
  \label{lem:a:expansion}
  Let $B$ be symmetric with $Bu=cu$ for a unit vector $u$ with positive entries, $D_u=\diag(u)$, $\psi\in u^\perp$, $s\in\Real$,
  and $T=(V,E)$ a tree on $v$ vertices with distance $d_T$ and degrees $\deg$. Then
  \[
    \begin{aligned}
      \hom\big(T;D_u(u+s\psi),D_u^{-1}BD_u^{-1}\big)&=c^{v-1}+s^2\sum_{\{a,b\}}c^{v-1-d_T(a,b)}\psi^\top B^{d_T(a,b)}\psi\\
      &\quad+\sum_{S\subseteq V,|S|\ge3}s^{|S|}h_S(\psi,B),
    \end{aligned}
  \]
  where the first sum runs over unordered pairs of distinct vertices and
  \[
    h_S(\psi,B):=\sum_{z\in[3]^V}\prod_{x\in V}u_{z_x}^{2-\deg(x)}\prod_{x\in S}\frac{\psi_{z_x}}{u_{z_x}}\prod_{xy\in E}B_{z_xz_y}
  \]
  is homogeneous of degree $|S|$ in $\psi$. If $R$ is orthogonal with $Ru=u$, the terms with $|S|\le2$ are invariant under
  $(\psi,B)\mapsto(R\psi,RBR^\top)$. For a connected unicyclic graph on $v$ vertices whose cycle has length $k$, the
  term without $s$ in the same expansion is $c^{v-k}\operatorname{tr}(B^k)$.
\end{lemma}

\begin{proof}
  With $\theta=D_u(u+s\psi)$ and $M=D_u^{-1}BD_u^{-1}$, we have $\theta_{z_x}=u_{z_x}^2(1+s\psi_{z_x}/u_{z_x})$ and
  $\prod_EM=\prod_EB\cdot\prod_xu_{z_x}^{-\deg(x)}$; expanding $\prod_x(1+s\psi_{z_x}/u_{z_x})$ gives
  $\hom=\sum_Ss^{|S|}h_S$. If $y\notin S$ is a leaf attached to $x$, then $\sum_{z_y}u_{z_y}B_{z_xz_y}=(Bu)_{z_x}=cu_{z_x}$, which is
  $c$ times the same expression without $y$, the degree of $x$ dropping by one. Repeatedly removing leaves outside $S$
  yields the minimal subtree spanning $S$ (one vertex if $S=\emptyset$). For $S=\emptyset$ this gives $c^{v-1}\sum_zu_z^2=c^{v-1}$;
  for $S=\{x\}$ it gives $c^{v-1}\langle u,\psi\rangle=0$; for $S=\{a,b\}$ it leaves a path of length $k=d_T(a,b)$ whose
  inner vertices have exponent $0$ and whose ends contribute $\psi$, giving $c^{v-1-k}\psi^\top B^k\psi$. These terms are
  invariant under the simultaneous rotation because $(R\psi)^\top(RBR^\top)^k(R\psi)=\psi^\top B^k\psi$. For a unicyclic graph the
  $s$-free term is computed by removing the $v-k$ tree edges hanging off the cycle, each contributing $c$, and the
  cycle then gives $\sum_z\prod B=\operatorname{tr}(B^k)$ because its vertices have degree $2$ and exponent $0$.
\end{proof}

With $B=\tM$, $s=1$ and $\psi=x$, \Cref{lem:a:expansion} gives \eqref{eq:expansion}, because
$\tM^de^\perp_i=\lambda_i^de^\perp_i$ and $e^\perp_1\perp e^\perp_2$. For the edge, $F_{K_2}(\theta^*+D_ux)=c+x^\top\tM x$ exactly. For the
wedge, $F_{P_3}(\theta^*+D_ux)=c^2+2cx^\top\tM x+x^\top\tM^2x+\tau(x)$ with the cubic
$\tau(x):=\sum_b(\tM x)_b^2x_b/u_b$, and no higher terms.

\begin{lemma}[stationarity on the faces]
  \label{lem:a:face}
  Let $f\in\Simplex$ be supported on $\{i,j\}$ with $(Mf)_i=(Mf)_j=:c_f$. Then for every tree $T$,
  $\frac{\mathrm d}{\mathrm dt}F_T(f+t(e_i-e_j))\big|_{t=0}=0$.
\end{lemma}

\begin{proof}
  By multilinearity the derivative is $\sum_{x\in V(T)}\sum_z(e_i-e_j)_{z_x}\prod_{y\ne x}f_{z_y}\prod_{yy'\in E(T)}M_{z_yz_{y'}}$. Fix $x$
  and root $T$ at $x$. For $y\ne x$ with subtree $T_y$ below it, let $\mu_y(a):=\sum_bM_{ab}f_b\prod_{y'\text{ child of }y}\mu_{y'}(b)$,
  an empty product being $1$. By induction from the leaves, $\mu_y(a)=c_f^{|T_y|}$ for $a\in\{i,j\}$: $f_b\ne0$ only for
  $b\in\{i,j\}$, where the product equals $c_f^{|T_y|-1}$, so $\mu_y(a)=c_f^{|T_y|-1}(Mf)_a$. The term of $x$ is
  $\sum_a(e_i-e_j)_a\prod_{y\text{ child of }x}\mu_y(a)$, which has two nonzero terms, $\pm c_f^{v-1}$, that cancel.
\end{proof}

By (F0) the face equitable points satisfy the hypothesis, so $\phi^T_{10}=0$ for every tree in (U2).

\subsection{Entropy and Lipschitz bounds}

\begin{lemma}[continuity of the entropy]
  \label{lem:a:entropy}
  For $x,x'\in[0,1]$ with $\delta:=|x-x'|$ we have $|\varphi(x)-\varphi(x')|\le\omega(\delta)$. The function $\omega$ is increasing on
  $[0,1]$ with $\omega(\delta)\ge2\delta$. Hence $|\Ent(\theta)-\Ent(\theta')|\le3\,\omega(|\theta-\theta'|)$ for
  $\theta,\theta'\in\Simplex$ with $|\theta-\theta'|\le1$.
\end{lemma}

\begin{proof}
  Let $x<x'$. Since $|\varphi'(\xi)|\le1+\log(1/\xi)$ and the right side decreases on $(0,1]$, the integral over an
  interval of length $\delta$ is largest on $[0,\delta]$:
  $|\varphi(x')-\varphi(x)|\le\int_0^\delta(1+\log(1/\xi))\,\mathrm d\xi=\omega(\delta)$. Further $\omega'(\delta)=1+\log(1/\delta)>0$ and
  $\omega(\delta)-2\delta=\delta\log(1/\delta)\ge0$. The last statement follows by summing over the three coordinates.
\end{proof}

\begin{lemma}[Lipschitz bound]
  \label{lem:a:lip}
  For a tree $T$ on $v$ vertices and $\vartheta,\vartheta'\in\Simplex$,
  $|F_T(\vartheta)-F_T(\vartheta')|\le vU^{v-1}\|\vartheta-\vartheta'\|_1$; in particular $L_F=5\bar U^4$ is a Lipschitz constant in $\ell^1$
  for every $T\in\cT_5$.
\end{lemma}

\begin{proof}
  The multilinear form $\Psi_T(a^{(1)},\dots,a^{(v)}):=\sum_z\prod_xa^{(x)}_{z_x}\prod_{xy\in E(T)}M_{z_xz_y}$ satisfies
  $|\Psi_T|\le U^{v-1}\prod_x\|a^{(x)}\|_1$ and $F_T(\vartheta)=\Psi_T(\vartheta,\dots,\vartheta)$. Replacing one slot at a time,
  $F_T(\vartheta)-F_T(\vartheta')$ is a sum of $v$ terms, each at most $U^{v-1}\|\vartheta-\vartheta'\|_1$.
\end{proof}

\section{The cubic entropy modulus}
\label{app:modulus}

\subsection{The conditions in coordinates}

Let $L:=\big(\begin{smallmatrix}\lambda_1&\lambda_2\\ \lambda_1^2+2c\lambda_1&\lambda_2^2+2c\lambda_2\end{smallmatrix}\big)$, the
matrix of the coefficients $\kappa_{K_2,i},\kappa_{P_3,i}$ in \eqref{eq:expansion}; $\det L=\lambda_1\lambda_2(\lambda_2-\lambda_1)$, so $L$ is
invertible under (N3) and (N4), and $(\alpha_T,\beta_T)$ in (U1) is the unique solution of
$\alpha_T\kappa_{K_2,i}+\beta_T\kappa_{P_3,i}=\kappa_{T,i}$, $i=1,2$.

\paragraph{Independence of the choices.}
The pair $(\alpha_T,\beta_T)$ depends only on $\lambda_1,\lambda_2,c$. Replacing $e^\perp_1$ by $-e^\perp_1$ maps the coefficients
$(\gamma_{T,0},\gamma_{T,1},\gamma_{T,2},\gamma_{T,3})$ of $\tFthree_T$ to $(-\gamma_{T,0},\gamma_{T,1},-\gamma_{T,2},\gamma_{T,3})$, and replacing
$e^\perp_2$ by $-e^\perp_2$ maps them to $(\gamma_{T,0},-\gamma_{T,1},\gamma_{T,2},-\gamma_{T,3})$; neither change affects (U1)(i) to (iii).
Exchanging the labels of $e^\perp_1$ and $e^\perp_2$ exchanges (ii) and (iii). In (U2) the face coordinates are fixed.

\paragraph{Quantitative forms.}
(U1)(i) is equivalent to $m_1(M):=\min_{|\zeta|=1}\max_T|\tFthree_T(\zeta)|>0$, (U1)(ii) to
$m_2(M):=\min_{X,Y\ge0,X+Y=1}\max_T|\gamma_{T,0}X+\gamma_{T,2}Y|>0$, (U1)(iii) to the analogous $m_3(M)>0$, and (U2)(ii) to
$\min_{X,Y\ge0,X+Y=1}\max_T|\psi^T_{11}X+\psi^T_{30}Y|>0$, where $T$ ranges over $\{P_4,K_{1,3}\}$.

\paragraph{Continuous frames.}
Let $\bar M$ satisfy (N0), (N1), (N3) and (N4). These are strict conditions and hold near $\bar M$, where $\theta^*$, $c$, $u$,
$D_u$ and $\tM$ are continuous in $M$ (rational functions, and square roots of positive ones). Projecting two fixed
vectors onto $u(M)^\perp$ and applying Gram--Schmidt gives an orthonormal basis of $u(M)^\perp$ that is continuous near
$\bar M$; in it $\tM$ restricted to $u^\perp$ is a continuous $2\times2$ matrix with distinct eigenvalues, so the unit
eigenvectors $e^\perp_i(M)$ can be chosen continuously near $\bar M$. Then $\lambda_i$, $L^{-1}$, $\alpha_T$, $\beta_T$, $\gamma_{T,j}$ and
the coefficients of every polynomial map below are continuous near $\bar M$, and so are $m_1,m_2,m_3$. Near a channel
satisfying (F0), the face equitable points $f_k(M)$ and the coefficients $\phi^T_{ab}(M)$ are continuous as well, and near a channel
satisfying (F0) and (U2)(i) so are $J_k(M)^{-1}$, $\alpha'_T,\beta'_T$ and $\psi^T_{ab}$.

\subsection{Points of rank two}

\begin{lemma}[rank two]
  \label{lem:b:rank}
  Let $\bar M\in\Sym$ and $\bar\theta\in\Simplex$ with $\rank DF^{\bar M}_{\le4}(\bar\theta)|_{\one^\perp}=2$. There are $\eps>0$, $r\in(0,1/2]$ and
  $c_R>0$ such that for $\|M-\bar M\|\le\eps$ and $\theta,\theta'\in\Simplex\cap\bar B(\bar\theta,r)$:
  (a) $\cD^M_4(\theta,\theta')\ge c_R|\theta-\theta'|$; (b) $\rank DF^M_{\le4}(\theta)|_{\one^\perp}=2$;
  (c) $|\Ent(\theta)-\Ent(\theta')|^3\le C_R\,\cD^M_4(\theta,\theta')$ with $C_R:=27\Xi_2/c_R$.
\end{lemma}

\begin{proof}
  Let $\sigma:=\sigma_{\min}(DF^{\bar M}_{\le4}(\bar\theta)|_{\one^\perp})>0$. Since $(M,\theta)\mapsto DF^M_{\le4}(\theta)|_{\one^\perp}$ is polynomial,
  there are $\eps$, $r$ such that $\|DF^M_{\le4}(\theta)|_{\one^\perp}-DF^{\bar M}_{\le4}(\bar\theta)|_{\one^\perp}\|_{\rm op}\le\sigma/2$ for
  $\|M-\bar M\|\le\eps$, $\theta\in\aff\Simplex$ and $|\theta-\bar\theta|\le r$. This gives (b). On the convex ball,
  $F^M_{\le4}(\theta')-F^M_{\le4}(\theta)=DF^{\bar M}_{\le4}(\bar\theta)(\theta'-\theta)+\int_0^1[DF^M_{\le4}(\theta+\rho(\theta'-\theta))-DF^{\bar M}_{\le4}(\bar\theta)](\theta'-\theta)\,\mathrm d\rho$,
  so $|F^M_{\le4}(\theta')-F^M_{\le4}(\theta)|\ge(\sigma/2)|\theta'-\theta|$, and $|w|_\infty\ge|w|/2$ on $\Real^4$ gives (a) with
  $c_R:=\sigma/4$. For (c), $\delta:=|\theta-\theta'|\le2r\le1$, and by \Cref{lem:a:entropy},
  $|\Ent(\theta)-\Ent(\theta')|^3\le27\,\omega(\delta)^3=27\,\delta\cdot\delta^2(2+\log(1/\delta))^3\le27\Xi_2\delta\le(27\Xi_2/c_R)\cD^M_4$.
\end{proof}

\subsection{The equitable point}

\needspace{14\baselineskip}
\begin{lemma}[equitable point]
  \label{lem:b:star}
  Let $\bar M$ satisfy (N0), (N1), (N3), (N4) and (U1). There are $\eps,r>0$, $c_S>0$ and $C_S<\infty$ such that every $M$ with
  $\|M-\bar M\|\le\eps$ satisfies these four conditions of (N), $\bar B(\theta^*(M),r)\subset\{\theta\ge\theta^*(M)/2\}$, and for
  $\theta,\theta'\in\bar B(\theta^*(M),r)$:
  \begin{enumerate}[label=(\alph*),leftmargin=*]
    \item $\cD^M_4(\theta,\theta')\ge c_S\max\{|\theta-\theta^*|,|\theta'-\theta^*|\}^2|\theta-\theta'|$;
    \item $\rank DF^M_{\le4}(\theta)|_{\one^\perp}=2$ if $\theta\ne\theta^*(M)$;
    \item $|\Ent(\theta)-\Ent(\theta')|^3\le C_S\,\cD^M_4(\theta,\theta')$.
  \end{enumerate}
\end{lemma}

\begin{proof}
  \emph{Target coordinates.} Use continuous frames near $\bar M$. By \eqref{eq:expansion}, let
  \[
    X(x):=L^{-1}\begin{pmatrix}F_{K_2}(\theta^*+D_ux)-c\\F_{P_3}(\theta^*+D_ux)-c^2\end{pmatrix}=\begin{pmatrix}x_1^2\\x_2^2\end{pmatrix}+L^{-1}\begin{pmatrix}0\\F^{(3)}_{P_3}(x)\end{pmatrix},
  \]
  \[
    \begin{aligned}
    G_T(x)&:=F_T(\theta^*+D_ux)-c^3-\alpha_T\big(F_{K_2}-c\big)-\beta_T\big(F_{P_3}-c^2\big)\\
    &=\tFthree_T(x)+F^{(4)}_T(x),\qquad T\in\{P_4,K_{1,3}\},
    \end{aligned}
  \]
  where we used that the edge function has no terms beyond the second order, that the wedge function has none
  beyond the third, and that for a tree on four vertices the terms of degree at least four in
  \Cref{lem:a:expansion} form a single quartic form $F^{(4)}_T$. Then $(X,G_{P_4},G_{K_{1,3}})=\Lambda(F_{\le4}(\theta^*+D_ux)-F_{\le4}(\theta^*))$
  for an invertible linear map $\Lambda$ of $\Real^4$.

  \emph{Scaling.} For $s\in(0,1]$ and $\zeta$ in the closed unit disc $\bar{\mathbb D}$ (in the coordinates of the frame), let
  $\Phi_s(\zeta):=(s^{-2}X(s\zeta),s^{-3}G_{P_4}(s\zeta),s^{-3}G_{K_{1,3}}(s\zeta))$. By homogeneity
  \[
    \begin{gathered}
    \Phi_s=\Phi_0+s\Psi,\qquad\Phi_0(\zeta)=\big(\zeta_1^2,\zeta_2^2,\tFthree_{P_4}(\zeta),\tFthree_{K_{1,3}}(\zeta)\big),\\
    \Psi(\zeta)=\Big(L^{-1}\begin{pmatrix}0\\F^{(3)}_{P_3}(\zeta)\end{pmatrix},F^{(4)}_{P_4}(\zeta),F^{(4)}_{K_{1,3}}(\zeta)\Big),
    \end{gathered}
  \]
  with $\Psi$ a polynomial map that does not depend on $s$.

  \emph{The limit map.} Let $\mathcal S:=\{(\zeta,\zeta')\in\bar{\mathbb D}^2:\max\{|\zeta|,|\zeta'|\}=1\}$, a compact set. We claim that at
  $M=\bar M$ there is $c_0>0$ with $|\Phi_0(\zeta)-\Phi_0(\zeta')|\ge c_0|\zeta-\zeta'|$ on $\mathcal S$.
  \emph{Injectivity:} if $\Phi_0(\zeta)=\Phi_0(\zeta')$ with $(\zeta,\zeta')\in\mathcal S$ and $\zeta\ne\zeta'$, the first two coordinates give
  $\zeta_1'^2=\zeta_1^2$ and $\zeta_2'^2=\zeta_2^2$, so $|\zeta|=|\zeta'|=1$ and $\zeta'=(\epsilon_1\zeta_1,\epsilon_2\zeta_2)$ with signs $\epsilon_i$.
  If $\zeta'=(-\zeta_1,\zeta_2)$ with $\zeta_1\ne0$, then $\tFthree_T(\zeta)-\tFthree_T(\zeta')=2\zeta_1(\gamma_{T,0}\zeta_1^2+\gamma_{T,2}\zeta_2^2)=0$ for both
  $T$, against (U1)(ii) with $(X,Y)=(\zeta_1^2,\zeta_2^2)$. If $\zeta'=(\zeta_1,-\zeta_2)$ with $\zeta_2\ne0$, the difference is
  $2\zeta_2(\gamma_{T,3}\zeta_2^2+\gamma_{T,1}\zeta_1^2)$, against (U1)(iii) with $(X,Y)=(\zeta_2^2,\zeta_1^2)$. If $\zeta'=-\zeta$, the
  difference is $2\tFthree_T(\zeta)$, against (U1)(i). \emph{Immersion on the circle:} at $|\bar\zeta|=1$ the rows of
  $D\Phi_0(\bar\zeta)$ are $(2\bar\zeta_1,0)$, $(0,2\bar\zeta_2)$ and $\nabla\tFthree_T(\bar\zeta)$. If $\bar\zeta_1\bar\zeta_2\ne0$ the first two
  rows are independent. At $\bar\zeta=(0,\pm1)$ we have $\partial_1\tFthree_T(\bar\zeta)=\gamma_{T,2}$, and some $\gamma_{T,2}\ne0$ by (U1)(ii) with
  $(X,Y)=(0,1)$, so that row is independent of the second; at $\bar\zeta=(\pm1,0)$ we have $\partial_2\tFthree_T(\bar\zeta)=\gamma_{T,1}$, and some
  $\gamma_{T,1}\ne0$ by (U1)(iii). \emph{Compactness:} if the claim failed, there would be $(\zeta_n,\zeta'_n)\in\mathcal S$ with
  $|\Phi_0(\zeta_n)-\Phi_0(\zeta'_n)|<n^{-1}|\zeta_n-\zeta'_n|$; a subsequence converges to $(\bar\zeta,\bar\zeta')\in\mathcal S$ with
  $\Phi_0(\bar\zeta)=\Phi_0(\bar\zeta')$, so $\bar\zeta=\bar\zeta'$ by injectivity and $|\bar\zeta|=1$. Writing
  $\Phi_0(\zeta'_n)-\Phi_0(\zeta_n)=\int_0^1D\Phi_0(\zeta_n+\rho(\zeta'_n-\zeta_n))\,\mathrm d\rho\,(\zeta'_n-\zeta_n)$, the averaged Jacobian
  converges to $D\Phi_0(\bar\zeta)$, whose smallest singular value is positive, a contradiction.

  \emph{Uniformity in $M$.} The difference $\Phi^M_0-\Phi^{\bar M}_0$ is a polynomial map whose coefficients tend to zero as $M\to\bar M$, so its
  Lipschitz constant on the convex set $\bar{\mathbb D}$ tends to zero; the Lipschitz constant of $\Psi^M$ on $\bar{\mathbb D}$ is
  bounded near $\bar M$ by some $\Lambda_\Psi$. Shrinking $\eps$ so that $\mathrm{Lip}(\Phi^M_0-\Phi^{\bar M}_0)\le c_0/4$ and taking
  $s_0:=\min\{1,c_0/(4\Lambda_\Psi)\}$, we obtain $|\Phi^M_s(\zeta)-\Phi^M_s(\zeta')|\ge(c_0/2)|\zeta-\zeta'|$ on $\mathcal S$ for $s\le s_0$. In the
  same way $\sigma_1:=\min_{|\zeta|=1}\sigma_{\min}(D\Phi^{\bar M}_0(\zeta))>0$, and after shrinking $\eps$ and $s_0$ we have
  $\sigma_{\min}(D\Phi^M_s(\zeta))\ge\sigma_1/2$ for $|\zeta|=1$ and $s\le s_0$.

  \emph{Back to the simplex.} Let $r$ be a positive lower bound of $\min\{s_0\min_cu_c(M),\min_c\theta^*_c(M)/2\}$ near $\bar M$. For
  $\theta=\theta^*+D_ux$ and $\theta'=\theta^*+D_ux'$ in $\bar B(\theta^*,r)$ we have $|x|\le|\theta-\theta^*|/\min_cu_c\le s_0$. If
  $\theta=\theta'=\theta^*$ there is nothing to prove. Otherwise let $s:=\max\{|x|,|x'|\}\in(0,s_0]$, $\zeta:=x/s$, $\zeta':=x'/s$, so
  $(\zeta,\zeta')\in\mathcal S$. Since $s\le1$,
  \[
    \big|\Lambda\big(F_{\le4}(\theta)-F_{\le4}(\theta')\big)\big|\ge s^3|\Phi_s(\zeta)-\Phi_s(\zeta')|\ge\tfrac{c_0}2s^2|x-x'|,
  \]
  so $\cD_4(\theta,\theta')\ge\frac{c_0}{4\|\Lambda\|_{\rm op}}s^2|x-x'|$. Since $u_c\le1$, we have $|\theta-\theta^*|\le|x|$ and $|\theta-\theta'|\le|x-x'|$, so
  $s\ge\max\{|\theta-\theta^*|,|\theta'-\theta^*|\}$, which gives (a) with $c_S:=c_0/(4\sup\|\Lambda\|_{\rm op})$. For (c), the ball lies in
  $\{\theta\ge\theta^*/2\}$, where $|\partial_c\Ent|\le L_H:=1+\log(2/\theta_{\min})$ with $\theta_{\min}>0$ a lower bound of $\min_c\theta^*_c(M)$; so
  $|\Ent(\theta)-\Ent(\theta')|\le\sqrt3L_H|x-x'|$, and with $|x-x'|\le2s$ we have
  $|\Ent(\theta)-\Ent(\theta')|^3\le3\sqrt3L_H^3(2s)^2|x-x'|\le(48\sqrt3L_H^3\sup\|\Lambda\|_{\rm op}/c_0)\cD_4(\theta,\theta')$. For (b), with
  $s:=|x|>0$ the identity $(X,G)(x')=(s^2\Phi_{s,1:2}(x'/s),s^3\Phi_{s,3:4}(x'/s))$ for all $x'$ gives, at $x'=x$,
  $D(X,G)(x)=\diag(s,s,s^2,s^2)D\Phi_s(x/s)$, which has rank two.
\end{proof}

\subsection{The face equitable points}

For $(t,y)$ with $y\ge0$ let $\nu(t,y):=\max\{|t|,\sqrt y\}$; the monomial $t^ay^b$ has weighted degree $a+2b$.

\begin{lemma}[face equitable points]
  \label{lem:b:face}
  Let $\bar M$ satisfy (F0) and (U2) on the face $k$. There are $\eps,\rho>0$, $c_F>0$ and $C_F<\infty$ such that for
  $\|M-\bar M\|\le\eps$ (such $M$ also satisfy (F0)), with $V_k(M):=\{\theta_k(t,y;M):|t|\le\rho,\ 0\le y\le\rho^2\}\subset\Simplex$,
  for $\theta=\theta_k(t,y)$ and $\theta'=\theta_k(t',y')$ in $V_k(M)$ and $s:=\max\{\nu(t,y),\nu(t',y')\}$:
  \begin{enumerate}[label=(\alph*),leftmargin=*]
    \item $\cD^M_4(\theta,\theta')\ge c_F\max\{s^2|t-t'|,\ s|y-y'|\}$;
    \item $\rank DF^M_{\le4}(\theta)|_{\one^\perp}=2$ if $\theta\ne f_k(M)$;
    \item $|\Ent(\theta)-\Ent(\theta')|^3\le C_F\,\cD^M_4(\theta,\theta')$.
  \end{enumerate}
  Moreover there is $r_k>0$ with $\Simplex\cap\bar B(f_k(\bar M),r_k)\subset V_k(M)$ for all $\|M-\bar M\|\le\eps$.
\end{lemma}

\begin{proof}
  \emph{Weighted expansion.} We have $F_T(\theta_k(t,y))-F_T(f_k)=\sum_{(a,b)\ne(0,0)}\phi^T_{ab}t^ay^b$ with $\phi^T_{10}=0$ by \Cref{lem:a:face}. The
  remaining monomials have weighted degree at least two: $t^2$ and $y$ have degree two, $t^3$ and $ty$ degree three. Let
  \[
    \begin{gathered}
    X:=J_k^{-1}\begin{pmatrix}F_{K_2}-F_{K_2}(f_k)\\F_{P_3}-F_{P_3}(f_k)\end{pmatrix}=\begin{pmatrix}y\\t^2\end{pmatrix}+X^{[\ge3]},\\
    G_T:=F_T-F_T(f_k)-\alpha'_T\big(F_{K_2}-F_{K_2}(f_k)\big)-\beta'_T\big(F_{P_3}-F_{P_3}(f_k)\big),
    \end{gathered}
  \]
  so that $G_T=\psi^T_{11}ty+\psi^T_{30}t^3+G_T^{[\ge4]}$, the superscripts denoting weighted degrees; $(X,G)=\Lambda_k(F_{\le4}-F_{\le4}(f_k))$
  with $\Lambda_k$ invertible.

  \emph{Scaling.} Let $\mathcal R:=[-1,1]\times[0,1]$ and, for $s\in(0,1]$,
  $\Phi_s(\tau,\upsilon):=(s^{-2}X(s\tau,s^2\upsilon),s^{-3}G(s\tau,s^2\upsilon))$. Each monomial $t^ay^b$ scales as $s^{a+2b}$, so
  $\Phi_s=\Phi_0+s\Psi_s$ with
  \[
    \Phi_0(\tau,\upsilon)=\big(\upsilon,\ \tau^2,\ \psi^{P_4}_{11}\tau\upsilon+\psi^{P_4}_{30}\tau^3,\ \psi^{K_{1,3}}_{11}\tau\upsilon+\psi^{K_{1,3}}_{30}\tau^3\big),
  \]
  and $\Psi_s$ a polynomial in $(s,\tau,\upsilon)$ that, with its derivatives, is bounded on $[0,1]\times\mathcal R$.

  \emph{The limit map.} Let $\mathcal S_w:=\{(\zeta,\zeta')\in\mathcal R^2:\max\{\nu(\zeta),\nu(\zeta')\}=1\}$, compact. At $M=\bar M$ there is $c_0>0$
  with $|\Phi_0(\zeta)-\Phi_0(\zeta')|\ge c_0|\zeta-\zeta'|$ on $\mathcal S_w$. \emph{Injectivity:} $\Phi_0(\zeta)=\Phi_0(\zeta')$ gives $\upsilon=\upsilon'$ and
  $\tau'^2=\tau^2$; if $\zeta\ne\zeta'$ then $\tau'=-\tau\ne0$, and the last two coordinates differ by $2\tau(\psi^T_{11}\upsilon+\psi^T_{30}\tau^2)$, which
  cannot vanish for both $T$ by (U2)(ii) with $(X,Y)=(\upsilon,\tau^2)$. \emph{Immersion at $\nu(\bar\zeta)=1$:} the rows of $D\Phi_0$ are
  $(0,1)$, $(2\bar\tau,0)$ and $(\psi^T_{11}\bar\upsilon+3\psi^T_{30}\bar\tau^2,\ \psi^T_{11}\bar\tau)$; if $\bar\tau\ne0$ the first two are independent;
  if $\bar\tau=0$ then $\bar\upsilon=1$, the last rows are $(\psi^T_{11},0)$, and some $\psi^T_{11}\ne0$ by (U2)(ii) with $(X,Y)=(1,0)$.
  Compactness concludes as in \Cref{lem:b:star}; $\mathcal R$ is convex.

  \emph{Uniformity and scaling back.} As in \Cref{lem:b:star} there is $s_0$ such that for $s\le s_0$ and $\|M-\bar M\|\le\eps$ we have
  $|\Phi^M_s(\zeta)-\Phi^M_s(\zeta')|\ge\frac{c_0}2|\zeta-\zeta'|$ on $\mathcal S_w$, and, with
  $\sigma_w:=\min_{\nu(\zeta)=1}\sigma_{\min}(D\Phi^{\bar M}_0(\zeta))>0$ by the immersion above, $\sigma_{\min}(D\Phi^M_s(\zeta))\ge\sigma_w/2$ for
  $\nu(\zeta)=1$. Let $\underline p>0$ be a lower bound of $\min\{p_k(M),1-p_k(M)\}$ for $\|M-\bar M\|\le\eps$, and let
  $\rho\le\min\{s_0,\underline p/4\}$. For $\theta\ne\theta'$ in $V_k$ we have $s\in(0,\rho]$, and $\zeta:=(t/s,y/s^2)$ and $\zeta':=(t'/s,y'/s^2)$ lie in
  $\mathcal R$ with $(\zeta,\zeta')\in\mathcal S_w$, so
  \[
    \big|\Lambda_k\big(F_{\le4}(\theta)-F_{\le4}(\theta')\big)\big|\ge s^3|\Phi_s(\zeta)-\Phi_s(\zeta')|\ge\tfrac{c_0}2s^3\max\Big\{\frac{|t-t'|}s,\frac{|y-y'|}{s^2}\Big\},
  \]
  which gives (a) with $c_F:=c_0/(4\sup\|\Lambda_k\|_{\rm op})$. For (b), with $s=\nu(t,y)>0$, the matrix
  $D(X,G)(t,y)=\diag(s^2,s^2,s^3,s^3)D\Phi_s(\zeta)\diag(1/s,1/s^2)$ has rank two.

  \emph{Entropy.} We have $\Ent(\theta_k(t,y))=\varphi(p_k+t-y/2)+\varphi(1-p_k-t-y/2)+\varphi(y)$. By the choice of $\rho$ the first two arguments
  lie in $[\underline p/2,1]$, where $|\varphi'|\le L_p:=1+\log(2/\underline p)$, so the first two terms change by at most
  $2L_p(|t-t'|+|y-y'|)$, and the third by at most $\omega(|y-y'|)$ (\Cref{lem:a:entropy}). Since $2|y-y'|\le\omega(|y-y'|)$, we have
  $|\Ent(\theta)-\Ent(\theta')|\le2L_p|t-t'|+(L_p+1)\omega(|y-y'|)$, and with $(a+b)^3\le4(a^3+b^3)$ this gives
  $|\Ent(\theta)-\Ent(\theta')|^3\le32L_p^3|t-t'|^3+4(L_p+1)^3\omega(|y-y'|)^3$. First term: $|t-t'|\le2s$, so
  $|t-t'|^3\le4s^2|t-t'|\le(4/c_F)\cD_4$. Second term: $\delta:=|y-y'|\le s^2$, because $0\le y,y'\le s^2$, and $\delta\le\cD_4/(c_Fs)$, so
  \[
    \omega(\delta)^3=\delta\cdot\delta^{3/2}\cdot\delta^{1/2}\big(2+\log(1/\delta)\big)^3\le\frac{\cD_4}{c_Fs}\cdot s^3\cdot \Xi_{1/2}\le\frac{\Xi_{1/2}}{c_F}\cD_4 .
  \]
  This gives (c) with $C_F:=(128L_p^3+4(L_p+1)^3\Xi_{1/2})/c_F$. The logarithm of the normal entropy term $-y\log y$ is absorbed
  by the extra power of $s$ supplied by $\delta\le s^2$ and the weighted lower bound $\cD_4\gtrsim s|y-y'|$.

  \emph{Neighborhood.} The map $\theta\mapsto(t,y)$ is an affine isomorphism from $\aff\Simplex$ onto $\Real^2$ with $y=\theta_k$ and
  $t=\frac12(\theta_i-\theta_j-(2p_k-1))$, so $|t|,|y|\le|\theta-f_k(M)|$. Take $r_k:=\rho^2/2$ and shrink $\eps$ so that
  $|f_k(M)-f_k(\bar M)|\le\rho^2/2$; then $\theta\in\Simplex\cap\bar B(f_k(\bar M),r_k)$ gives $|t|\le\rho^2\le\rho$ and $0\le y\le\rho^2$.
\end{proof}

\subsection{Proof of \texorpdfstring{\Cref{thm:modulus}(a)}{the cubic modulus}}

Fix a compact $\mathcal K\subset\cU$. \emph{Near pairs.} For $(\bar M,\bar\theta)\in\mathcal K\times\Simplex$ choose $\eps(\bar M,\bar\theta)>0$, a relatively
open $W(\bar M,\bar\theta)\ni\bar\theta$ in $\Simplex$ and a constant $C(\bar M,\bar\theta)$ such that $|\Ent(\theta)-\Ent(\theta')|^3\le C\,\cD^M_4(\theta,\theta')$
for $\|M-\bar M\|<\eps$ and $\theta,\theta'\in W$: if $\bar\theta=\theta^*(\bar M)$, use \Cref{lem:b:star}, shrink $\eps$ so that
$|\theta^*(M)-\bar\theta|<r/2$, and take $W:=\Simplex\cap B(\bar\theta,r/2)\subset\bar B(\theta^*(M),r)$; if $\bar\theta=f_k(\bar M)$, use
\Cref{lem:b:face} with $W:=\Simplex\cap B(f_k(\bar M),r_k)$; otherwise $\bar\theta$ has rank two by (U3) and \Cref{lem:b:rank} applies
with $W:=\Simplex\cap B(\bar\theta,r)$. The open sets $B(\bar M,\eps)\times W$ cover the compact set $\mathcal K\times\Simplex$; take a finite
subcover, its Lebesgue number $\delta_0>0$ for the metric $\max\{\|M-M'\|,|\theta-\theta'|\}$, and $C_{\rm near}$ the largest of the finitely
many constants. For $M\in\mathcal K$ and $|\theta-\theta'|<\delta_0$ the set $\{(M,\theta),(M,\theta')\}$ has diameter less than $\delta_0$ and lies in
one element of the subcover, so $|\Ent(\theta)-\Ent(\theta')|^3\le C_{\rm near}\cD^M_4(\theta,\theta')$.

\emph{Far pairs.} The set $Z:=\{(M,\theta,\theta')\in\mathcal K\times\Simplex^2:|\theta-\theta'|\ge\delta_0,\ \cD^M_4(\theta,\theta')=0\}$ is compact, and $\cD^M_5>0$ on
$Z$ by (U4). Let $\eta_1:=\min_Z\cD_5$ ($\eta_1:=1$ if $Z=\emptyset$) and $\eta:=\eta_1/2$. The compact set
$K':=\{(M,\theta,\theta')\in\mathcal K\times\Simplex^2:|\theta-\theta'|\ge\delta_0,\ \cD^M_5\le\eta\}$ is disjoint from $Z$, so $d_4:=\min_{K'}\cD_4>0$
($d_4:=1$ if $K'=\emptyset$). Since $0\le\Ent\le\log3$, on $K'$ we have $|\Ent(\theta)-\Ent(\theta')|^3\le(\log3)^3\cD_4/d_4$. Every pair
with $\cD^M_5\le\eta$ is either near or in $K'$, so $C:=\max\{C_{\rm near},(\log3)^3/d_4\}$ works. \qed

\subsection{Openness}

\begin{proposition}[openness]
  \label{prop:b:open}
  $\cU$ and $\cU_4$ are open in $\Sym$.
\end{proposition}

\begin{proof}
  Let $\bar M\in\cU$. (N), (F0), (U1) and (U2) are strict conditions on quantities that are continuous near $\bar M$ (the
  quantitative forms above and $\det J_k$), so they hold on a neighborhood. For $\bar\theta\in\Simplex$ take $(\eps,W)$ as in the
  proof of \Cref{thm:modulus}(a), now using only parts (a) and (b) of the local lemmas; by compactness of $\Simplex$
  finitely many $W_1,\dots,W_L$ cover $\Simplex$; let $\eps$ be the smallest of the $\eps_l$ and $\delta_0$ the Lebesgue number of the
  cover. (U3): if $\|M-\bar M\|<\eps$ and $\theta\in(\Simplex\setminus\{\theta^*(M),f_1(M),f_2(M),f_3(M)\})\cap W_l$, then part (b) of the
  lemma that produced $W_l$ gives rank two. (U4): suppose $M_n\to\bar M$ and $\theta_n\ne\theta'_n$ with
  $F^{M_n}_{\le5}(\theta_n)=F^{M_n}_{\le5}(\theta'_n)$; pass to a subsequence with $(\theta_n,\theta'_n)\to(\bar\theta,\bar\theta')$. If
  $\bar\theta\ne\bar\theta'$, continuity gives $F^{\bar M}_{\le5}(\bar\theta)=F^{\bar M}_{\le5}(\bar\theta')$, against (U4) at $\bar M$. If
  $\bar\theta=\bar\theta'$, for large $n$ both points lie in one $W_l$, and part (a) of the corresponding lemma gives
  $\cD^{M_n}_4(\theta_n,\theta'_n)>0$, a contradiction. The same argument with $F_{\le4}$ in place of $F_{\le5}$ shows that $\cU_4$ is
  open.
\end{proof}

\subsection{Proof of \texorpdfstring{\Cref{thm:modulus}(b) and (c)}{the remaining parts}}

(b) The near part of the proof of (a) uses only $\mathcal K\subset\cU$. For far pairs, the set
$K'':=\{(M,\theta,\theta')\in\mathcal K\times\Simplex^2:|\theta-\theta'|\ge\delta_0\}$ is compact and $\cD_4>0$ on it because $F^M_{\le4}$ is injective, so
$d''_4:=\min_{K''}\cD_4>0$ ($d''_4:=1$ if $K''=\emptyset$), and $C_4:=\max\{C_{\rm near},(\log3)^3/d''_4\}$ works.

(c) If $F^M_{\le4}(\theta)=F^M_{\le4}(\theta')$ and $\Ent(\theta)\ne\Ent(\theta')$, the left side of the inequality in (b) is positive and
the right side vanishes. For the second statement let $M\in\Omega$, $\eta>0$ and $p\in(0,3)$, and take the points $\theta_\pm=\theta_\pm(s)$
of \Cref{lem:rotation} for one fixed $r\ge4$. The edge and wedge functions agree at them. For a tree $T$ on four vertices,
both endpoints carry the channel $M$, so by \Cref{lem:rotation}(ii),
$|F^M_T(\theta_+)-F^M_T(\theta_-)|=|\int_0^1\partial_tF_T(\theta_t;M_t)\,\mathrm dt|\le Ls^3$; hence $\cD^M_4(\theta_+,\theta_-)\le Ls^3$, while
$|\Ent(\theta_+)-\Ent(\theta_-)|\ge s|\ell(\zeta_0)|$. Both points tend to $\theta^*$ as $s\to0$, so $\cD^M_5(\theta_+,\theta_-)\le\eta$ for small $s$,
and $|\Ent(\theta_+)-\Ent(\theta_-)|^p/\cD^M_4(\theta_+,\theta_-)\ge|\ell(\zeta_0)|^ps^{p-3}/L\to\infty$ (if $\cD^M_4=0$ the ratio is infinite).
So no bound $|\Ent(\theta)-\Ent(\theta')|\le C'\cD^M_4(\theta,\theta')^{1/p}$ with $p\in(0,3)$ holds on pairs with $\cD^M_5\le\eta$. \qed

\section{The upper bound}
\label{app:upper}

Throughout, $W_{ij}:=\Exp[X_{ij}\mid Z]=M_{Z_iZ_j}$ and $\sigma^2_{ij}:=\Var(X_{ij}\mid Z)=p_{Z_iZ_j}(1-p_{Z_iZ_j})/\Delta^2\le\rho$, where
$p_{ab}:=q+\Delta M_{ab}$ and $\rho:=\max_{ab}p_{ab}(1-p_{ab})/\Delta^2$. For injective maps $\varphi,\varphi'$ let
$E_\varphi:=\{\{\varphi(x),\varphi(y)\}:xy\in E(T)\}$, $Y_\varphi:=\prod_{xy\in E(T)}X_{\varphi(x)\varphi(y)}$, $s:=|\operatorname{im}\varphi\cap\operatorname{im}\varphi'|$
and $t:=|E_\varphi\cap E_{\varphi'}|$. We write $\delta_m:=m\Delta^2$, so $N_v(m)=m\delta_m^{v-1}$.

\subsection{Proof of \texorpdfstring{\Cref{lem:ustat}}{the tree-statistic lemma}}

\emph{Unbiasedness.} For injective $\varphi$ the $v-1$ pairs in $E_\varphi$ are distinct and not diagonal. Given $Z$ they are
independent with means $W_e$, so $\Exp[Y_\varphi\mid Z]=\prod_{xy\in E(T)}M_{Z_{\varphi(x)}Z_{\varphi(y)}}$; the labels $Z_{\varphi(1)},\dots,Z_{\varphi(v)}$ are
i.i.d.\ from $\theta$, so $\Exp Y_\varphi=F_T(\theta;M)$.

\emph{Covariance of two embeddings.} If $\operatorname{im}\varphi\cap\operatorname{im}\varphi'=\emptyset$, then $Y_\varphi$ and $Y_{\varphi'}$ are functions of
$(Z_{S_1},A_{S_1})$ and $(Z_{S_2},A_{S_2})$ for disjoint $S_1,S_2$, where $A_S:=(A_{ij})_{i<j,\,i,j\in S}$. Their joint law is obtained
from $\prod_i\theta_{Z_i}\prod_{i<j}\Bern(A_{ij};p_{Z_iZ_j})$ by summing out the other verdicts and labels, each Bernoulli factor and
each $\theta$ factor summing to one, and it factorizes; so the two are independent and the covariance vanishes. If
$s\ge1$, then $E_\varphi\cap E_{\varphi'}$ is a set of edges of the tree $\varphi(T)$, hence a forest, whose edges have both ends in
$\operatorname{im}\varphi\cap\operatorname{im}\varphi'$; a forest with $t\ge1$ edges has at least $t+1$ vertices, so $t\le s-1$, which also holds for
$t=0$. Given $Z$, we have $Y_\varphi Y_{\varphi'}=\prod_{e\in E_\varphi\triangle E_{\varphi'}}X_e\prod_{e\in E_\varphi\cap E_{\varphi'}}X_e^2$ with distinct pairs, so
$\Exp[Y_\varphi Y_{\varphi'}\mid Z]=\prod_{e\in E_\varphi\triangle E_{\varphi'}}W_e\prod_{e\in E_\varphi\cap E_{\varphi'}}(W_e^2+\sigma^2_e)$, and
\[
  \big|\Exp[Y_\varphi Y_{\varphi'}]\big|\le U^{2(v-1-t)}(U^2+\rho)^t\le U^{2(v-s)}(U^2+\rho)^{s-1},
\]
because the middle expression does not decrease in $t$ and $t\le s-1$. Also $|\Exp Y_\varphi\,\Exp Y_{\varphi'}|\le U^{2(v-1)}$, which is
at most the same bound. Hence $|\Cov(Y_\varphi,Y_{\varphi'})|\le2U^{2(v-s)}(U^2+\rho)^{s-1}$.

\emph{Counting.} For $1\le s\le v$ the number of ordered pairs $(\varphi,\varphi')$ with $|\operatorname{im}\varphi\cap\operatorname{im}\varphi'|=s$ is
$(m)_v\binom vs(v)_s(m-v)_{v-s}$: choose $\varphi$, the $s$ vertices $x$ with $\varphi'(x)\in\operatorname{im}\varphi$, their images in
$\operatorname{im}\varphi$, and the images of the other $v-s$ vertices outside $\operatorname{im}\varphi$. Since $(m-v)_{v-s}\le m^{v-s}$, and
$(m)_v\ge(m-v+1)^v\ge(m/2)^v$ for $m\ge2v-2$, their fraction among all $(m)_v^2$ pairs is at most
$\binom vs(v)_s2^vm^{-s}$.

\emph{Variance.} Summing over $s\ge1$,
\[
  \Var(\widehat U_T[S])\le2^{v+1}\sum_{s=1}^v\binom vs(v)_s\frac{U^{2(v-s)}(U^2+\rho)^{s-1}}{m^s}.
\]
By \eqref{eq:P}, $\Delta U\le1$ and $\rho\le1/(4\Delta^2)$, so $U^2+\rho\le5/(4\Delta^2)$ and
\[
  \frac{U^{2(v-s)}(U^2+\rho)^{s-1}}{m^s}\le\bar U^{2v-2}\Big(\frac54\Big)^{v-1}\frac1{m\delta_m^{s-1}}\le\bar U^{2v-2}\Big(\frac54\Big)^{v-1}\max\Big\{\frac1m,\frac1{N_v(m)}\Big\},
\]
using $1/(m\delta_m^{s-1})\le1/m$ if $\delta_m\ge1$ and $\le1/N_v(m)$ if $\delta_m<1$. With $\sum_{s=1}^v\binom vs(v)_s\le2^vv!$ this gives
the bound with $B_v=2^{2v+1}v!(5/4)^{v-1}\bar U^{2v-2}$. Numerically $B_2=80\bar U^2$, $B_3=1200\bar U^4$, $B_4=24000\bar U^6$,
$B_5=600000\bar U^8$, and $B_v$ increases in $v$.

\emph{Attainment.} The pairs $\varphi'=\varphi\circ\pi$ with $\pi\in\mathrm{Aut}(T)$ have $Y_{\varphi'}=Y_\varphi$ and contribute
$|\mathrm{Aut}(T)|(\Exp\prod_{e\in E_\varphi}(W_e^2+\sigma^2_e)-F_T^2)/(m)_v$. Relative to this term, the other overlap patterns are of
order $\delta_m^{v-s}$ for $s<v$ and $\Delta^2$ for $s=v$, $t<v-1$. For fixed $q\in(0,1)$ and $M$, with $\Delta\to0$ and $\delta_m\to0$ we have
$\sigma^2_e\Delta^2\to q(1-q)$, so the variance is asymptotic to $|\mathrm{Aut}(T)|(q(1-q))^{v-1}/((m)_v\Delta^{2v-2})$. \qed

\subsection{Proof of \texorpdfstring{\Cref{lem:md}}{the moderate-deviation lemma}}

Fix $Z$ and write $\xi_e:=A_e-p_{Z_iZ_j}$ for $e=\{i,j\}$. Given $Z$ the $\xi_e$ are independent, centered and bounded
by $1$, so $\|\xi_e\|_{\psi_2}\le L_0:=(\log2)^{-1/2}$, where $\|Y\|_{\psi_2}:=\inf\{u>0:\Exp e^{Y^2/u^2}\le2\}$. Since
$X_e=W_e+\xi_e/\Delta$ and the $v-1$ pairs of an injective copy are distinct,
\[
  \hF_T=f(\xi):=\frac1{(n)_v}\sum_{\varphi}\prod_{xy\in E(T)}\Big(W_{\varphi(x)\varphi(y)}+\frac{\xi_{\varphi(x)\varphi(y)}}\Delta\Big)
\]
is a multilinear polynomial of degree $v-1\le4$ in $\xi$, and its conditional mean is
\[
  \Exp[f\mid Z]=V_T(Z):=\frac1{(n)_v}\sum_\varphi\prod_{xy\in E(T)}M_{Z_{\varphi(x)}Z_{\varphi(y)}} .
\]

\emph{The concentration inequality.} By \citet[Theorem~1.4]{adamczak2015concentration}, for a polynomial $f$ of degree
$D_f$ in independent variables with $\psi_2$-norms at most $L_0$ and every $t>0$,
\[
  \Prob\big(|f-\Exp f|\ge t\big)\le2\exp\Big(-\frac1{C_{D_f}}\min_{1\le d\le D_f}\ \min_{\mathcal J\in P_d}
  \Big(\frac t{L_0^d\|\Exp D^df\|_{\mathcal J}}\Big)^{2/\#\mathcal J}\Big),
\]
where $C_{D_f}$ depends only on $D_f$, $P_d$ is the set of partitions of $[d]$ into nonempty blocks, and for a tensor
$a=(a_{i_1\dots i_d})$ and $\mathcal J=\{J_1,\dots,J_k\}$,
$\|a\|_{\mathcal J}:=\sup\{\sum_ia_i\prod_{l=1}^kx^{(l)}_{i_{J_l}}:\|x^{(l)}\|_2\le1,\ l\le k\}$. For $\mathcal J=\{[d]\}$ this
is the Hilbert--Schmidt norm, and for two blocks it is the operator norm of the corresponding matricization. If $\mathcal J'$
is coarser than $\mathcal J$, then $\|a\|_{\mathcal J}\le\|a\|_{\mathcal J'}$, because tensor products of unit vectors on the blocks
of $\mathcal J$ are unit vectors on the blocks of $\mathcal J'$. Let $C_{\rm AW}:=\max_{D_f\le4}C_{D_f}$.

\emph{Derivative tensors.} Write $f=\sum_Sa_S\prod_{e\in S}\xi_e$ over sets $S$ of at most $v-1$ pairs, with
\[
  a_S=\Delta^{-|S|}(n)_v^{-1}\sum_{\varphi:S\subseteq E_\varphi}\prod_{e\in E_\varphi\setminus S}W_e .
\]
For distinct pairs
$e_1,\dots,e_d$ we have $(\Exp[D^df\mid Z])_{e_1\dots e_d}=a_{\{e_1,\dots,e_d\}}$, because every other term keeps a centered factor,
and the entries with a repeated index vanish. A \emph{pattern} is an ordered tuple $\pi=(f_1,\dots,f_d)$ of distinct edges
of $T$; $F_\pi:=\{f_1,\dots,f_d\}$ is a forest with $s_\pi$ vertices and $c_\pi$ components, so $s_\pi=d+c_\pi$. With
$\varphi(xy):=\{\varphi(x),\varphi(y)\}$ let
\[
  a^\pi_{e_1\dots e_d}:=\frac{\Delta^{-d}}{(n)_v}\sum_{\varphi:\ \varphi(f_j)=e_j,\ j\le d}\ \prod_{xy\in E(T)\setminus F_\pi}W_{\varphi(x)\varphi(y)} .
\]
Each $\varphi$ with $\{e_1,\dots,e_d\}\subseteq E_\varphi$ has exactly one pattern with $\varphi(f_j)=e_j$, so
$\Exp[D^df\mid Z]=\sum_\pi a^\pi$, a sum over at most $4^4$ patterns. For fixed $\pi$ and $(e_j)$, the maps $\varphi$ with
$\varphi(f_j)=e_j$ are determined on $V(F_\pi)$ up to the orientation of each edge, at most $2^d$ choices, and the other
$v-s_\pi$ vertices have at most $n$ images each. With $(n)_v\ge(n/2)^v$ for $n\ge2v-2$,
\[
  |a^\pi_e|\le2^{v+d}\bar U^{v-1}\Delta^{-d}n^{-s_\pi},
\]
and at most $n^{s_\pi}$ tuples $e$ have $a^\pi_e\ne0$.

\emph{One block.} Hence $\|a^\pi\|_{\{[d]\}}\le2^{v+d}\bar U^{v-1}\Delta^{-d}n^{-s_\pi/2}\le2^{2v}\bar U^{v-1}N_{d+1}^{-1/2}$,
because $s_\pi\ge d+1$, and $N_{d+1}=n\delta^d\ge N_v$ for $d\le v-1$ and $\delta\le1$. So
$\|\Exp[D^df\mid Z]\|_{\{[d]\}}\le K_1\bar U^{v-1}N_v^{-1/2}$ with $K_1:=4^42^{2v}$.

\emph{Two or more blocks.} Let $\mathcal J=\{J_1,J_2\}$, $V_l:=V(\{f_j:j\in J_l\})$ and $i:=|V_1\cap V_2|$, so that
$|V_1|+|V_2|=s_\pi+i$. In the matricization with rows $(e_j)_{j\in J_1}$, a row determines $\varphi$ on $V_1$ up to
$2^{|J_1|}$ choices, and the supported columns of that row are then determined by the images of $V_2\setminus V_1$; so
the absolute row sums are at most $2^d\cdot2^{v+d}\bar U^{v-1}\Delta^{-d}n^{-s_\pi}n^{|V_2\setminus V_1|}=
2^{v+2d}\bar U^{v-1}\Delta^{-d}n^{-|V_1|}$, and the absolute column sums at most the same with $|V_2|$ in place of $|V_1|$.
The Schur test gives
\[
  \|a^\pi\|_{\mathcal J}\le2^{v+2d}\bar U^{v-1}\Delta^{-d}n^{-(s_\pi+i)/2}.
\]
We claim that $\Delta^{-d}n^{-(s_\pi+i)/2}\le N_v^{-1/2}n^{-1/2}$, that is, $n^{i-u-1}\Delta^{-2e'}\ge1$ with
$u:=v-s_\pi$ and $e':=v-1-d$. If $F_\pi$ is connected, then $e'=u$, and $i\ge1$ because two nonempty sets of edges of a
connected graph whose union is all its edges share a vertex; so $n^{i-u-1}\Delta^{-2e'}=n^{i-1}\delta^{-u}\ge1$. If
$c_\pi\ge2$, then $e'=u+c_\pi-1\ge u+1$, and $i\ge0$ and $\Delta\le1$ give $n^{i-u-1}\Delta^{-2e'}\ge
n^{-u-1}\Delta^{-2(u+1)}=\delta^{-(u+1)}\ge1$. A partition with three or more blocks has a two-block coarsening, to which
the same bound applies. So $\|\Exp[D^df\mid Z]\|_{\mathcal J}\le K_2\bar U^{v-1}N_v^{-1/2}n^{-1/2}$ for every $\mathcal J$
with $\#\mathcal J\ge2$, where $K_2:=4^42^{v+8}$.

\emph{The exponent.} Let $t:=x/\sqrt{N_v}$ and $\gamma:=1/(L_0^4\max\{K_1,K_2\}\bar U^{v-1})\le1$. The one-block terms of
the inequality are at least $(\gamma x)^2$, and the $k$-block terms, $k\in\{2,3,4\}$, at least $(\gamma x\sqrt n)^{2/k}$.
The latter are at least $\gamma^2\min\{x^2,n^{1/3}\}$: if $1\le x\le n^{1/6}$, then $\sqrt n\ge x^3$ and
$(\gamma x\sqrt n)^{2/k}\ge\gamma^{2/k}x^{8/k}\ge\gamma^2x^2$; if $x<1$, then $(\gamma x\sqrt n)^{2/k}\ge(\gamma x)^{2/k}\ge
(\gamma x)^2$; if $x>n^{1/6}$, then $(\gamma x\sqrt n)^{2/k}\ge\gamma^{2/k}n^{4/(3k)}\ge\gamma^2n^{1/3}$. Hence
\[
  \Prob\big(|\hF_T-V_T(Z)|\ge x/\sqrt{N_v}\ \big|\ Z\big)\le2\exp\big(-(\gamma^2/C_{\rm AW})\min\{x^2,n^{1/3}\}\big).
\]

\emph{The labels.} The quantity $V_T(Z)$ has mean $F_T(\theta)$, and changing one label $Z_i$ changes $v(n-1)_{v-1}$ of its $(n)_v$
terms, each by at most $2\bar U^{v-1}$, hence changes $V_T$ by at most $2v\bar U^{v-1}/n$. McDiarmid's inequality
\citep{mcdiarmid1989method} gives $\Prob(|V_T(Z)-F_T(\theta)|\ge t)\le2\exp(-t^2n/(2v^2\bar U^{2v-2}))$, which at
$t=x/\sqrt{N_v}$ is at most $2\exp(-x^2/(2v^2\bar U^{2v-2}))$, since $N_v\le n$ for $\delta\le1$.

\emph{Combination.} The two bounds give $\Prob(|\hF_T-F_T(\theta)|\ge2x/\sqrt{N_v})\le4\exp(-c'\min\{x^2,n^{1/3}\})$ with
$c':=\min\{\gamma^2/C_{\rm AW},1/(2v^2\bar U^{2v-2})\}$. Replacing $x$ by $x/2$ and using
$\min\{x^2/4,n^{1/3}\}\ge\frac14\min\{x^2,n^{1/3}\}$ proves the lemma with $c_C:=\frac14\min_{2\le v\le5}c'$, which is
nonincreasing in $\bar U$. \qed

\subsection{Proof of \texorpdfstring{\Cref{thm:upper}}{the upper bounds}}

Let $\hat\eps_4:=\max_{T\in\cT_4}|\hF_T-F_T(\theta)|$, $\hat\eps_5:=\max_{T\in\cT_5\setminus\cT_4}|\hF_T-F_T(\theta)|$ and
$r_n:=\max\{1/n,1/N_4\}$, and take $B_v$ and $c_C$ for $\bar U=\bar U_{\mathcal K}$, which dominate their values for
$M\in\mathcal K$. Then $\Phi_w(\theta)=\max\{\hat\eps_4,\lambda\hat\eps_5\}$ and $\Phi_w(\htheta)\le\Phi_w(\theta)$. For $T\in\cT_4$,
$|F_T(\htheta)-F_T(\theta)|\le|F_T(\htheta)-\hF_T|+|\hF_T-F_T(\theta)|\le\Phi_w(\htheta)+\Phi_w(\theta)\le2\Phi_w(\theta)$, and for a
five-vertex tree the same inequality multiplied by $\lambda$ gives $\lambda|F_T(\htheta)-F_T(\theta)|\le2\Phi_w(\theta)$. Since
$0<\lambda\le1$,
\[
  \cD_4(\htheta,\theta)\le2\Phi_w(\theta),\qquad\cD_5(\htheta,\theta)\le2\Phi_w(\theta)/\lambda .
\]

\emph{Second moment.} \Cref{lem:ustat} applies on all $n\ge8$ answers. For $T\in\cT_4$ with $v$ vertices,
$B_v\le B_4$ and $\max\{1/n,1/N_v\}\le r_n$: if $\delta\le1$, then $N_v\ge N_4$ and $N_4\le n$; if $\delta>1$, then
$N_v\ge n$. So $\Exp\hat\eps_4^2\le\sum_{T\in\cT_4}\Var(\hF_T)\le4B_4r_n$. For the five-vertex trees,
$\lambda^2\max\{1/n,1/N_5\}\le r_n$: if $\delta\le1$, then $\lambda^2=\delta$ and $\delta/N_5=1/N_4$; if $\delta>1$, then
$\lambda=1$ and $N_5\ge n$. So $\lambda^2\Exp\hat\eps_5^2\le3B_5r_n$, and
\[
  \Exp\Phi_w(\theta)^2\le\Exp\hat\eps_4^2+\lambda^2\Exp\hat\eps_5^2\le(4B_4+3B_5)r_n .
\]

\emph{Part (a).} Take $(\eta,C)$ from \Cref{thm:modulus}(a), and let $\delta\le1$, so that $\lambda=\sqrt\delta$, $N_4\le n$ and
$r_n=1/N_4$. On $\mathcal G:=\{\Phi_w(\theta)\le\lambda\eta/2\}$ we have $\cD_5(\htheta,\theta)\le\eta$, the modulus gives
$|\hH_n-\Ent(\theta)|^3\le C\cD_4(\htheta,\theta)\le2C\Phi_w(\theta)$, and Jensen's inequality gives
\[
  \Exp\big[(\hH_n-\Ent(\theta))^2\mathbf1_{\mathcal G}\big]\le(2C)^{2/3}(4B_4+3B_5)^{1/3}N_4^{-1/3}.
\]
On $\mathcal G^c$ some $T\in\cT_4$ with $v$ vertices has $|\hF_T-F_T(\theta)|>\lambda\eta/2=x/\sqrt{N_v}$ with
$x=\frac\eta2\sqrt{\delta N_v}\ge\frac\eta2\sqrt{\delta N_4}=\frac\eta2\sqrt{N_5}$, or some five-vertex tree has
$|\hF_T-F_T(\theta)|>\eta/2=x/\sqrt{N_5}$ with $x=\frac\eta2\sqrt{N_5}$. The bound of \Cref{lem:md} decreases in $x$, so
\[
  \Prob(\mathcal G^c)\le28\exp\big(-c_C\min\{\eta^2N_5/4,\,n^{1/3}\}\big)\le28\big(e^{-c_UN_5}+e^{-c_Cn^{1/3}}\big),\qquad
  c_U:=c_C\eta^2/4 .
\]
Since $ye^{-c_Cy}\le1/(ec_C)$ for $y>0$ and $N_4\le n$, we have $e^{-c_Cn^{1/3}}\le N_4^{-1/3}/(ec_C)$. The loss is at most $(\log3)^2$
on $\mathcal G^c$, so part (a) holds with $C_U:=(2C)^{2/3}(4B_4+3B_5)^{1/3}+28(\log3)^2(1+1/(ec_C))$.

\emph{Part (b).} \Cref{thm:modulus}(b) gives $|\hH_n-\Ent(\theta)|^3\le C_4\cD_4(\htheta,\theta)\le2C_4\Phi_w(\theta)$ for every
observation, and Jensen's inequality with the second moment gives the bound; $(\log3)^2$ bounds the risk because
$\Ent\in[0,\log3]$.

\emph{Grid versions.} Let $\Theta_{3,n}:=\Simplex\cap(\mathbb Z/n)^3$ and let $\htheta^{\rm G}$ be the lexicographically first
minimizer of $\Phi_w$ over it. Rounding $\theta$ to $\vartheta^0$ with $\vartheta^0_c:=\lfloor n\theta_c\rfloor/n$ for $c=1,2$ and
$\vartheta^0_3:=1-\vartheta^0_1-\vartheta^0_2$ gives $\vartheta^0\in\Theta_{3,n}$ with $\|\vartheta^0-\theta\|_1<4/n$, so \Cref{lem:a:lip} and
$\lambda\le1$ give $\Phi_w(\htheta^{\rm G})\le\Phi_w(\vartheta^0)\le\Phi_w(\theta)+4L_F/n$, with $L_F:=5\bar U_{\mathcal K}^4$. The triangle
inequalities then give $\cD_4(\htheta^{\rm G},\theta)\le2\Phi_w(\theta)+4L_F/n$ and
$\cD_5(\htheta^{\rm G},\theta)\le(2\Phi_w(\theta)+4L_F/n)/\lambda$, and subadditivity of $x\mapsto x^{2/3}$ with
$n^{-2/3}\le r_n^{1/3}$ handles the extra term. In (b) this gives the bound with $C_4^{2/3}[2^{2/3}(4B_4+3B_5)^{1/3}+
(4L_F)^{2/3}]$ in place of the constant. In (a) use $\mathcal G':=\{\Phi_w(\theta)\le\lambda\eta/4\}$. If
$\lambda\ge8L_F/(\eta n)$, then $\cD_5(\htheta^{\rm G},\theta)\le\eta$ on $\mathcal G'$, and the argument above, with $\eta/4$ in place
of $\eta/2$, gives the bound with $c_C\eta^2/16$ in place of $c_U$ and another constant. If $\lambda<8L_F/(\eta n)$, then
$N_5=n\lambda^8<(8L_F/\eta)^8$, so $e^{-c_C\eta^2N_5/16}$ is bounded below by a positive constant depending only on
$\mathcal K$, and the trivial bound $(\log3)^2$ gives the claim after enlarging the constant. For the operation count,
each statistic enumerates $(n)_v=O(n^5)$ injective maps, and the grid has $(n+1)(n+2)/2$ points, at which the closed
forms \eqref{eq:closed} cost $O(1)$ each. \qed

\subsection{Proof of \texorpdfstring{\Cref{thm:main}}{the minimax theorem}}

\emph{Part (a).} Let $c_U,C_U$ be the constants of \Cref{thm:upper}(a) and $n_L,c_L$ those of \Cref{thm:lower} for
$(\mathcal K,q_0,\kappa)$, which applies because $\mathcal K\subset\cU\subset\Omega$. Take $C_{\log}:=1/(3c_U)$,
$n_A:=\max\{n_L,\lceil(\bar U_{\mathcal K}/q_0)^2\rceil,8,\lceil e^{1/C_{\log}}\rceil\}$, $c_A:=c_L$ and $C_A:=2C_U$. Since
$\delta\le n^{-\kappa}\le1$, we have $\Delta\le n^{-1/2}$ and $\Delta\bar U_{\mathcal K}\le q_0$, so \eqref{eq:P} holds. As $N_5\ge C_{\log}\log n$, we have
$e^{-c_UN_5}\le n^{-1/3}\le N_4^{-1/3}$, and \Cref{thm:upper}(a) gives $\sup_\theta\Exp_\theta(\hH_n-\Ent)^2\le2C_UN_4^{-1/3}$, so
$R^*_n(M)$ does not exceed this. \Cref{thm:lower} gives $R^*_n(M)\ge c_L\min\{1,N_4^{-1/3}\}$, and
$N_4=N_5/\delta\ge N_5\ge C_{\log}\log n\ge1$, so the minimum is $N_4^{-1/3}$. The two conditions on $\Delta$ are equivalent
to $(C_{\log}\log n/n)^{1/4}\le\delta\le n^{-\kappa}$, which is nonempty if and only if $C_{\log}\log n\le n^{1-4\kappa}$: this
holds for large $n$ when $\kappa<1/4$, and fails for $\kappa\ge1/4$ once $C_{\log}\log n>1$.

\emph{Part (b).} Take $n'_A:=\max\{n_L,\lceil(\bar U_{\mathcal K}/q_0)^2\rceil,8\}$, $c'_A:=c_L$, $C'':=(2C_4)^{2/3}(4B_4+3B_5)^{1/3}$ and
$C'_A:=\max\{(\log3)^2,C''\}$. As in (a), \eqref{eq:P} holds and $r_n=1/N_4$, so \Cref{thm:upper}(b) gives
$\sup_\theta\Exp_\theta(\hH_n-\Ent)^2\le\min\{(\log3)^2,C''N_4^{-1/3}\}\le C'_A\min\{1,N_4^{-1/3}\}$, and \Cref{thm:lower} gives the
lower bound; no condition on $N_5$ is used. \qed

\section{The lower bound}
\label{app:lower}

Throughout, $M\in\Omega$, $\ell$ and $\zeta_0=(e^\perp_1+\sigma e^\perp_2)/\sqrt2$ as in
\Cref{sec:lower}, and $z_1=1/\sqrt2$, $z_2=\sigma/\sqrt2$ the coordinates of $\zeta_0$. Let $\tau(\psi,B):=\sum_b(B\psi)_b^2\psi_b/u_b$, the cubic
term of the wedge function in \Cref{lem:a:expansion}.

\subsection{Proof of \texorpdfstring{\Cref{lem:move}}{the moving-channel lemma}}

By \Cref{lem:a:expansion}, in the coordinates $x=x_1e^\perp_1+x_2e^\perp_2$ we have
$F_{K_2}-c=\lambda_1x_1^2+\lambda_2x_2^2$ and $F_{P_3}-c^2=(\lambda_1^2+2c\lambda_1)x_1^2+(\lambda_2^2+2c\lambda_2)x_2^2+\tau(x,\tM)$. The Jacobian with
respect to $(x_1,x_2)$ has rows $(2\lambda_1x_1,2\lambda_2x_2)$ and
$(2(\lambda_1^2+2c\lambda_1)x_1+\partial_1\tau,\,2(\lambda_2^2+2c\lambda_2)x_2+\partial_2\tau)$, where $\partial_i\tau=O(|x|^2)$. Its determinant is
$4x_1x_2[\lambda_1(\lambda_2^2+2c\lambda_2)-\lambda_2(\lambda_1^2+2c\lambda_1)]+O(|x|^3)=4x_1x_2\lambda_1\lambda_2(\lambda_2-\lambda_1)+O(|x|^3)$, which at
$x=s\zeta_0$ equals $\pm2s^2\lambda_1\lambda_2(\lambda_2-\lambda_1)+O(s^3)$ and is nonzero for small $s>0$ by (N3) and (N4). Since
$x\mapsto\theta^*+D_ux$ is an affine isomorphism onto $\aff\Simplex$, the inverse function theorem makes $\theta_+$ an isolated
point of the set $\{\theta:F^M_{K_2}(\theta)=F^M_{K_2}(\theta_+),\ F^M_{P_3}(\theta)=F^M_{P_3}(\theta_+)\}$. For a continuous path with constant values
starting at $\theta_+$, the set $\{t:\theta_t=\theta_+\}$ is closed, and open by isolation and continuity, so it is the whole
interval. \qed

\subsection{Proof of \texorpdfstring{\Cref{lem:rotation}}{the rotation-path lemma}}

\emph{Basic facts.} The channel $M_t$ is symmetric; $\tM_tu=cu$, so $M_t\theta^*=c\one$; $\sum_c\theta_{t,c}=\langle u,u+sR_t\chi\rangle=1$ for $\chi\in u^\perp$;
$R_0=I$ and $R_1=2uu^\top-I$, so $R_1\tM R_1=\tM$ because $\tM=cuu^\top+\tM|_{u^\perp}$ and $R_1=-I$ on $u^\perp$. Hence
the channel of the path is $M$ at $t=0$ and at $t=1$.

\emph{Path equations.} For $\rho\in u^\perp$ let $\chi:=\zeta_0+s\rho$. By \Cref{lem:a:expansion} with $B=\tM_t$ and $\psi=R_t\chi$, and the
rotation invariance of the terms with $|S|\le2$,
\[
  F_{K_2}(\theta_t;M_t)=c+s^2\chi^\top\tM\chi,\qquad F_{P_3}(\theta_t;M_t)=c^2+s^2\big(2c\chi^\top\tM\chi+\chi^\top\tM^2\chi\big)+s^3\tau(R_t\chi,\tM_t).
\]
Requiring both to equal their values at $t=0$ and $\rho=0$ is equivalent, for $s\ne0$, to $\mathbf F(\rho;t,s)=0$ with
\[
  \mathbf F_1:=2\zeta_0^\top\tM\rho+s\rho^\top\tM\rho,\qquad
  \mathbf F_2:=2\zeta_0^\top\tM^2\rho+s\rho^\top\tM^2\rho+\tau\big(R_t(\zeta_0+s\rho),\tM_t\big)-\tau(\zeta_0,\tM);
\]
the edge equation is $s^3\mathbf F_1=0$ and, given it, the wedge equation is $s^3\mathbf F_2=0$. The map $\mathbf F$ is polynomial in $(\rho,s)$
with coefficients trigonometric in $t$.

\emph{Existence.} Let $J_0:=2\big(\begin{smallmatrix}\zeta_0^\top\tM\\ \zeta_0^\top\tM^2\end{smallmatrix}\big)$ as a linear map $u^\perp\to\Real^2$; in the basis
$(e^\perp_1,e^\perp_2)$, $\det J_0=4\lambda_1\lambda_2z_1z_2(\lambda_2-\lambda_1)=\pm2\lambda_1\lambda_2(\lambda_2-\lambda_1)\ne0$ by (N3) and (N4). Let
$b(t):=(0,\tau(R_t\zeta_0,\tM_t)-\tau(\zeta_0,\tM))$ and $\rho^{(0)}(t):=-J_0^{-1}b(t)$, so that $\mathbf F(\rho;t,0)=J_0\rho+b(t)$. Writing
$\rho=\rho^{(0)}(t)+\eta$, we have $\mathbf F=J_0\eta+s\mathbf G(\eta;t,s)$ with $\mathbf G$ polynomial, $|\mathbf G|\le C_G$ and $|\partial_\eta\mathbf G|\le C'_G$ for
$|\eta|\le1$, $t\in[-1,2]$, $|s|\le1$. For $|s|\le s_P:=\min\{1,1/(\|J_0^{-1}\|C_G),1/(2\|J_0^{-1}\|C'_G)\}$ the map
$\eta\mapsto-sJ_0^{-1}\mathbf G(\eta;t,s)$ sends the unit ball into itself and is a contraction, so it has a unique fixed point
with $|\eta|\le\|J_0^{-1}\|C_G|s|$; since $\partial_\rho\mathbf F=J_0+O(s)$ is invertible, the analytic implicit function theorem gives a
solution $\rho(t,s)$ that is real-analytic in $(t,s)$. At $t=0$ we have $b(0)=0$, so $\rho^{(0)}(0)=0$, and $\rho=0$ solves
$\mathbf F(0;0,s)=0$; by uniqueness $\rho(0,s)=0$. Shrinking $s_P$ so that $s(1+s\sup|\rho|)\max_cu_c\le\min_c\theta^*_c/2$ gives
$\theta_t\ge\theta^*/2$ along the path. This proves (i) and the endpoint statements.

\emph{Derivatives.} (ii) By \Cref{lem:a:expansion} and rotation invariance, for a tree on $4\le v\le r$ vertices,
$F_T(\theta_t;M_t)=c^{v-1}+s^2Q_T(\chi_t)+s^3G_T(t,s)$ with $Q_T(\chi):=\sum_{\{a,b\}}c^{v-1-d_T(a,b)}\chi^\top\tM^{d_T(a,b)}\chi$,
$\chi_t:=\zeta_0+s\rho(t,s)$ and $G_T(t,s):=\sum_{|S|\ge3}s^{|S|-3}h_S(R_t\chi_t,\tM_t)$ analytic. Since $\partial_t\chi_t=s\partial_t\rho$, we have
$\partial_tF_T=s^3[2Q_T(\chi_t,\partial_t\rho)+\partial_tG_T]$, and the bracket is bounded on $[0,1]\times[0,s_P]$. (iii) For a connected
unicyclic graph with cycle length $k$, the $s$-free term $c^{v-k}\operatorname{tr}(\tM_t^k)=c^{v-k}\operatorname{tr}(\tM^k)$ does not depend on $t$,
and every other term carries a factor $s$ and has a bounded derivative. (iv) $\hom(G;\theta_t,M_t)$ is a polynomial in
$(\theta_t,M_t)$, whose derivatives are bounded by (v). (v) $\dot\theta_t=sD_u(\dot R_t\chi_t+sR_t\partial_t\rho)$,
$\dot M_t=D_u^{-1}(\dot R_t\tM R_t^\top+R_t\tM\dot R_t^\top)D_u^{-1}$, and $\theta_{t,c}\ge\theta^*_c/2$.

\emph{Entropy gap.} We have $\nabla\Ent(\theta^*)\cdot D_u\zeta=\ell(\zeta)$ for $\zeta\in u^\perp$, because $\sum_cu_c\zeta_c=0$. With
$\theta_\pm=\theta^*+v_\pm$, $v_+=sD_u\zeta_0$ and $v_-=-sD_u(\zeta_0+s\rho(1,s))$, a second-order Taylor expansion gives
$\Ent(\theta_+)-\Ent(\theta_-)=2s\ell(\zeta_0)+s^2\ell(\rho(1,s))+O(s^3)$, the quadratic terms at $v_+$ and $v_-$ differing by $O(s^3)$. So
$|\Ent(\theta_+)-\Ent(\theta_-)|\ge s|\ell(\zeta_0)|$ for $s$ below some $s_H>0$; we replace $s_P$ by $\min\{s_P,s_H\}$.

\emph{Uniformity.} All constants depend on $M$ only through $\min_c\theta^*_c$, $|\lambda_1-\lambda_2|$, $\min_i|\lambda_i|$, a lower bound of
$\|\ell|_{u^\perp}\|$, upper bounds of $\|M\|$ and $\|\tM\|$, and bounds on the coefficients of the polynomials above. By (N1)
to (N4) and continuity, the lower bounds are positive and the upper bounds finite on a compact subset of $\Omega$. Near each
channel the frames can be chosen continuously (\Cref{app:modulus}), the four sign changes of a frame preserve all these
bounds, and finitely many such neighborhoods cover $\mathcal K$. The choice of $\sigma$ enters only through
$|\ell(\zeta_0)|\ge\|\ell|_{u^\perp}\|/\sqrt2$ and $|z_1z_2|=1/2$, so the constants hold for every admissible choice of frame and $\sigma$. \qed

\subsection{Proof of \texorpdfstring{\Cref{prop:info}}{the information inequality}}

The proof is that of the information inequality for diagonal channels \citep[Lemma~A19]{meaningdependent2026}, which
combines a sweep lemma \citep[Lemma~A3]{meaningdependent2026} with eight steps \citep[Lemmas~A11 to
A18]{meaningdependent2026}. We define the objects, state each step in the form used here, with its hypotheses and the
dependence of its constants, prove what a general channel changes, and write out the counting estimates of the
pairing and remainder steps; a part stated without proof is proved as in the reference, with $w_c\Ind{z_i=z_j}$ replaced
by $M_t[z_i,z_j]$ and $w_{\rm hi}$ by $M_{\rm hi}$.

Fix one piece of the path and $t$ in it, with one-sided derivatives at its ends, and write $\theta=\theta_t$, $M=M_t$,
$\ell_c:=\dot\theta_c/\theta_c$ and $\eta:=\|\dot\theta_t\|_1+\|\dot M_t\|_\infty\le L$. We assume
\begin{equation}
  \tag{A}
  \label{eq:d:A}
  q\in[q_0,1-q_0],\qquad\Delta\le\Delta_0:=q_0/(8\max\{1,M_{\rm hi}\}),
\end{equation}
which holds for $n\ge n_I$ and $\delta\le1$ once $n_I\ge\Delta_0^{-2}$, since then $\Delta=\sqrt{\delta/n}\le n^{-1/2}$.

\emph{Objects.} Let $\eps_{ij}:=\Delta(A_{ij}-q)/(q(1-q))$ and $g_{ij}(z):=M[z_i,z_j]$. For labels $z$ let
$f_{ic}:=\prod_{j\ne i}(1+M[c,z_j]\eps_{ij})$, $f_i:=(f_{ic})_c$, and define the \emph{locally averaged score}
\[
  \SG:=\sum_i\Phi(f_i)+\sum_{i<j}\chi_{ij},\qquad\Phi(f):=\frac{\sum_c\dot\theta_cf_c}{\sum_c\theta_cf_c},\qquad
  \chi_{ij}:=\frac{\dot M[z_i,z_j]\eps_{ij}}{1+M[z_i,z_j]\eps_{ij}},
\]
evaluated at $(Z,A)$; it replaces each label term $\ell_{Z_i}$ of the complete-data score
$\sum_i\ell_{Z_i}+\sum_{i<j}\chi_{ij}$ by its conditional expectation given $(A,Z_{-i})$ (Step 3). For $U\subseteq[n]$ let
$\Pu_Uf:=\Exp_t[f\mid A,Z_{-U}]$, for $S\subseteq[n]$ let $\DS_S:=\sum_{U\subseteq S}(-1)^{|S|-|U|}\Pu_U$, and let
$\Sw_r:=\sum_{|S|\le r}\DS_S\SG$ be the sweep of order $r$. With $\mathsf w(z;\eps):=\prod_v\theta_{z_v}\prod_{k<l}(1+g_{kl}(z)\eps_{kl})$, we have
$\Pu_Uf(z)=\sum_{z'_U}\mathsf w(z'_U,z_{-U})f(z'_U,z_{-U})/\sum_{z'_U}\mathsf w(z'_U,z_{-U})$ at $\eps=\eps(A)$ (Step~3), so $\SG$ and each
$\DS_S\SG$ are rational functions of $\eps$; after the factors that depend only on the fixed $z_{-U}$ cancel, each
conditional normalizer equals $1$ at $\eps=0$, and $\Sw_r=\mathcal S_r(\eps;z)$ is analytic near $\eps=0$. A \emph{motif} $\Gamma$ is
a multiset of pairs with multiplicities $m_e$; $\eps^\Gamma:=\prod_e\eps_e^{m_e}$, $|\Gamma|:=\sum_em_e$, $V(\Gamma)$ is the set of
vertices it touches, $v(\Gamma):=|V(\Gamma)|$, $d(\Gamma)$ the number of distinct pairs, $G_\Gamma$ the underlying simple graph and
$\beta$ its cyclomatic number; $B^{(r)}_\Gamma(z):=[\eps^\Gamma]\,\mathcal S_r(\cdot;z)$. The motif system on a vertex set $V$ is the same
construction with only the pairs inside $V$, marked by a superscript $(V)$. Finally
$\Pi_D:=\sum_{1\le|\Gamma|\le D}\eps^\Gamma B^{(r)}_\Gamma(Z)$, the Taylor polynomial of degree $D$ of $\Sw_r$ because $\SG=0$ at $\eps=0$,
$\Rem_D:=\Sw_r-\Pi_D$, $\varsigma:=\sqrt\delta\log n$, and
$\mathcal G:=\{\max_{v\in[n],c\in[3]}|\sum_{u\ne v}\eps_{vu}\Ind{Z_u=c}|\le\varsigma\}$.

\emph{Step 1 (sweep; \citealp[Lemma~A3]{meaningdependent2026}).} For any function $f$ of $(Z,A)$ and $\theta>0$:
$\Exp_t[\DS_Sf\mid A]=0$ for $S\ne\emptyset$, hence $\Exp_t[\Sw_r\mid A]=\Exp_t[\SG\mid A]$ and $\Exp_t(\Exp_t[\SG\mid A])^2\le\Exp_t\Sw_r^2$,
and $\sum_{S\subseteq T}\DS_S=\Pu_T$ for every $T\subseteq[n]$. The lemma does not involve the channel.

\emph{Step 2 (edge variables; \citealp[Lemma~A11]{meaningdependent2026}).} Under \eqref{eq:d:A}: (i)
$P_t(A\mid Z)=\prod_{i<j}\Bern(A_{ij};q)\prod_{i<j}(1+g_{ij}(Z)\eps_{ij})$; (ii) $|\eps_{ij}|\le2\Delta/q_0$ and
$|g_{ij}\eps_{ij}|\le2M_{\rm hi}\Delta/q_0\le1/4$; (iii) given $Z$ the $\eps_{ij}$ are independent, $\Exp_t[\eps_{ij}\mid Z]=\Delta^2g_{ij}/(q(1-q))$,
and $|\Exp_t[\eps_{ij}^a\mid Z]|\le(C_\eps\Delta)^{c(a)}$ for integers $a\ge1$, with $c(1):=2$, $c(a):=a$ for $a\ge2$ and
$C_\eps:=4\max\{1,M_{\rm hi}\}/q_0$, so that $C_\eps\Delta\le1/2$. For (i), $\Bern(b;q+\Delta g)/\Bern(b;q)=1+\Delta g(b-q)/(q(1-q))$ holds for
every real $g$ with $q+\Delta g\in[0,1]$, of either sign.

\emph{Step 3 (Glauber identity and score; \citealp[Lemma~A12]{meaningdependent2026}).} (i) For $U\subseteq[n]$ the
conditional law of $Z_U$ given $(A,Z_{-U})$ is proportional to $z'_U\mapsto\prod_{k\in U}\theta_{z'_k}\prod_{e\cap U\ne\emptyset}(1+g_e(z)\eps_e)$,
$z=(z'_U,Z_{-U})$; in particular $P_t(Z_i=c\mid A,Z_{-i})=\theta_cf_{ic}/\sum_b\theta_bf_{ib}$. (ii) $\Sobs:=\partial_t\log P_t(A)=\Exp_t[\SG\mid A]$,
$\Exp_t\Sobs=0$ and $I(t)=\Exp_t\Sobs^2$; at $\eps=0$ we have $f_i=\one$, $\Phi(f_i)=0$ and $\chi=0$; $|\Phi(f)|\le\max_c|\ell_c|$ for positive $f$; and
$|\chi_{ij}|\le2\|\dot M_t\|_\infty|\eps_{ij}|$. Changes: for $U=\{i\}$ the pairs at $i$ contribute
$\prod_{j\ne i}(1+M[z'_i,Z_j]\eps_{ij})=f_{i,z'_i}$, a product over all $j$ rather than over the $j$ with $Z_j=z'_i$, and $f_i$ does
not depend on $z'_i$. Differentiating the finite sum $P_t(A)=\sum_z\mathsf w(z;\eps(A))\prod_e\Bern(A_e;q)$ with $\partial_tg_{ij}=\dot M[z_i,z_j]$
gives the channel term $\chi_{ij}$ without an indicator. By Step 2(ii), $f_{ic}\ge(3/4)^{n-1}>0$, so $\Phi(f)$ is an average of the
$\ell_c$ with positive weights, and $|1+M[z_i,z_j]\eps_{ij}|\ge3/4$.

\emph{Step 4 (locality; \citealp[Lemma~A13]{meaningdependent2026}).} Let $\Gamma$ be a motif and $V:=V(\Gamma)$. (i) If
$S\not\subseteq V$, then $[\eps^\Gamma]\,\DS_S\SG=0$. (ii) If $S\subseteq V$, then $[\eps^\Gamma]\,\DS_S\SG(z)=[\eps^\Gamma]\,\DS^{(V)}_S\SG^{(V)}(z_V)$.
(iii) Hence $B^{(r)}_\Gamma$ depends only on $z_V$ and $B^{(r)}_\Gamma=[\eps^\Gamma]\sum_{S\subseteq V,|S|\le r}\DS^{(V)}_S\SG^{(V)}$. (iv) If $\Gamma$ is not
connected, $B^{(r)}_\Gamma=0$. Change: the coefficient depends only on the restriction to $\{\eps_e=0,\ e\notin\Gamma\}$; there a
vertex $u\notin V$ has $f_u=\one$, $\Phi(f_u)=0$ and $\chi_{iu}=0$, and for $i\in V$ the factor $1+M[c,z_u]\eps_{iu}$ of $f_{ic}$ equals $1$,
so $f_i$ does not depend on $z_u$ and $\mathsf w(z;\eps)=\prod_{u\notin V}\theta_{z_u}\cdot\mathsf w^{(V)}(z_V;\eps)$, which is what the
reference uses. For (iv), split $V$ into nonempty disjoint sets $V_1,V_2$ with no pair of $\Gamma$ between them. On the subspace
every pair between $V_1$ and $V_2$ has $\eps_e=0$, whatever the entries of $M$. So for $i\in V_1$ the factors $1+M[c,z_k]\eps_{ik}$,
$k\in V_2$, of $f_{ic}$ equal $1$ and $\Phi(f_i)$ involves only $z_{V_1}$ and the pairs inside $V_1$, and symmetrically for $i\in V_2$;
$\chi_{ij}=0$ for $i\in V_1$, $j\in V_2$; and the conditional weight of $z'_U$, $U\subseteq V$, is the product of a factor in the
labels of $V_1$ and a factor in the labels of $V_2$. Hence $\SG^{(V)}=\SG^{(V_1)}+\SG^{(V_2)}$, and $\Pu^{(V)}_Uh=\Pu^{(V_1)}_{U\cap V_1}h$ for
every function $h$ of $z_{V_1}$ and the pairs inside $V_1$. If $S$ meets $V_2$, pick $u\in S\cap V_2$: in
$\DS^{(V)}_S=\sum_{U\subseteq S}(-1)^{|S|-|U|}\Pu^{(V)}_U$ the terms for $U$ and $U\triangle\{u\}$ have opposite signs and, since both sets
meet $V_1$ in $U\cap V_1$, equal values on $\SG^{(V_1)}$, so $\DS^{(V)}_S\SG^{(V_1)}=0$. If $S\subseteq V_1$, then
$\DS^{(V)}_S\SG^{(V_1)}=\DS^{(V_1)}_S\SG^{(V_1)}$, which involves no pair meeting $V_2$, while $\eps^\Gamma$ contains a pair inside $V_2$, since every
vertex of $V_2$ lies on a pair of $\Gamma$ and that pair lies inside $V_2$; so $[\eps^\Gamma]\DS^{(V)}_S\SG^{(V_1)}=0$. The same holds with
$V_1$ and $V_2$ exchanged, and (iii) gives $B^{(r)}_\Gamma=0$.

\emph{Step 5 (exact identity; \citealp[Lemma~A14]{meaningdependent2026}).} For every motif $\Gamma$, $V:=V(\Gamma)$, $r\ge0$ and $z$,
\[
  B^{(r)}_\Gamma(z)=\partial_tC_\Gamma-\sum_{S\subseteq V,|S|>r}[\eps^\Gamma]\big(\DS^{(V)}_S\SG^{(V)}\big)(z),\qquad
  C_\Gamma:=[\eps^\Gamma]\log\Exp_\theta\prod_{e\subseteq V}\big(1+g_e(Z)\eps_e\big),
\]
with $Z\sim\theta^{\otimes V}$, and $C_\Gamma$ is a polynomial, with coefficients depending only on $\Gamma$, in the functions
$\hom(Q;\theta,M)$ of the connected subgraphs $Q$ of $G_\Gamma$; only trees occur if $G_\Gamma$ is a tree, and only trees and
unicyclic graphs if $\beta(G_\Gamma)=1$. Changes: $\Exp\prod_{e\in B}g_e(Z)=\prod_Q\hom(Q;\theta,M)$ over the connected components $Q$ of
$(V(B),B)$, whose labels are independent; for a diagonal channel only monochromatic components contribute. Here every
component does. The identity $\Pu^{(V)}_V\SG^{(V)}=\partial_t\log\sum_z\mathsf w_V(z)$ uses only that, as a function of $z_i$, the weight $\mathsf w_V$ is
$\theta_{z_i}f_{i,z_i}$ times a factor free of $z_i$ (Step 3) and that $\partial_t\mathsf w_V=\mathsf w_V(\sum_i\ell_{z_i}+\sum_e\chi_e)$. Since
$\log\sum_z\mathsf w_V$ is jointly analytic in $t$ on a neighborhood of the piece and in $\eps$ near $0$, the operators $[\eps^\Gamma]$ and $\partial_t$ commute.

\emph{Step 6 (small motifs; \citealp[Lemma~A15]{meaningdependent2026}).} If $v(\Gamma)\le r$, then $B^{(r)}_\Gamma=\partial_tC_\Gamma$, which does not
depend on $z$, and: $B^{(r)}_\Gamma=0$ if $G_\Gamma$ is a tree on at most $m-1$ vertices; $|B^{(r)}_\Gamma|\le\kappa_\Gamma L\eps_{\rm tree}$ if
$G_\Gamma$ is a tree on $m\le v(\Gamma)\le r$ vertices; $|B^{(r)}_\Gamma|\le\kappa_\Gamma L\eps_{\rm uni}$ if $\beta(G_\Gamma)=1$; and
$|B^{(r)}_\Gamma|\le\kappa_\Gamma L$ otherwise, with $\kappa_\Gamma$ depending only on $\Gamma$ and $M_{\rm hi}$. Changes: the values $\hom(Q)$
satisfy $|\hom(Q)|\le M_{\rm hi}^{e(Q)}\le B_r:=\max\{1,M_{\rm hi}\}^{\binom r2}$ and may be negative, so we bound the partial derivatives of
the polynomial on the symmetric box $[-B_r,B_r]$. The subgraphs of a tree on at most $m-1$ vertices are trees on at most
$m-1$ vertices, whose functions are constant; for a tree on $m\le v(\Gamma)\le r$ vertices the chain rule and
$|\partial_tF_T|\le L\eps_{\rm tree}$ give the bound; for $\beta(G_\Gamma)=1$ the unicyclic subgraphs add $|\partial_t\hom|\le L\eps_{\rm uni}$, and
$\eps_{\rm tree}\le\eps_{\rm uni}$; otherwise $|\partial_t\hom(Q)|\le(v(Q)+e(Q))\max\{1,M_{\rm hi}\}^{e(Q)}\eta$.

\emph{Step 7 (uniform coefficient bound; \citealp[Lemma~A16]{meaningdependent2026}).} For every motif $\Gamma$ with $v:=v(\Gamma)\ge2$,
every $r\ge0$ and $z\in[3]^V$, $|B^{(r)}_\Gamma(z)|\le c_v\varrho_v^{-|\Gamma|}\eta$ with $\varrho_v:=\log(3/2)/(\binom v2\max\{1,M_{\rm hi}\})$ and
$c_v:=3^{v+1}v^2$. Changes: on the polydisc $\{|\eps_e|\le\varrho_v\}$ each product $\prod_e(1+g_e\eps_e)$ is within $1/2$ of $1$,
$|f_{ic}-1|\le e^{(v-1)M_{\rm hi}\varrho_v}-1\le1/2$, and $M_{\rm hi}\varrho_v\le\log(3/2)$ gives
$|\chi_e|\le\|\dot M_t\|_\infty\varrho_v/(1-M_{\rm hi}\varrho_v)\le\|\dot M_t\|_\infty$; so $|\SG^{(V)}|\le(v+\binom v2)\eta\le(c_v/3^{v+1})\eta$, and the Cauchy
estimate with $\sum_{S\subseteq V}2^{|S|}=3^v$ and Step 4(iii) concludes. The reference's factor $\max\{1,1/w_{\rm hi}\}$ came from dividing
by a positive diagonal entry; here no entry of $M_t$ is divided by, and signed or zero entries are allowed.

\emph{Step 8 (pairing; \citealp[Lemma~A17]{meaningdependent2026}).} Under \eqref{eq:d:A}, the hypotheses of \Cref{prop:info} and
$\delta\le1$ we have $\Var_t(\Pi_D)\le C_D[N_m\eps_{\rm tree}^2+\delta^3\eps_{\rm uni}^2+\delta^5/n+n\delta^r]$ with $C_D$ depending only on
$(q_0,r,D,L,M_{\rm hi})$. Proof: split the motifs into five classes: \emph{zero} (disconnected, or tree motifs with
$v(\Gamma)\le\min\{r,m-1\}$; coefficient $0$ by Steps 4(iv) and 6), \emph{tree} (tree motifs with $m\le v(\Gamma)\le r$), \emph{unicyclic}
($v(\Gamma)\le r$, $\beta=1$), \emph{multicyclic} ($v(\Gamma)\le r$, $\beta\ge2$) and \emph{large} (connected, $v(\Gamma)\ge r+1$; coefficient at most
$c_v\varrho_v^{-|\Gamma|}L$ by Step 7), so that $\Var_t(\Pi_D)\le4\sum_{\rm classes}\Var_t(\Pi_{\rm class})$. For two placed motifs $\Gamma,\Gamma'$ of one
class let $X_\Gamma:=\eps^\Gamma B^{(r)}_\Gamma(Z)$. If their vertex sets are disjoint, $X_\Gamma$ and $X_{\Gamma'}$ are independent. Otherwise their
union $U$, with multiplicities $a_e=m_e+m'_e$, is connected, and Step 2(iii) with $C_\eps\Delta\le1$ gives
$|\Cov(X_\Gamma,X_{\Gamma'})|\le2(C_\eps\Delta)^{2d(U)}\|B_\Gamma\|_\infty\|B_{\Gamma'}\|_\infty$, because $c(a)+c(a')\ge c(a+a')$ and $c(a)\ge2$. For fixed
isomorphism types and overlap pattern, finitely many depending only on $D$, there are at most $n^{v(U)}$ placements. Tree class:
$d(U)\ge m-1$ and $v(U)\le d(U)+1$, so $n^{v(U)}\Delta^{2d(U)}\le n\delta^{d(U)}\le n\delta^{m-1}=N_m$, and Step 6 bounds the coefficients by
$CL^2\eps_{\rm tree}^2$. Unicyclic class: $U$ contains the cycle of $G_\Gamma$, so $v(U)\le d(U)$ and $d(U)\ge3$, hence
$n^{v(U)}\Delta^{2d(U)}\le\delta^{d(U)}\le\delta^3$, with coefficients $CL^2\eps_{\rm uni}^2$. Multicyclic class: $\beta(U)\ge2$, so $v(U)\le d(U)-1$, and a
connected simple graph with cyclomatic number $2$ has at least $5$ edges, hence $n^{v(U)}\Delta^{2d(U)}\le\delta^{d(U)}/n\le\delta^5/n$, with
coefficients $CL^2$. Large class: $d(U)\ge v(\Gamma)-1\ge r$, so the placements contribute at most $n\delta^r$, with coefficients $CL^2$.

\emph{Step 9 (remainder; \citealp[Lemma~A18]{meaningdependent2026}).} Under \eqref{eq:d:A}, the hypotheses of \Cref{prop:info},
$\delta\le1$ and $D\ge r+2$: (i) there is $n_0$ depending only on $(q_0,M_{\rm hi})$ with $P_t(\mathcal G^c)\le6n\exp(-q_0^2\log^2n/36)$ for
$n\ge n_0$; (ii) there are $C$ and $n_R$ depending only on $(q_0,r,D,L,M_{\rm hi})$, and $c_R\in(0,1]$ depending only on $(r,M_{\rm hi})$, such
that for $n\ge n_R$ and $\varsigma\le c_R/2$ we have $|\Rem_D|\le C\eta\,n(n\Delta)^{r+1}\varsigma^{D-r}$ on $\mathcal G$ and $|\Rem_D|\le Cn^{2D}$ everywhere;
(iii) hence $\Exp_t\Rem_D^2\le C[n^{r+3}\delta^{D+1}(\log n)^{2(D-r)}+n^{4D+1}e^{-q_0^2\log^2n/36}]$. Changes and the reason the other
parts are unchanged:
\begin{itemize}[leftmargin=*]
  \item In (i) each field $\sum_u\eps_{vu}\Ind{Z_u=c}$ has, given $Z$, conditional mean of modulus at most $2M_{\rm hi}\delta/q_0$; Bernstein's
    inequality and the union over the $3n$ pairs $(v,c)$ are as in the reference.
  \item \emph{Local decomposition.} We have $\Sw_r=\sum_i\sum_{|S|\le r}\DS_S\Phi(f_i)+\sum_{i<j}\sum_{|S|\le r}\DS_S\chi_{ij}$. For a term with root set $R$
    ($\{i\}$ or $\{i,j\}$) and swept set $S$ let $V_{R,S}:=R\cup S$, with internal variables $x:=(\eps_e)_{e\subseteq V_{R,S}}$ and external
    factors $\phi_{vc}:=\prod_{u\notin V_{R,S}}(1+M[c,z_u]\eps_{vu})$, $v\in V_{R,S}$, a product over all outside vertices rather than over
    those with label $c$. Given the outside labels, the term is a function $g_{R,S}(x,\phi;z_{V_{R,S}})$: the conditional weight of
    $z'_U$, $U\subseteq S$, is proportional to $\prod_{k\in U}\theta_{z'_k}\phi_{k,z'_k}\prod_{e\subseteq V_{R,S},e\cap U\ne\emptyset}(1+g_e\eps_e)$,
    $f_{ic}=\phi_{ic}\prod_{k\in V_{R,S}\setminus\{i\}}(1+M[c,z_k]\eps_{ik})$, and $\chi_{ij}$ depends only on $(z_i,z_j,\eps_{ij})$.
  \item \emph{Connected decomposition.} If $(V_{R,S},\mathrm{supp}\,x)$ is disconnected, then $g_{R,S}=0$. If $R=\{i,j\}$ is split
    between components, then $x_{ij}=0$ and $\chi_{ij}=0$. Otherwise $R$ lies in one component and another component $W$ avoids $R$,
    so $W\subseteq S$. Given the outside labels, $\phi_{k,c}$ depends only on $(k,c)$: it is a one-vertex potential and couples no two
    labels. No pair of $\mathrm{supp}\,x$ joins $W$ to $V_{R,S}\setminus W$, so the conditional weight of $z'_U$, $U\subseteq S$, is the
    product of a factor in the labels of $W$ and a factor in the labels of $V_{R,S}\setminus W$. The swept function, $\Phi(f_i)$ or
    $\chi_{ij}$, does not depend on $z_W$: the factors $1+M[c,z_k]\eps_{ik}$, $k\in W$, of $f_{ic}$ equal $1$, and $\chi_{ij}$ involves only
    $(z_i,z_j,\eps_{ij})$ with $i,j\notin W$. Summing out the labels of $W$ shows that $\Pu_U$ and $\Pu_{U\setminus W}$ agree on the swept
    function for every $U\subseteq S$. For $u\in W$ the terms of $\DS_S$ for $U$ and $U\triangle\{u\}$ have opposite signs and, since
    $(U\triangle\{u\})\setminus W=U\setminus W$, equal values, so $g_{R,S}=0$. Inclusion--exclusion over edge sets then gives
    $g_{R,S}=\sum_Bx^Bh_B(x_B,\phi)$ over edge sets $B$ with $(V_{R,S},B)$ connected, so that $|B|\ge|V_{R,S}|-1$, with
    $h_B:=\int_{[0,1]^B}\partial_{x_B}g_{R,S}(\tau\odot x_B,0,\phi)\,\mathrm d\tau$: the difference over a disconnected $B$ is a signed sum
    of values of $g_{R,S}$ at points supported in $B$, all of which vanish.
  \item \emph{Radial continuation.} Replace the edge variables by $\lambda\eps$, $\lambda\in\Cpx$, $|\lambda|\le\mathcal R:=c_R/\varsigma$. On $\mathcal G$,
    $|\lambda M[c,z_u]\eps_{vu}|\le2c_RM_{\rm hi}/(q_0\sqrt n\log n)\le1/2$ for $n\ge n_R$, and
    $\log\phi_{vc}(\lambda)=\lambda\sum_{c'}M[c,c']\sum_{u\notin V_{R,S}:z_u=c'}\eps_{vu}+\rho_{vc}$ with $|\rho_{vc}|\le4c_R^2M_{\rm hi}^2/(q_0^2\log^2n)$.
    Each inner sum differs from a field of $\mathcal G$ by at most $(r+2)2\Delta/q_0\le\varsigma$, so the linear part has modulus at most
    $6c_RM_{\rm hi}$, one factor $3$ more than for a diagonal channel, because the sum now runs over the three labels. If $c_R$,
    depending only on $(r,M_{\rm hi})$, is chosen so small that every conditional normalizer has modulus at least $1/2$, then $g_{R,S}$ is analytic
    with modulus at most $C\eta$ on a fixed complex neighborhood, by the bounds of Step 7, and Cauchy estimates give
    $|h_B(\lambda x_B,\phi(\lambda))|\le C'\eta$ for $|\lambda|\le\mathcal R$.
  \item \emph{Taylor tail.} The factor $x^B$ is a monomial of degree $|B|$, so the Taylor polynomial of degree $D$ of $x^Bh_B$ in $\eps$ is
    $x^BT_{D-|B|}h_B$, with $T_j$ the Taylor polynomial of degree $j$. For $G$ analytic on $|\lambda|\le\mathcal R$, the homogeneous part of
    degree $j$ of $G(\eps)$ is $[\lambda^j]G(\lambda\eps)$, of modulus at most $\sup_{|\lambda|=\mathcal R}|G|\,\mathcal R^{-j}$; since $\mathcal R\ge2$, we have
    $|(h_B-T_{D-|B|}h_B)(\eps)|\le2C'\eta\mathcal R^{-(D+1-|B|)}$. With $|x^B|\le(2\Delta/q_0)^{|B|}$ and $2\Delta/q_0\le\varsigma/c_R$ for $n\ge n_R$, both
    for $|B|\le D$ and for $|B|\ge D+1$ (then $T_{D-|B|}=0$), we have $|x^B(h_B-T_{D-|B|}h_B)|\le C\eta\,\Delta^{s'}\varsigma^{D+1-s'}$ with $s':=|V_{R,S}|-1$.
  \item \emph{Counting.} For each $s'\le r$ there are at most $2n^{s'+1}$ pairs $(i,S)$ with $|\{i\}\cup S|=s'+1$, and for $1\le s'\le r+1$ at
    most $4n^{s'+1}$ triples $(i<j,S)$ with $|\{i,j\}\cup S|=s'+1$; each term has at most $2^{\binom{r+3}2}$ sets $B$. Since
    $n\Delta/\varsigma=\sqrt n/\log n\ge1$, the largest $s'$ dominates, and on $\mathcal G$ we have
    $|\Rem_D|\le C\eta\sum_{s'=0}^{r+1}n^{s'+1}\Delta^{s'}\varsigma^{D+1-s'}\le C''\eta\,n(n\Delta)^{r+1}\varsigma^{D-r}$.
  \item \emph{Everywhere.} For real $\eps$, $|\DS_S\Phi(f_i)|\le2^r\max_c|\ell_c|\le2^rL$ and $|\DS_S\chi_{ij}|\le2^r\cdot4L\Delta/q_0$ by Step 3, so
    $|\Sw_r|\le Cn^{r+2}$; there are at most $n^{2D}$ motifs with $|\Gamma|\le D$, $|\eps^\Gamma|\le1$, and Step 7 bounds their coefficients, so
    $|\Pi_D|\le Cn^{2D}$ and, as $D\ge r+2$, we have $|\Rem_D|\le Cn^{2D}$. For (iii),
    $n^2(n\Delta)^{2r+2}\varsigma^{2(D-r)}=n^{r+3}\delta^{r+1}\cdot\delta^{D-r}(\log n)^{2(D-r)}$; on $\mathcal G^c$ use the bound everywhere and (i), and
    $\eta\le L$.
\end{itemize}

\emph{Step 10 (conclusion; \citealp[Lemma~A19]{meaningdependent2026}).} By Steps 1 and 3, $I(t)=\Exp_t\Sobs^2\le\Exp_t\Sw_r^2$ and
$\Exp_t\Sw_r=\Exp_t\Sobs=0$, so $\Exp\Pi_D=-\Exp\Rem_D$ and $\Exp_t\Sw_r^2\le2\Var_t(\Pi_D)+4\Exp_t\Rem_D^2$. Steps 8 and 9 give the
inequality with $\varsigma_0:=c_R/2$ and $n_I\ge\max\{n_0,n_R,\Delta_0^{-2}\}$. The constants depend only on $(M_{\rm hi},L,q_0,r,D)$ and
not on $\eps_{\rm tree},\eps_{\rm uni}$, which enter only through Steps 6 and 8. \qed

\subsection{Proof of \texorpdfstring{\Cref{thm:lower}}{the lower bound}}

Let $r:=\max\{4,\lfloor1/\kappa\rfloor+1\}$ and $D:=\max\{r+2,\lceil(r+4)/\kappa\rceil\}$, and take $s_P$, $L$ from \Cref{lem:rotation} for $\mathcal K$ and $r$.
The rotation path satisfies the hypotheses of \Cref{prop:info} with $m=4$, $\eps_{\rm tree}=s^3$, $\eps_{\rm uni}=s$,
$\theta_{\min}=\min_{\mathcal K}\min_c\theta^*_c/2$, $M_{\rm hi}:=\max_{\mathcal K}\max_{t}\|M_t\|\le\max_{\mathcal K}\|\tM\|_{\rm op}/\min_c\theta^*_c$ and the constant $L$ of
\Cref{lem:rotation}, because $s\le1$ gives $s^3\le s\le1$ and $Ls\le L$; let $C_I,n_I,\varsigma_0$ be its constants. Let
$s_1:=\min\{s_P,1\}$, $a_0:=\min\{1,(8C_I)^{-1/6}\}$ and $s:=\min\{s_1,a_0N_4^{-1/6}\}$. Then $C_IN_4s^6\le C_Ia_0^6\le1/8$; on the window $\delta\le n^{-\kappa}$ we have $\delta^3s^2\le n^{-3\kappa}$,
$\delta^5/n\to0$, $n\delta^r\le n^{1-\kappa r}\to0$, $n^{r+3}\delta^{D+1}(\log n)^{2(D-r)}\le n^{r+3-\kappa(D+1)}(\log n)^{2D}\to0$, and the last term
decays faster than any power of $n$; also $\varsigma=\sqrt\delta\log n\to0$. So there is $n_L\ge n_I$, depending only on
$(\mathcal K,q_0,\kappa)$, with $\sup_tI(t)\le1/4$ for $n\ge n_L$. Each $P_t$ is a positive distribution on a finite set, because
$\theta_t>0$ and $q+\Delta M_t[a,b]\in[q-q_0/8,q+q_0/8]\subset(0,1)$, and $P_t(A)$ is a polynomial in $(\theta_t,M_t)$. \Cref{lem:a:hellinger}
gives $\TV(P_0,P_1)\le\frac12(1/4)^{1/2}=1/4$. Here $P_0$ and $P_1$ are the laws under the channel $M$ at $\theta_+$ and $\theta_-$, whose
entropies differ by at least $s|\ell(\zeta_0)|$, and \Cref{lem:a:lecam} gives
\[
  R^*_n(M)\ge\frac{s^2\ell(\zeta_0)^2}{16}=\frac{\ell(\zeta_0)^2}{16}\min\{s_1^2,a_0^2N_4^{-1/3}\}\ge c_L\min\{1,N_4^{-1/3}\},
\]
with $c_L:=\inf_{\mathcal K}\ell(\zeta_0)^2\min\{s_1^2,a_0^2\}/16$, which is positive because $|\ell(\zeta_0)|\ge\|\ell|_{u^\perp}\|/\sqrt2$, which is
bounded below on $\mathcal K$ (proof of \Cref{lem:rotation}). \qed

\section{The five-vertex threshold}
\label{app:threshold}

\subsection{Proof of \texorpdfstring{\Cref{lem:projection}}{the projection lemma}}

The map $X\mapsto X\theta$ from $\Sym$ to $\Real^3$ is onto: for $y\in\Real^3$,
$X:=(y\theta^\top+\theta y^\top)/|\theta|^2-(\theta^\top y)\theta\theta^\top/|\theta|^4$ satisfies $X\theta=y$. The fourth coordinate is independent of the
first three if and only if some $X$ has $X\theta=0$ and $g^\top Xg\ne0$. Write $g=\lambda\theta+g_\perp$ with $g_\perp\perp\theta$. If $g_\perp=0$,
then $X\theta=0$ implies $g^\top Xg=\lambda^2\theta^\top X\theta=0$; if $g_\perp\ne0$, then $X:=g_\perp g_\perp^\top$ has $X\theta=0$ and
$g^\top Xg=|g_\perp|^4\ne0$. Since $\theta>0$, the identity $g=\lambda\theta$ holds if and only if $d_a=\lambda$ for every $a$. So the map is onto when $d$ is
not constant, and the closest point to $M$ in the Frobenius inner product $\langle X,Y\rangle=\operatorname{tr}(XY)$ is the orthogonal
projection $X^*=M+\mathcal A^*(\mathcal A\mathcal A^*)^{-1}((d,c_P)-\mathcal AM)$, with $\mathcal A X:=(X\theta,g^\top Xg)$; it is a rational function of
$(\theta,d)$ with nonvanishing denominator, hence real-analytic. \qed

\subsection{Gauss nodes and three-point configurations}

\begin{lemma}[Gauss nodes]
  \label{lem:e:gauss}
  Let $v=m_2-m_1^2>0$. The coefficients $\sigma_1,\sigma_2$ of $\pi_2$ exist and are unique, $\pi_2(m_1)=-v$, and $\pi_2$ has real roots
  $\xi_1<m_1<\xi_2$. The measure $W_1\delta_{\xi_1}+W_2\delta_{\xi_2}$ with $W_1:=(\xi_2-m_1)/(\xi_2-\xi_1)$ and $W_2:=(m_1-\xi_1)/(\xi_2-\xi_1)$ has
  positive weights and moments $1,m_1,m_2,m_3$.
\end{lemma}

\begin{proof}
  The equations $\mathsf L(\pi_2)=m_2-\sigma_1m_1+\sigma_2=0$ and $\mathsf L(z\pi_2)=m_3-\sigma_1m_2+\sigma_2m_1=0$ form a linear system with determinant
  $m_2-m_1^2=v\ne0$. Writing $\pi_2(z)=(z-m_1)^2-\alpha(z-m_1)-\beta$, the identities $\mathsf L(z-m_1)=0$ and $\mathsf L((z-m_1)^2)=v$ give
  $0=\mathsf L(\pi_2)=v-\beta$, so $\pi_2(m_1)=-v<0$, and a monic quadratic that is negative at one point has two real roots, one on
  either side of that point. The two-point measure $\nu$ has total mass $1$ and mean $m_1$, and $\nu(\pi_2)=\nu(z\pi_2)=0=\mathsf L(\pi_2)=\mathsf L(z\pi_2)$.
  The polynomials $1,z,\pi_2,z\pi_2$ form a basis of the polynomials of degree at most three, so $\nu=\mathsf L$ on that space.
\end{proof}

\begin{proof}[Proof of \Cref{lem:fibers}(i)]
  Let $\mu:=\sum_c\theta_c\delta_{d_c}$ and $p:=\prod_c(z-d_c)$. If $\mu$ has moments $m_1,m_2,m_3$, then $\mu=\mathsf L$ on polynomials of
  degree at most three and $\mathsf L(p)=\mu(p)=0$. A monic cubic can be written uniquely as $z\pi_2+A\pi_2+B\pi_1+C$, and
  $\mathsf L(p)=C$, so $C=0$ and $p=p_{A,B}$. For the weights, the Lagrange polynomials $\ell_c(z)=p(z)/((z-d_c)p'(d_c))$ are quadratic
  with $\ell_c(d_{c'})=\Ind{c=c'}$, so $\theta_c=\mu(\ell_c)=\mathsf L(\ell_c)$. For a root $x$ of $p=p_{a,b}$,
  \[
    \frac{p(z)}{z-x}=\frac{p(z)-p(x)}{z-x}=\pi_2(z)+(x+a)\frac{\pi_2(z)-\pi_2(x)}{z-x}+b=\pi_2(z)+(x+a)(z+x-\sigma_1)+b,
  \]
  and applying $\mathsf L$ gives $\mathsf L(\ell_x)=w_{a,b}(x)$. Conversely, if $p_{a,b}$ has distinct real roots $x_i$, then
  $\sum_iw_{a,b}(x_i)f(x_i)=\mathsf L(f)$ for every quadratic $f=\sum_if(x_i)\ell_{x_i}$, and since $z^3-p_{a,b}$ is quadratic and
  $\mathsf L(p_{a,b})=\mathsf L(z\pi_2)+a\mathsf L(\pi_2)+b\mathsf L(\pi_1)=0$, the measure $\sum_iw_{a,b}(x_i)\delta_{x_i}$ has third moment $m_3$. When the
  weights are positive it is a configuration in $\Int\times\Real^3$ with the given moments.
\end{proof}

\begin{proof}[Proof of \Cref{lem:fibers}(ii)]
  Since $\pi_2(\xi_i)=0$, we have $p_{a,b}(\xi_i)=b(\xi_i-m_1)$. If $b<0$, then $p(\xi_1)>0>p(\xi_2)$, and a monic cubic is negative at $-\infty$
  and positive at $+\infty$, so each of the three intervals cut by $\xi_1,\xi_2$ contains a root; the roots are simple and depend
  analytically on $(a,b)$. If a weight vanished at a root $x$, then $(x+a)(x+m_1-\sigma_1)=-b$; substituting into
  $p(x)=(x+a)\pi_2(x)+b(x-m_1)=0$ gives $(x+a)[\pi_2(x)-(x-m_1)(x+m_1-\sigma_1)]=0$, and the bracket equals $\sigma_2-\sigma_1m_1+m_1^2=\pi_2(m_1)=-v\ne0$,
  so $x=-a$ and then $b=0$, a contradiction. The half-plane is connected and each weight is continuous and nonvanishing on it,
  so its sign is constant.

  Now let $(\theta,d)$ be a positive configuration with distinct atoms and the given moments, with parameter $(a,b)$. If $b=0$,
  then $p=(z+a)\pi_2$ and the roots are $\xi_1,\xi_2,-a$ with $-a\notin\{\xi_1,\xi_2\}$, and the numerator of $w(-a)$ vanishes, against
  positivity. Suppose $b>0$. Then $p(\xi_1)<0<p(\xi_2)$, so $(-\infty,\xi_1)$ and $(\xi_2,\infty)$ each contain an even number of roots and
  $(\xi_1,\xi_2)$ an odd number, none equal to a node. The positive measure $\mu$ satisfies $\mu(\pi_2)=\mathsf L(\pi_2)=0$ and $\mu(z\pi_2)=0$,
  and $\pi_2$ is negative on $(\xi_1,\xi_2)$ and positive outside. If all three roots lie in $(\xi_1,\xi_2)$, then $\mu(\pi_2)<0$. Otherwise
  exactly one root $x_m$ lies in $(\xi_1,\xi_2)$ and the other two lie on the same side; $\mu((z-x_m)\pi_2)=\mu(z\pi_2)-x_m\mu(\pi_2)=0$,
  while $(x-x_m)\pi_2(x)$ vanishes at $x_m$ and has the same strict sign at the other two roots. Both cases contradict positivity,
  so $b<0$.
\end{proof}

\begin{proof}[Proof of \Cref{lem:fibers}(iii)]
  Let $\xi^o$ be the other node. Since $\pi_2(\xi+\alpha\zeta)=\pi_2'(\xi)\alpha\zeta+\alpha^2\zeta^2$ exactly,
  \[
    \begin{gathered}
      p_{-\xi+\alpha,-c\alpha^2}(\xi+\alpha\zeta)=\alpha(\zeta+1)\pi_2(\xi+\alpha\zeta)-c\alpha^2(\xi+\alpha\zeta-m_1)=\alpha^2\tilde p(\zeta,\alpha),\\
      \tilde p(\zeta,\alpha):=(\zeta+1)(\pi_2'(\xi)\zeta+\alpha\zeta^2)-c(\xi-m_1)-c\alpha\zeta .
    \end{gathered}
  \]
  Here $\pi_2'(\xi_2)=\xi_2-\xi_1>0$ and $\pi_2'(\xi_1)=-(\xi_2-\xi_1)$, so $\tilde p(\zeta,0)=\pi_2'(\xi)(\zeta^2+\zeta-k)$ with
  $k:=c|\xi-m_1|/(\xi_2-\xi_1)>0$ in both cases. Its roots $\zeta_\pm(0)=(-1\pm\sqrt{1+4k})/2$ are simple, with $\zeta_-(0)<-1<0<\zeta_+(0)$; by the
  implicit function theorem, $\zeta_\pm(\alpha)$ are analytic near $0$, and for $\alpha\ne0$ the point lies in the open half-plane where
  the roots are simple, so the branches $\xi+\alpha\zeta_\pm(\alpha)$ are analytic on the whole real line. By Vieta's formulas the third
  root $x_o(\alpha)=\sigma_1-(-\xi+\alpha)-2\xi-\alpha(\zeta_+(\alpha)+\zeta_-(\alpha))$ is analytic too, with $x_o(0)=\sigma_1-\xi=\xi^o$.

  For the weights, at $x=\xi+\alpha\zeta$ the numerator of $w$ is $\alpha[(\zeta+1)(\xi+m_1-\sigma_1+\alpha\zeta)-c\alpha]$ and
  $p'(x)=\alpha\,\partial_\zeta\tilde p(\zeta,\alpha)$, so $w=[(\zeta+1)(\xi+m_1-\sigma_1+\alpha\zeta)-c\alpha]/\partial_\zeta\tilde p(\zeta,\alpha)$, analytic
  at simple roots. At $\alpha=0$ we have $w_\pm=(\zeta_\pm+1)(\xi+m_1-\sigma_1)/(\pi_2'(\xi)(2\zeta_\pm+1))$, and $\xi+m_1-\sigma_1=m_1-\xi^o$, so
  $(\xi+m_1-\sigma_1)/\pi_2'(\xi)=|\xi^o-m_1|/(\xi_2-\xi_1)>0$; $\zeta_+>0$ makes $(\zeta_++1)/(2\zeta_++1)>0$, and $\zeta_-<-1$ makes both
  $\zeta_-+1$ and $2\zeta_-+1=-\sqrt{1+4k}$ negative, so both weights are positive. At $x_o(0)=\xi^o$ we have $p=(z-\xi)^2(z-\xi^o)$ and
  $w(\xi^o)=(m_1-\xi)/(\xi^o-\xi)$, the Gauss weight of $\xi^o$, which is positive. For $\alpha\ne0$ the weights are positive by (ii), once
  one point of the half-plane carries positive weights, which holds in the application. For $\alpha>0$ we have
  $\xi+\alpha\zeta_-<\xi<\xi+\alpha\zeta_+$, and the order is reversed for $\alpha<0$; in the half-plane the roots interlace with the nodes,
  so the two roots near $\xi_2$ occupy positions $2,3$, and those near $\xi_1$ positions $1,2$.
\end{proof}

\subsection{Proof of \texorpdfstring{\Cref{lem:fiber-path}}{the fiber-path lemma}}

Let $d^i:=M\theta^i$. The edge, wedge and star functions agree at the two points, so $(\theta^0,d^0)$ and $(\theta^1,d^1)$ have the same
moments $(m_1,m_2,m_3)$, with $v>0$. By \Cref{lem:fibers}(i) they correspond to parameters $(a^i,b^i)$ and to assignments $\pi^i$ of the
three meanings to the three positions; by (ii), $b^i<0$, and since $(a^0,b^0)$ carries positive weights, every point of the
half-plane does. For each assignment $\pi$, the map $(a,b)\mapsto(\theta,d)$ with $d_{\pi(k)}:=x_k(a,b)$ and $\theta_{\pi(k)}:=w_{a,b}(x_k)$ is
real-analytic on the half-plane, with values in the positive configurations with distinct atoms and the given moments, and
$\sum_c\theta_c=\mathsf L(1)=1$.

Write $\pi^1=\pi^0\tau_1\cdots\tau_k$ with adjacent position exchanges $\tau_j\in\{(1\,2),(2\,3)\}$ and $k\le3$. Move along a straight
segment in the convex half-plane from $(a^0,b^0)$ to the point with $\alpha=-1$ on the parabola of \Cref{lem:fibers}(iii) for the
node $\xi_2$ if $\tau_1=(2\,3)$ and $\xi_1$ if $\tau_1=(1\,2)$, with $c=1$; cross along $\alpha\in[-1,1]$, which changes the assignment to
$\pi^0\tau_1$; repeat for $\tau_2,\dots,\tau_k$; finish with a straight segment to $(a^1,b^1)$. Each piece is linearly reparametrized to
an interval $[t_{j-1},t_j]$. On straight pieces, $(\theta,d)$ is analytic by \Cref{lem:fibers}(ii); on crossings we follow the analytic
branches of \Cref{lem:fibers}(iii), which for $\alpha\ne0$ coincide with the configurations of the half-plane with assignment $\pi$
or $\pi\tau$, so consecutive pieces match. There are at most $2k+1\le7$ pieces.

Let $M_t:=X^*(\theta_t,d_t)$ from \Cref{lem:projection}, with base point $M$ and $c_P:=F^M_{P_4}(\theta^0)$. Along the path the variance is
$v>0$, so $d_t$ is not constant and $M_t$ is analytic on each piece and continuous. At the ends, $M$ itself is feasible, since
$M\theta^i=d^i$ and $(g^i)^\top Mg^i=F^M_{P_4}(\theta^i)=c_P$, so the channel of the path is $M$ at both ends. Along the path $M_t\theta_t=d_t$, so
$F^{M_t}_{K_2},F^{M_t}_{P_3},F^{M_t}_{K_{1,3}}$ at $\theta_t$ equal $m_1,m_2,m_3$, and $F^{M_t}_{P_4}(\theta_t)=g_t^\top M_tg_t=c_P$. The entries of
$\theta_t$ are positive weights. \qed

For the channel $M_0$ the construction in \Cref{fig:fiber} uses the parabolas with $c=3/(4W)$ at each node, $W$ the Gauss weight
of the other node, and crossings over $\alpha\in[-0.05,0.05]$; the lemma allows any $c>0$ and any crossing interval.

\subsection{The likelihood ratio along the fiber path}
\label{app:e:llr}

Take the path of \Cref{lem:fiber-path}. Each piece is real-analytic on a neighborhood of its closed parameter
interval: the straight pieces run between points of the open half-plane, and the crossings are analytic for all $\alpha$.
By compactness, $\theta_{\min}:=\min_{t,c}\theta_{t,c}>0$ and $M_{\rm hi}:=\max_t\|M_t\|<\infty$, and for every integer $r$ the
derivatives of $\theta_t$, of $M_t$ and of $\hom(G;\theta_t,M_t)$ for the connected graphs $G$ on at most $r$ vertices are bounded
on each piece; let $L\ge1$ be one bound for all of them that also bounds $\max_c|\dot\theta_{t,c}|/\theta_{t,c}$. The tree
functions on at most four vertices are constant. So the path satisfies the hypotheses of \Cref{prop:info} with
$(m,\eps_{\rm tree},\eps_{\rm uni})=(5,1,1)$, and the objects and Steps 1 to 9 of its proof are available on
each piece, with constants depending only on $(M_{\rm hi},L,q_0,r,D)$. Let $P_t$ be the law of the verdicts under
$(\theta_t,M_t)$, $\Exp_t$ the expectation of $(Z,A)$ under that parameter, and $\ell_t(A):=\log(P_t(A)/P_0(A))$; then $P_0$
and $P_1$ are the laws under the channel $M$ at $\theta^0$ and at $\theta^1$.

The sweep order $r$ and the Taylor degree $D$ are integers. Call a motif $\Gamma$ small if $v(\Gamma)\le r$ and
$1\le|\Gamma|\le D$, and let
\[
  \Lambda_t(A):=\sum_{\Gamma\ \text{small}}\eps^\Gamma C_\Gamma(t),\qquad X_t:=\Pi^{>r}_D+\Rem_D,\qquad
  \Pi^{>r}_D:=\sum_{v(\Gamma)>r,\ 1\le|\Gamma|\le D}\eps^\Gamma B^{(r)}_\Gamma(Z),
\]
with $C_\Gamma(t)$ the coefficient of Step 5 at $(\theta_t,M_t)$ and $X_t$ formed at the parameter $t$; the sums run over all
motifs on $[n]$.

\begin{lemma}[decomposition]
  \label{lem:e:decomp}
  Under \eqref{eq:d:A}, for every $A$ and $t\in[0,1]$,
  \[
    \ell_t(A)=\Lambda_t(A)-\Lambda_0(A)+\int_0^t\Exp_s[X_s\mid A]\,\mathrm ds .
  \]
\end{lemma}

\begin{proof}
  Fix a piece and $t$ in it. By Steps 1 and 3, $\partial_t\log P_t(A)=\Exp_t[\Sw_r\mid A]=\Exp_t[\Pi_D\mid A]+\Exp_t[\Rem_D\mid A]$.
  A disconnected motif has $B^{(r)}_\Gamma=0$ by Step 4(iv), and $C_\Gamma=0$: on the subspace $\{\eps_e=0,\ e\notin\Gamma\}$, which
  determines the coefficient, the expectation in Step 5 factors over two parts of $V(\Gamma)$ between which $\Gamma$ has no
  pair, because their labels are independent, and its logarithm has no monomial that involves both parts. For a
  connected small motif, Steps 5 and 6 give $B^{(r)}_\Gamma=\partial_tC_\Gamma(t)$, which does not depend on $Z$ and leaves the
  conditional expectation. Hence $\partial_t\log P_t(A)=\partial_t\Lambda_t(A)+\Exp_t[X_t\mid A]$. Under \eqref{eq:d:A} each $P_t(A)$
  is positive and a polynomial in $(\theta_t,M_t)$, and $C_\Gamma(t)$ is a polynomial in the values $\hom(Q;\theta_t,M_t)$; so
  $\log P_t(A)$ and $\Lambda_t(A)$ are continuous on $[0,1]$ and continuously differentiable on each piece. Integrating piece
  by piece and joining the pieces by continuity gives the identity.
\end{proof}

\begin{lemma}[error term]
  \label{lem:e:error}
  Assume \eqref{eq:d:A}, $D\ge r+2$, $n\ge n_I$, $\delta\le1$ and $\varsigma\le\varsigma_0$. There is $C_X$, depending only on
  $(M_{\rm hi},L,q_0,r,D)$, such that
  \[
    \sup_{s\in[0,1]}\Exp_sX_s^2\le\eps_X(n)^2:=C_X\Big[n\delta^r+n^2\delta^{2r}+n^{r+3}\delta^{D+1}(\log n)^{2(D-r)}
    +n^{4D+1}e^{-q_0^2\log^2n/36}\Big].
  \]
\end{lemma}

\begin{proof}
  We have $\Exp_sX_s^2\le2\Exp_s(\Pi^{>r}_D)^2+2\Exp_s\Rem_D^2$, and Step 9(iii) bounds the second term. The first is
  $\Var_s(\Pi^{>r}_D)+(\Exp_s\Pi^{>r}_D)^2$, and the large class in Step 8 bounds the variance by $Cn\delta^r$. For the mean,
  only connected motifs contribute (Step 4(iv)). Such a motif with $v=v(\Gamma)\ge r+1$ vertices has $d(\Gamma)\ge v-1$
  distinct pairs, so conditioning on $Z$ and Step 2(iii) give
  $|\Exp_s[\eps^\Gamma B^{(r)}_\Gamma(Z)]|\le\|B^{(r)}_\Gamma\|_\infty(C_\eps\Delta)^{2(v-1)}$, and Step 7 bounds
  $\|B^{(r)}_\Gamma\|_\infty\le c_v\varrho_v^{-D}L$, since $\varrho_v<1$. There are finitely many motif types with $|\Gamma|\le D$,
  each with at most $n^v$ placements and $v\le2D$, so $|\Exp_s\Pi^{>r}_D|\le C\sum_{v=r+1}^{2D}n\delta^{v-1}\le C'n\delta^r$.
\end{proof}

Two motifs have the same type if a permutation of $[n]$ maps one onto the other with the multiplicities. $C_\Gamma(t)$
depends only on the type, because the labels are i.i.d., and we write $C_\gamma(t)$, $S_\gamma(A):=\sum_{\Gamma\in\gamma}\eps^\Gamma$
and $b_\gamma:=\sup_{t\in[0,1]}|C_\gamma(t)-C_\gamma(0)|$. The types of connected small motifs fall into five classes:
$\mathfrak M_0$, those whose $G_\Gamma$ is a tree on at most four vertices; $\mathfrak M_1$, a tree on five vertices with all
multiplicities in $\{1,2\}$; $\mathfrak M_2$, a tree on five vertices with some multiplicity at least $3$; $\mathfrak M_3$, a tree
on $6\le v\le r$ vertices; and $\mathfrak M_4$, a graph with a cycle.

\begin{lemma}[main term]
  \label{lem:e:main}
  Assume \eqref{eq:d:A}, $\delta\le1$, $r\ge5$ and $D\ge8$.
  \begin{enumerate}[label=(\roman*),leftmargin=*]
    \item $b_\gamma=0$ for $\gamma\in\mathfrak M_0$. Every $b_\gamma$ is finite, and $\mathfrak M_1$ is a finite set of types whose
      $b_\gamma$ depend only on the type and the path, not on $(r,D)$.
    \item For every type there is $K_\gamma$, depending only on $\gamma$, $q_0$ and $M_{\rm hi}$, such that $\Exp_0|S_\gamma|$ is at
      most $K_\gamma(N_5+\sqrt{N_5})$ on $\mathfrak M_1$, $K_\gamma(N_5\Delta+\sqrt{N_5\Delta})$ on $\mathfrak M_2$,
      $K_\gamma(N_5\delta+\sqrt{N_5\delta})$ on $\mathfrak M_3$ and $K_\gamma(\delta^3+\delta^{3/2})$ on $\mathfrak M_4$.
    \item For every $A$, $\sup_t|\Lambda_t(A)-\Lambda_0(A)|\le\Sigma_{\rm main}(A)+\Sigma_{\rm other}(A)$, where
      $\Sigma_{\rm main}:=\sum_{\gamma\in\mathfrak M_1}b_\gamma|S_\gamma|$ and $\Sigma_{\rm other}:=\sum_{\gamma\in\mathfrak M_2\cup\mathfrak M_3\cup\mathfrak M_4}b_\gamma|S_\gamma|$.
      Moreover $\Exp_0\Sigma_{\rm main}\le C_*(N_5+\sqrt{N_5})$ with $C_*:=\sum_{\gamma\in\mathfrak M_1}b_\gamma K_\gamma$, which does not
      depend on $(r,D)$, and $\Exp_0\Sigma_{\rm other}\le C_S\eps_S(n)$ with
      $\eps_S(n):=N_5\Delta+\sqrt{N_5\Delta}+N_5\delta+\sqrt{N_5\delta}+\delta^{3/2}$ and $C_S$ depending also on $(r,D)$.
  \end{enumerate}
\end{lemma}

\begin{proof}
  (i) For $\gamma\in\mathfrak M_0$ every connected subgraph of $G_\Gamma$ with an edge is a tree on at most four vertices, whose
  function $\hom(Q;\theta_t,M_t)=F_Q(\theta_t;M_t)$ is constant along the path, so by Step 5, $C_\gamma$ does not depend on $t$. In
  general $|\hom(Q;\theta_t,M_t)|\le M_{\rm hi}^{e(Q)}$, and $C_\gamma$ is a fixed polynomial in these values, so
  $b_\gamma\le2\sup_t|C_\gamma(t)|<\infty$. The types in $\mathfrak M_1$ are given by the three trees on five vertices and
  multiplicities in $\{1,2\}^4$, so their degree is at most $8$; they are all small once $r\ge5$ and $D\ge8$, and $C_\gamma$
  is defined by $\gamma$ alone.

  (ii) Under $P_0$, two placed motifs with disjoint vertex sets give independent $\eps^\Gamma$, by the factorization in the
  proof of \Cref{lem:ustat}. Otherwise their union $\mathcal V:=\Gamma\uplus\Gamma'$, with multiplicities $a_e=m_e+m'_e$, is
  connected, and Step 2(iii) at $t=0$ gives
  \[
    |\Cov_0(\eps^\Gamma,\eps^{\Gamma'})|\le|\Exp_0\eps^\Gamma\eps^{\Gamma'}|+|\Exp_0\eps^\Gamma||\Exp_0\eps^{\Gamma'}|\le2(C_\eps\Delta)^{\sum_ec(a_e)},
  \]
  where the product of means uses $c(a)+c(a')\ge c(a+a')$ with $c(0):=0$ and $C_\eps\Delta\le1$. Since $c(a)\ge2$, we have
  $\sum_ec(a_e)\ge2d(\mathcal V)$. For a fixed type and overlap pattern, of which there are finitely many for each type, there
  are at most $n^{v(\mathcal V)}$ placements, and $|\Exp_0S_\gamma|\le n^{v(\gamma)}(C_\eps\Delta)^{\sum_ec(m_e)}$. On $\mathfrak M_1$, the identity
  $\sum_ec(m_e)=8$ gives $|\Exp_0S_\gamma|\le C_\eps^8N_5$, and $d(\mathcal V)\ge4$ and $v(\mathcal V)\le d(\mathcal V)+1$ give
  $n^{v(\mathcal V)}\Delta^{2d(\mathcal V)}\le n\delta^{d(\mathcal V)}\le N_5$. On $\mathfrak M_2$, with $m_{e^*}\ge3$, we have $\sum_ec(m_e)\ge9$, and in an
  overlap $c(a_{e^*})$ exceeds $2$ by at least one whether or not $e^*$ is shared, so $\sum_ec(a_e)\ge2d(\mathcal V)+1$: both
  bounds gain a factor $\Delta$. On $\mathfrak M_3$ we have $|\Exp_0S_\gamma|\le C\,n\delta^{v-1}\le CN_5\delta$, and $d(\mathcal V)\ge5$ gives
  $n\delta^5=N_5\delta$. On $\mathfrak M_4$, the graph $G_\Gamma$ has a cycle, so $d\ge v\ge3$ and $|\Exp_0S_\gamma|\le Cn^v\Delta^{2v}\le C\delta^3$,
  while $\mathcal V$ contains that cycle, so $v(\mathcal V)\le d(\mathcal V)$ and $n^{v(\mathcal V)}\Delta^{2d(\mathcal V)}\le\delta^{d(\mathcal V)}\le\delta^3$.
  Finally $\Exp_0|S_\gamma|\le|\Exp_0S_\gamma|+\Var_0(S_\gamma)^{1/2}$.

  (iii) $\Lambda_t-\Lambda_0=\sum_\gamma(C_\gamma(t)-C_\gamma(0))S_\gamma$ over the types of connected small motifs; the terms of
  $\mathfrak M_0$ vanish, and every other term is at most $b_\gamma|S_\gamma|$ for every $t$. Take expectations and use (ii).
\end{proof}

\begin{lemma}[stopping time]
  \label{lem:e:stop}
  Under the hypotheses of \Cref{lem:e:error}, for $K>0$ let
  $\mathcal E_2(K):=\{A:\sup_{t\in[0,1]}|\Lambda_t(A)-\Lambda_0(A)|<K\log n-1\}$. Then
  \[
    P_0\Big(\mathcal E_2(K)\cap\Big\{\ell_1<-\sup_{t\in[0,1]}|\Lambda_t-\Lambda_0|-1\Big\}\Big)\le n^K\eps_X(n).
  \]
\end{lemma}

\begin{proof}
  Fix $A$. The map $s\mapsto\ell_s(A)$ is continuous with $\ell_0(A)=0$. Let $\sigma(A):=\inf\{s\in[0,1]:\ell_s(A)\le-K\log n\}$,
  and $\sigma(A):=1$ if this set is empty. If it is not empty, it is closed and $\ell_{\sigma(A)}(A)=-K\log n$. In either case
  $P_0(A)<n^KP_s(A)$ for $s<\sigma(A)$. Let $\mathcal E_1:=\{A:\int_0^{\sigma(A)}|\Exp_s[X_s\mid A]|\,\mathrm ds\le1\}$. Markov's
  inequality, Fubini's theorem on the finite observation space and $|\Exp_s[X_s\mid A]|\le\Exp_s[|X_s|\mid A]$ give
  \begin{align*}
    P_0(\mathcal E_1^c)&\le\int_0^1\sum_AP_0(A)\Ind{s<\sigma(A)}\big|\Exp_s[X_s\mid A]\big|\,\mathrm ds\\
    &\le n^K\int_0^1\sum_AP_s(A)\,\Exp_s\big[|X_s|\bigm|A\big]\,\mathrm ds\le n^K\sup_s\Exp_s|X_s|\le n^K\eps_X(n).
  \end{align*}
  On $\mathcal E_1$, \Cref{lem:e:decomp} gives $|\ell_t-(\Lambda_t-\Lambda_0)|\le1$ for all $t\le\sigma(A)$. If $A\in\mathcal E_1\cap\mathcal E_2(K)$
  and the set defining $\sigma(A)$ were not empty, then $-K\log n=\ell_{\sigma(A)}\ge\Lambda_{\sigma(A)}-\Lambda_0-1>-K\log n$, a
  contradiction. So $\sigma(A)=1$, and $\ell_1\ge\Lambda_1-\Lambda_0-1\ge-\sup_t|\Lambda_t-\Lambda_0|-1$. The event of the lemma is
  therefore contained in $\mathcal E_1^c$.
\end{proof}

\subsection{Proof of \texorpdfstring{\Cref{thm:threshold}}{the threshold theorem}}

\emph{Part (a).} Fix $c'>0$ and let $C_*$ be the constant of \Cref{lem:e:main}, which does not depend on $(r,D)$. Let
$K:=\lceil8C_*c'+1\rceil$, $r:=10K+15$ and $D:=5r+10K+25$, integers with $D\ge r+2$, $r\ge5$ and $D\ge8$; with this $r$ take
$L$, then $C_X,n_I,\varsigma_0$ from \Cref{lem:e:error} and $C_S$ from \Cref{lem:e:main}. Let $N_5\le c'\log n$. Then
$\delta=(N_5/n)^{1/4}\le(c'\log n/n)^{1/4}$, so for $n$ large, $\delta\le n^{-1/5}$, $\varsigma\le\varsigma_0$, $n\ge n_I$ and
$\Delta=\sqrt{\delta/n}\le n^{-1/2}\le\Delta_0$, which gives \eqref{eq:d:A}. Then $n\delta^r\le n^{-2K-2}$,
$n^2\delta^{2r}\le n^{-4K-4}$ and $n^{r+3}\delta^{D+1}(\log n)^{2(D-r)}\le n^{-2K-11/5}(\log n)^{2D}$, and the last term of
$\eps_X(n)^2$ decays faster than every power of $n$; so for $n$ large, $\eps_X(n)^2\le n^{-2K-1}$ and
$n^K\eps_X(n)\le n^{-1/2}\le1/8$. Also $\eps_S(n)\to0$, so $C_S\eps_S(n)\le1$ for $n$ large. Let
\[
  \mathcal A:=\big\{\Sigma_{\rm main}\le8C_*(N_5+\sqrt{N_5})\big\}\cap\big\{\Sigma_{\rm other}\le8\big\}
  \cap\Big\{\ell_1\ge-\sup_t|\Lambda_t-\Lambda_0|-1\Big\}.
\]
By \Cref{lem:e:main} and Markov's inequality, each of the first two events fails with probability at most $1/8$. On
both, $\sup_t|\Lambda_t-\Lambda_0|\le8C_*(c'\log n+\sqrt{c'\log n})+8$, which is less than
$K\log n-1\ge8C_*c'\log n+\log n-1$ once $\log n>8C_*\sqrt{c'\log n}+9$. So for $n$ large the first two events lie in
$\mathcal E_2(K)$, and \Cref{lem:e:stop} bounds by $1/8$ the probability that the third fails on them. Hence
$P_0(\mathcal A)\ge5/8$, and on $\mathcal A$, by $\sqrt{N_5}\le(N_5+1)/2$,
\[
  \ell_1\ge-8C_*\big(N_5+\sqrt{N_5}\big)-9\ge-L_n,\qquad L_n:=12C_*N_5+4C_*+9 .
\]
For an estimator $\hH$, which may use auxiliary randomness independent of the verdicts, let
$\mathcal E_i:=\{|\hH-\Ent(\theta^i)|\ge\omega/2\}$; since the entropies differ by $\omega$, $\mathcal E_0^c\subseteq\mathcal E_1$. On $\mathcal A$,
$P_1(A)\ge e^{-L_n}P_0(A)$, so
\[
  \begin{aligned}
  P_0(\mathcal E_0)+P_1(\mathcal E_1)&\ge\sum_{A\in\mathcal A}P_0(A)\Pr(\mathcal E_0\mid A)+e^{-L_n}\sum_{A\in\mathcal A}P_0(A)\Pr(\mathcal E_0^c\mid A)\\
  &\ge e^{-L_n}P_0(\mathcal A)\ge\tfrac58e^{-L_n},
  \end{aligned}
\]
where $\Pr(\cdot\mid A)$ is over the auxiliary randomness, whose law is the same under both parameters. One of the two
probabilities is at least $\frac5{16}e^{-L_n}$, and by Markov's inequality the corresponding risk is at least
$(\omega/2)^2\frac5{16}e^{-L_n}$. This proves (a) with $c_1:=\frac{5\omega^2}{64}e^{-4C_*-9}$ and $C_1:=12C_*$, which depend
only on the path, $q_0$ and $\omega$; the threshold $n_1$ depends also on $c'$.

\emph{Part (b).} Let $M\in\cU$. The upper bound is \Cref{thm:upper}(a) for $\mathcal K=\{M\}$, with constants $c_U,C_U$; it
applies because the window gives $\delta\le1$ and \eqref{eq:P} holds once $\Delta\bar U\le q_0$. If $N_4<1$, the risk is at most
$(\log3)^2=(\log3)^2\min\{1,N_4^{-1/3}\}$; so $R^*_n\le C_T(\min\{1,N_4^{-1/3}\}+e^{-c_UN_5})$ with
$C_T:=\max\{C_U,(\log3)^2\}$. For the lower bound, which uses only $M\in\Omega$, part (a) with $c'=1$ covers
$N_5\le\log n$. For $N_5>\log n$, \Cref{thm:lower} for $\{M\}\subset\Omega$ gives
$R^*_n\ge c_L\min\{1,N_4^{-1/3}\}\ge c_Ln^{-1/3}\ge c_Le^{-N_5/3}$, because $N_4\le n$. So $R^*_n\ge c_Ee^{-C_EN_5}$ with
$c_E:=\min\{c_1,c_L\}$ and $C_E:=\max\{C_1,1/3\}$, and with \Cref{thm:lower}
once more, $R^*_n\ge\max\{c_L\min\{1,N_4^{-1/3}\},c_Ee^{-C_EN_5}\}\ge c_T(\min\{1,N_4^{-1/3}\}+e^{-C_EN_5})$ with
$c_T:=\frac12\min\{c_L,c_E\}$. Take $n_T$ at least $n_1$ for $c'=1$, $n_L$, $\lceil(\bar U/q_0)^2\rceil$ and $8$. \qed

\subsection{Proof of \texorpdfstring{\Cref{cor:phase}}{the phase diagram}}

With $\Delta=n^{-\beta}$ we have $\delta=n^{1-2\beta}$, $N_4=n\delta^3=n^{4-6\beta}$ and $N_5=n\delta^4=n^{5-8\beta}$. Let
$\kappa:=\min\{2\beta-1,1/2\}\in(0,1)$; then $n\Delta^2=n^{-(2\beta-1)}\le n^{-\kappa}$. Since $q_n\in[q_0,1-q_0]$ and
$\Delta\bar U\le q_0$ for large $n$, \eqref{eq:P} holds.

(i) The compact set $\{M\}\subset\cU_4$ and \Cref{thm:main}(b) give $R^*_n\asymp\min\{1,N_4^{-1/3}\}=\min\{1,n^{-(4/3-2\beta)}\}$,
which is $n^{-(4/3-2\beta)}$ for $\beta<2/3$ and $1$ for $\beta\ge2/3$.

(ii) If $\beta<5/8$, then $\kappa=2\beta-1<1/4$ and $N_5=n^{5-8\beta}\ge C_{\log}\log n$ for large $n$, so \Cref{thm:main}(a)
with $\mathcal K=\{M\}$ gives $R^*_n\asymp N_4^{-1/3}=n^{-(4/3-2\beta)}$. If $\beta\ge5/8$, then $N_5\le1\le\log n$, and
\Cref{thm:threshold}(a) with $c'=1$ gives $R^*_n\ge c_1e^{-C_1}$; for $5/8\le\beta<2/3$ we have $N_4=n^{4-6\beta}\to\infty$. Since
$\Ent\le\log3$, we have $R^*_n\le(\log3)^2$.

In both cases $R^*_n\asymp n^{-a(\beta)}$ with the stated $a(\beta)$ for every $\beta>1/2$, so $-\log R^*_n/\log n\to a(\beta)$.
In case (ii), $a(\beta)=4/3-2\beta\to1/12$ as $\beta\uparrow5/8$. \qed

\section{Channels with equal diagonal entries}
\label{app:symmetric}

\subsection{Strata and conditions}

For a face $k$ let $\{i,j\}=[3]\setminus\{k\}$ with $i<j$, let $\sigma_k$ exchange the coordinates $i$ and $j$, and let
$\mathbf d_k:=e_i-e_j$, $\mathbf n_k:=e_k-\frac12(e_i+e_j)$ and $\theta_k(t,y):=\frac12(e_i+e_j)+t\,\mathbf d_k+y\,\mathbf n_k$, so that the $k$-th coordinate of
$\theta_k(t,y)$ is $y$ and $\sigma_k\theta_k(t,y)=\theta_k(-t,y)$; on a face with $M_{ii}=M_{jj}$, where $f_k=\frac12(e_i+e_j)$ under (F0$\sigma$)
below, these are the face coordinates of (U2). Let $\ell_k:=\{\theta\in\aff\Simplex:\theta_k=0\}$. For $\cJ\subseteq[3]$,
$\cS_\cJ:=\{M\in\Sym:M_{ii}=M_{jj}\text{ for all }k\in\cJ\}$, and $\cJ(M):=\{k:M_{ii}=M_{jj}\}$, which is empty, a singleton or $[3]$. The
symmetric pairs of the face $k$ are $R_k:=\{(\varphi,\sigma_k\varphi):\varphi\in\ell_k\cap\Simplex\}$, and $R_\cJ:=\bigcup_{k\in\cJ}R_k$. Let
$q(t,y):=(t^2,y,ty)$ and, for $\varphi\in\ell_k$, let $K_k(\varphi)$ be the $4\times3$ matrix with columns $DF_{\le4}(\varphi)\mathbf d_k$,
$DF_{\le4}(\varphi)\mathbf n_k$ and $DF_{\le4}(\sigma_k\varphi)\mathbf n_k$.

For $M\in\cS_\cJ$ and $k\in\cJ$ consider:
\begin{description}[leftmargin=2.6em,style=nextline,itemsep=2pt]
  \item[(F0$\sigma$)] $M_{ii}\ne M_{ij}$. Then (F0) holds on the face $k$ with $p_k=1/2$, and $f_k=\frac12(e_i+e_j)=\theta_k(0,0)$.
  \item[(U2$\sigma$)] In the face coordinates of (U2), $\det J_k\ne0$ and $(\psi^{P_4}_{11},\psi^{K_{1,3}}_{11})\ne(0,0)$.
  \item[(T)] $\rank K_k(\varphi)=3$ for every $\varphi\in\ell_k\cap\Simplex$ with $\varphi\ne f_k$.
  \item[(U4$\sigma$)] If $\theta\ne\theta'$ and $F^M_{\le5}(\theta)=F^M_{\le5}(\theta')$, then $(\theta,\theta')\in R_\cJ$.
\end{description}
Let $\cU^\sigma_\cJ$ be the set of $M\in\cS_\cJ$ with $\cJ(M)=\cJ$ that satisfy (N), (F0) and (U1), (U2) on every face $k\notin\cJ$,
(F0$\sigma$) and (U2$\sigma$) on every face $k\in\cJ$, (U3), (T) and (U4$\sigma$), and let $\cU^\sigma_{\cJ,4}$ be the subset on which $\theta\ne\theta'$ and
$F^M_{\le4}(\theta)=F^M_{\le4}(\theta')$ imply $(\theta,\theta')\in R_\cJ$. For $\cJ=\emptyset$, (T) is empty, (U4$\sigma$) is (U4), and
$\cU^\sigma_\emptyset=\cU$; for $k\in\cJ$, (U2)(ii) always fails (\Cref{lem:f:sym}(d)).

\subsection{Symmetry}

\begin{lemma}[symmetry]
  \label{lem:f:sym}
  Let $M_{ii}=M_{jj}$. (a) For every graph $G$ and $\theta\in\ell_k$, $\hom(G;\sigma_k\theta,M)=\hom(G;\theta,M)$. (b) For every tree $T$,
  $F_T(\theta_k(t,y))=a_T(t^2,y)+t\,y\,b_T(t^2,y)$ for unique polynomials $a_T,b_T$ whose coefficients are polynomials in $M$.
  (c) $\Ent(\theta_k(t,y))=\Ent(\theta_k(-t,y))$. (d) In the face coordinates of (U2), $\phi^T_{10}=\phi^T_{30}=0$, hence $\psi^T_{30}=0$ and
  (U2)(ii) fails. (e) For $\varphi\in\ell_k$, $DF_T(\sigma_k\varphi)\mathbf d_k=-DF_T(\varphi)\mathbf d_k$. (f) If also $M_{ii}\ne M_{ij}$, then $p_k=1/2$.
\end{lemma}

\begin{proof}
  (a) Viewing $\sigma=\sigma_k$ as a permutation of $[3]$, the substitution $z\mapsto\sigma\circ z$ gives $\hom(G;\sigma\theta,M)=\hom(G;\theta,\sigma M\sigma)$.
  For $\theta_k=0$, only labelings with values in $\{i,j\}$ contribute, and $\sigma M\sigma$ agrees with $M$ on $\{i,j\}^2$:
  $(\sigma M\sigma)_{ii}=M_{jj}=M_{ii}$, $(\sigma M\sigma)_{ij}=M_{ij}$, $(\sigma M\sigma)_{jj}=M_{jj}$. (b) Split $F_T(\theta_k(t,y))$ into its even and odd
  parts in $t$, $A(t^2,y)+tB(t^2,y)$. By (a) and $\theta_k(-t,0)=\sigma\theta_k(t,0)$, the function $F_T(\theta_k(t,0))$ is even, so $B(\tau,0)\equiv0$ and
  $B=y\,b_T$. (c) $\Ent$ is invariant under permuting coordinates. (d) $\phi^T_{a0}$ is the coefficient of $t^a$ in $F_T(\theta_k(t,0))$,
  which vanishes for odd $a$; so $\psi^T_{30}=0$, and $(X,Y)=(0,1)$ solves the two equations of (U2)(ii). (e) Differentiate the
  identity $F_T\circ\sigma=F_T$ on $\ell_k$ along $\mathbf d_k$, using $\sigma\mathbf d_k=-\mathbf d_k$. (f) $p_k=(M_{jj}-M_{ij})/(M_{ii}-2M_{ij}+M_{jj})=1/2$.
\end{proof}

\subsection{The midpoint of a symmetric face}

For $M\in\cS_{\{k\}}$ let $\Psi^M_T(\tau,y,w):=a_T(\tau,y)+w\,b_T(\tau,y)$, so that $F^M_T(\theta_k(t,y))=\Psi^M_T(q(t,y))$. The columns of
$D\Psi^M(0)$ are $(\phi^T_{20})_T$, $(\phi^T_{01})_T$ and $(\phi^T_{11})_T$. If $\det J_k\ne0$, then in the rows of $P_4$ and $K_{1,3}$, subtracting $\alpha'_T$ times the edge row and
$\beta'_T$ times the wedge row makes the first two entries vanish and the third entry
$\psi^T_{11}$, while the edge and wedge rows carry the invertible $J_k$ in the first two columns. So (U2$\sigma$) implies
$\rank D\Psi^M(0)=3$, and under $\det J_k\ne0$ the two are equivalent; only the first implication is used.

\begin{lemma}[symmetric midpoint]
  \label{lem:f:mid}
  Let $\bar M\in\cS_{\{k\}}$ satisfy (F0$\sigma$) and (U2$\sigma$). There are $\eps>0$, $\rho\in(0,1/4]$, $c_F>0$ and $C_F<\infty$ such that for
  $M\in\cS_{\{k\}}$ with $\|M-\bar M\|\le\eps$ and $\theta=\theta_k(t,y)$, $\theta'=\theta_k(t',y')$ in
  $V_k(\rho):=\{\theta_k(t,y):|t|\le\rho,\ 0\le y\le\rho\}\subset\Simplex$:
  \begin{enumerate}[label=(\alph*),leftmargin=*]
    \item $\cD^M_4(\theta,\theta')\ge c_F|q(t,y)-q(t',y')|$;
    \item $\rank DF^M_{\le4}(\theta)|_{\one^\perp}=2$ if $\theta\ne f_k$;
    \item $|\Ent(\theta)-\Ent(\theta')|^3\le C_F\,\cD^M_4(\theta,\theta')$;
    \item $\rank K^M_k(\theta_k(t,0))=3$ for $0<|t|\le\rho$.
  \end{enumerate}
  Moreover $\Simplex\cap\bar B(f_k,\rho)\subset V_k(\rho)$.
\end{lemma}

\begin{proof}
  The coordinates $i,j$ of $\theta_k(t,y)$ are $\frac12\pm t-\frac y2\ge\frac18$ for $\rho\le1/4$, so $V_k(\rho)\subset\Simplex$. Let
  $\sigma_0:=\sigma_{\min}(D\Psi^{\bar M}(0))>0$. Since $(M,p)\mapsto D\Psi^M(p)$ is polynomial, there are $\eps,r_\Psi>0$ with
  $\|D\Psi^M(p)-D\Psi^{\bar M}(0)\|_{\rm op}\le\sigma_0/2$ for $\|M-\bar M\|\le\eps$ and $|p|\le r_\Psi$. Take $\rho\le\min\{1/4,r_\Psi/3\}$, so
  $|q(t,y)|\le t^2+y+|t|y\le3\rho\le r_\Psi$. (a) On the convex ball of radius $r_\Psi$, $|\Psi^M(p)-\Psi^M(p')|\ge\frac{\sigma_0}2|p-p'|$ as in
  \Cref{lem:b:rank}, and $|w|_\infty\ge|w|/2$ gives $c_F:=\sigma_0/4$. (b) $\partial_{(t,y)}F(\theta_k(t,y))=D\Psi(q)\,Dq$, where
  $Dq=\big(\begin{smallmatrix}2t&0\\0&1\\y&t\end{smallmatrix}\big)$, which has rank two for $(t,y)\ne0$, and $D\Psi(q)$ is injective. (d) With
  $\varphi=\theta_k(t,0)$ we have $DF(\varphi)\mathbf d_k=D\Psi(q)(2t,0,0)^\top$, $DF(\varphi)\mathbf n_k=D\Psi(q)(0,1,t)^\top$ and, since $q(-t,0)=q(t,0)$,
  $DF(\sigma_k\varphi)\mathbf n_k=D\Psi(q)(0,1,-t)^\top$; so $K_k(\varphi)=D\Psi(q)B(t)$ with $\det B(t)=-4t^2\ne0$.

  (c) $\Ent(\theta_k(t,y))=\psi_y(t)+\varphi(y)$ with $\psi_y(t):=\varphi(a+t)+\varphi(a-t)$ and $a:=\frac12-\frac y2\in[\frac38,\frac12]$, so $a\pm t\in[\frac18,\frac34]$.
  The function $\psi_y$ is even, $\psi_y'(0)=0$ and $|\psi_y''|\le16$ because $|\varphi''(x)|=1/x\le8$; hence $|\psi_y(t)-\psi_y(t')|\le8|t^2-t'^2|$. Also
  $|\partial_y\psi_y(t)|=\frac12|\varphi'(a+t)+\varphi'(a-t)|\le L_y:=1+\log8$ and $|\varphi(y)-\varphi(y')|\le\omega(|y-y'|)$. With
  $\delta:=|y-y'|\le\rho$ and $\omega(\delta)\ge2\delta$ we have $|\Ent(\theta)-\Ent(\theta')|\le8|t^2-t'^2|+(L_y/2+1)\omega(\delta)$, and with $(a+b)^3\le4(a^3+b^3)$,
  $|t^2-t'^2|\le\rho^2$ and $\omega(\delta)^3\le\Xi_2\delta$,
  \[
    \begin{aligned}
      |\Ent(\theta)-\Ent(\theta')|^3&\le2048\rho^4|t^2-t'^2|+4(L_y/2+1)^3\Xi_2\delta\\
      &\le C'\big(|t^2-t'^2|+|y-y'|\big)\le\sqrt2C'|q-q'|\le(\sqrt2C'/c_F)\cD_4,
    \end{aligned}
  \]
  with $C':=\max\{2048\rho^4,4(L_y/2+1)^3\Xi_2\}$. Finally $\theta-f_k=t\mathbf d_k+y\mathbf n_k$, which has $|\theta-f_k|^2=2t^2+\frac32y^2$, so $|\theta-f_k|\le\rho$ gives
  $|t|\le\rho$ and $0\le y\le\rho$; and $f_k$ does not depend on $M\in\cS_{\{k\}}$.
\end{proof}

\subsection{The curve of symmetric pairs}

\begin{lemma}[symmetric pairs]
  \label{lem:f:pairs}
  Let $\bar M\in\cS_{\{k\}}$ and $\bar\varphi\in\ell_k\cap\Simplex$ with $\rank K^{\bar M}_k(\bar\varphi)=3$. There are $\eps,r,c_T>0$ and $C_T<\infty$ such
  that for $M\in\cS_{\{k\}}$ with $\|M-\bar M\|\le\eps$, $\theta\in\Simplex\cap\bar B(\bar\varphi,r)$ and $\theta'\in\Simplex\cap\bar B(\sigma_k\bar\varphi,r)$, with
  $\theta'':=\sigma_k\theta'$, $\theta=\theta_k(t,y)$ and $\theta''=\theta_k(t'',y'')$:
  (a) $\cD^M_4(\theta,\theta')\ge c_T|(t-t'',y,y'')|$; (b) $|\Ent(\theta)-\Ent(\theta')|^3\le C_T\,\cD^M_4(\theta,\theta')$.
\end{lemma}

\begin{proof}
  The map $G^M(\theta,\theta''):=F^M_{\le4}(\theta)-F^M_{\le4}(\sigma_k\theta'')$ is polynomial in $(t,y,t'',y'',M)$, and $G(\varphi,\varphi)=0$ for $\varphi\in\ell_k$ by
  \Cref{lem:f:sym}(a). Let $\varphi:=\theta_k(\bar t,0)$ with $\bar t:=(t+t'')/2$ and $v:=(\theta-\varphi,\theta''-\varphi)$. By the chain rule,
  the identities $\sigma_k\mathbf d_k=-\mathbf d_k$ and $\sigma_k\mathbf n_k=\mathbf n_k$, and \Cref{lem:f:sym}(e),
  \[
    \begin{aligned}
    DG(\varphi,\varphi)v&=DF(\varphi)\big[\tfrac{t-t''}2\mathbf d_k+y\mathbf n_k\big]-DF(\sigma_k\varphi)\sigma_k\big[\tfrac{t''-t}2\mathbf d_k+y''\mathbf n_k\big]\\
    &=K_k(\varphi)\,(t-t'',\ y,\ -y'')^\top .
    \end{aligned}
  \]
  Since $(M,\varphi)\mapsto\sigma_{\min}(K^M_k(\varphi))$ is continuous, after shrinking $\eps,r$ it is at least $\sigma_1/2$, with
  $\sigma_1:=\sigma_{\min}(K^{\bar M}_k(\bar\varphi))>0$. Let $z:=(t-t'',y,y'')$. Since $\mathbf d_k\perp\mathbf n_k$, $|\mathbf d_k|^2=2$ and $|\mathbf n_k|^2=3/2$, we have
  $|v|^2=(t-t'')^2+\frac32(y^2+y''^2)\le\frac32|z|^2$. Taylor's formula gives $|G(\theta,\theta'')|\ge\frac{\sigma_1}2|z|-C_2|v|^2\ge\frac{\sigma_1}2|z|-\frac32C_2|z|^2$,
  with $C_2$ a bound on the second derivatives of $G$ near the relevant points; for $r$ small, $|z|\le\sigma_1/(6C_2)$, so
  $|G|\ge\frac{\sigma_1}4|z|$ and $\cD_4(\theta,\theta')=|G|_\infty\ge\frac{\sigma_1}8|z|$, which is (a). For (b), $\Ent(\theta')=\Ent(\theta'')$; since
  $\theta-\theta''=(t-t'')\mathbf d_k+(y-y'')\mathbf n_k$, we have $|\theta-\theta''|^2\le3|z|^2$ and, for $r$ small, $|\theta-\theta''|\le1$, so by
  \Cref{lem:a:entropy}, $|\Ent(\theta)-\Ent(\theta'')|^3\le27\Xi_2|\theta-\theta''|\le27\sqrt3\,\Xi_2|z|\le(216\sqrt3\,\Xi_2/\sigma_1)\cD_4$.
\end{proof}

The lemma also holds at the vertices $\bar\varphi\in\{e_i,e_j\}$, since the argument only uses Taylor expansions in $\aff\Simplex$.

\subsection{Proof of \texorpdfstring{\Cref{thm:symmetric}}{the theorem on equal diagonal entries}}

\emph{The modulus on a stratum.} Fix a compact $\mathcal K\subset\cU^\sigma_\cJ$; all neighborhoods are taken in $\cS_\cJ\times\Simplex$ or
$\cS_\cJ\times\Simplex^2$. \emph{Near pairs:} for $(\bar M,\bar\theta)\in\mathcal K\times\Simplex$ use \Cref{lem:b:star} at $\theta^*$, \Cref{lem:b:face} at
$f_k$ with $k\notin\cJ$, \Cref{lem:f:mid} at $f_k$ with $k\in\cJ$ (with $W:=\Simplex\cap B(f_k,\rho)$; note $\cS_\cJ\subset\cS_{\{k\}}$), and
\Cref{lem:b:rank} elsewhere by (U3); the first three lemmas hold on full neighborhoods of $\bar M$ and restrict to $\cS_\cJ$. As in
the proof of \Cref{thm:modulus}(a), a finite subcover and its Lebesgue number $\delta_0$ give $|\Delta\Ent|^3\le C_{\rm near}\cD_4$ for
$|\theta-\theta'|<\delta_0$. \emph{Pairs near symmetric pairs:} for $k\in\cJ$ and $(\bar M,\bar\varphi)\in\mathcal K\times(\ell_k\cap\Simplex)$, use
\Cref{lem:f:pairs} with the open set $B(\bar M,\eps)\times B(\bar\varphi,r)\times B(\sigma_k\bar\varphi,r)$ if $\bar\varphi\ne f_k$, which is possible by (T),
and \Cref{lem:f:mid} with $B(\bar M,\eps)\times(\Simplex\cap B(f_k,\rho))^2$ if $\bar\varphi=f_k$. These open sets cover the compact set
$\mathcal K\times R_k$; take a finite subcover $U_1,\dots,U_L$, and let $\delta_1>0$ be the smallest over $k$ of the distance from
$\mathcal K\times R_k$ to the complement of $\bigcup_lU_l$. If $M\in\mathcal K$ and $\operatorname{dist}((\theta,\theta'),R_k)<\delta_1$, the triple
$(M,\theta,\theta')$ lies in some $U_l$, where the bound holds for every triple: $|\Delta\Ent|^3\le C_{\rm refl}\cD_4$. \emph{Far pairs:}
the set $Z:=\{(M,\theta,\theta')\in\mathcal K\times\Simplex^2:|\theta-\theta'|\ge\delta_0,\ \operatorname{dist}((\theta,\theta'),R_\cJ)\ge\delta_1,\ \cD^M_4=0\}$ is compact
and $\cD_5>0$ on it by (U4$\sigma$); with $\eta:=\frac12\min_Z\cD_5$ ($\eta:=1$ if $Z=\emptyset$), the compact set $K'$ of triples in the same
region with $\cD_5\le\eta$ has $d_4:=\min_{K'}\cD_4>0$ ($d_4:=1$ if $K'=\emptyset$), and $|\Delta\Ent|^3\le(\log3)^3\cD_4/d_4$ there. So
$C:=\max\{C_{\rm near},C_{\rm refl},(\log3)^3/d_4\}$ works for all pairs with $\cD_5\le\eta$. On $\cU^\sigma_{\cJ,4}$ the same argument
without the $\cD_5$ condition, using $d''_4:=\min\cD_4>0$ on the triples with $|\theta-\theta'|\ge\delta_0$ and
$\operatorname{dist}((\theta,\theta'),R_\cJ)\ge\delta_1$, gives a constant $C_4$ for all pairs.

\emph{Relative openness.} Let $\bar M\in\cU^\sigma_\cJ$. For $k\notin\cJ$, the inequality $\bar M_{ii}\ne\bar M_{jj}$ persists nearby, so $\cJ(M)=\cJ$ near $\bar M$
in $\cS_\cJ$. (N), (F0), (F0$\sigma$), (U1) and (U2) for $k\notin\cJ$ are open as in \Cref{prop:b:open}, and (U2$\sigma$) is a strict
condition on quantities continuous on $\cS_{\{k\}}$. (U3) follows from part (b) of \Cref{lem:b:rank,lem:b:star,lem:b:face,lem:f:mid} as
in \Cref{prop:b:open}. (T): for $0<|t|\le\rho$ it holds near $\bar M$ by \Cref{lem:f:mid}(d), and for $\rho\le|t|\le\frac12$ by continuity of
$\sigma_{\min}(K^M_k)$ on a compact set. (U4$\sigma$): let $M_n\to\bar M$ in $\cS_\cJ$ with $\theta_n\ne\theta'_n$,
$F^{M_n}_{\le5}(\theta_n)=F^{M_n}_{\le5}(\theta'_n)$ and $(\theta_n,\theta'_n)\notin R_\cJ$, converging to $(\bar\theta,\bar\theta')$. If $\bar\theta\ne\bar\theta'$ and
$(\bar\theta,\bar\theta')\notin R_\cJ$, the limit contradicts (U4$\sigma$) at $\bar M$. If $\bar\theta=\bar\theta'$, for large $n$ both points lie in one
element of the near cover; \Cref{lem:b:rank,lem:b:star,lem:b:face} give $\cD_4>0$, and \Cref{lem:f:mid}(a) gives $\cD_4\ge c_F|q-q'|$, where
$q(t,y)=q(t',y')$ with $(t,y)\ne(t',y')$ holds only if $y=y'=0$ and $t'=-t\ne0$, that is, $(\theta,\theta')\in R_k$, which is excluded. If
$(\bar\theta,\bar\theta')\in R_k$ with $\bar\theta\ne\bar\theta'$, then $\bar\theta\in\ell_k\setminus\{f_k\}$, and \Cref{lem:f:pairs} gives $\cD_4\ge c_T|z|$, where
$z=0$ only if $\theta_n\in\ell_k$ and $\theta'_n=\sigma_k\theta_n$, again excluded. The same argument with $F_{\le4}$ shows that
$\cU^\sigma_{\cJ,4}$ is relatively open.

\emph{The rates.} The proof of \Cref{thm:upper}(a) uses $\cU$ only through \Cref{thm:modulus}(a), and the estimator does not depend
on $\mathcal K$; the modulus on the stratum therefore gives \Cref{thm:upper}(a) on compact subsets of $\cU^\sigma_\cJ$, with
constants that depend only on the compact set. In the same way the modulus on $\cU^\sigma_{\cJ,4}$ gives \Cref{thm:upper}(b).
\Cref{thm:lower} needs only (N), which is part of the definition. The arithmetic of the proof of \Cref{thm:main} does not involve
$M$. On a symmetric face the estimators may return the reflection $\sigma_k\theta$ of $\theta$, which has the same entropy. The
memberships of $M_3,\dots,M_9$ are proved in \Cref{app:certificates}. \qed

\section{Computer-assisted certificates}
\label{app:certificates}

\subsection{The channels}

\[
  M_0=\tfrac1{20}\!\begin{pmatrix}20&-6&1\\-6&17&4\\1&4&15\end{pmatrix}\!,\
  M_1=\tfrac1{20}\!\begin{pmatrix}21&-4&-2\\-4&25&6\\-2&6&26\end{pmatrix}\!,\
  M_2=\tfrac1{20}\!\begin{pmatrix}27&6&1\\6&23&-4\\1&-4&14\end{pmatrix}\!,
\]
\[
  M_3=\tfrac1{20}\!\begin{pmatrix}17&5&-4\\5&17&-2\\-4&-2&22\end{pmatrix}\!,\
  M_4=\tfrac1{20}\!\begin{pmatrix}23&-5&-4\\-5&17&3\\-4&3&23\end{pmatrix}\!,\
  M_5=\tfrac1{20}\!\begin{pmatrix}25&0&-6\\0&21&4\\-6&4&21\end{pmatrix}\!,
\]
\[
  M_6=\tfrac1{60}\!\begin{pmatrix}52&-18&3\\-18&52&12\\3&12&52\end{pmatrix}\!,\
  M_7=\tfrac1{20}\!\begin{pmatrix}18&-1&-5\\-1&18&1\\-5&1&18\end{pmatrix}\!,
\]
\[
  M_8=\tfrac1{20}\!\begin{pmatrix}28&2&-5\\2&28&-3\\-5&-3&28\end{pmatrix}\!,\
  M_9=\tfrac1{20}\!\begin{pmatrix}13&3&-2\\3&13&5\\-2&5&13\end{pmatrix}\!.
\]
The twenty grid channels were drawn with independent diagonal entries uniform on
$\{12,\dots,28\}/20$ and off-diagonal entries uniform on $\{-6,\dots,6\}/20$, conditioned on $\min_c\theta^*_c\ge1/20$; $M_0,\dots,M_5$ are six of
them. $M_3$, $M_4$, $M_5$ have one pair of equal diagonal entries, so $\cJ(M_3)=\{3\}$, $\cJ(M_4)=\{2\}$ and $\cJ(M_5)=\{1\}$. The channel $M_6$ has the
off-diagonal entries of $M_0$ and the common diagonal entry $13/15$, the mean of the diagonal of $M_0$; $M_7$, $M_8$, $M_9$ are the
first three channels of a random draw that satisfy (N) and (F0), with a common diagonal entry uniform on $\{12,\dots,28\}/20$ and
distinct off-diagonal entries uniform on $\{-6,\dots,6\}/20$. The random samples have independent
entries uniform on the grid of step $1/200$, diagonal entries in $[1/2,3/2]$ and off-diagonal entries in $[-1/2,1/2]$; in the
second sample one diagonal draw is shared by the three diagonal entries. Seeds and selection rules were fixed before any
channel was examined; the seeds and the matrices of the other $14$ grid channels and of all $300$ sampled
channels are in the repository given in \Cref{sec:conclusion}.

\subsection{Method}

All computations use exact rational arithmetic in SymPy, including resultants, factorization over $\mathbb Q$ and the exact
isolation of real roots, refined by rational bisection to width at most $2^{-330}$; only at irrational roots are candidate
points enclosed by interval arithmetic in mpmath at $400$-bit precision with outward rounding.

\emph{(N), (F0), (U1).} Since $\tM=D_u(M\Theta^*)D_u^{-1}$ with $\Theta^*=\diag(\theta^*)$, the matrices $\tM$ and $M\Theta^*$ have the same spectrum, whose
characteristic polynomial is $(x-c)\,\mathsf q(x)$ with $\mathsf q$ rational. The reduced cubic forms are computed in the coordinates
$v=\theta-\theta^*=v_1(e_1-e_3)+v_2(e_2-e_3)$; a linear change of coordinates preserves degrees, and $(\alpha_T,\beta_T)$ is determined by the
identity of quadratic forms, so $\tFthree_T$ is a cubic form on the tangent plane independent of coordinates. (U1)(i) holds if the
homogeneous resultant of the two forms is nonzero. For (U1)(ii) and (iii), the vectors $D_ue^\perp_i$ are the generalized eigendirections
of the pair of quadratic forms of the edge and wedge functions, with distinct generalized eigenvalues $\lambda_i+2c$, computed with
coordinates in $\mathbb Q(\sqrt{\mathrm{disc}})$; rescaling eigenvectors maps the closed quadrant of $(X,Y)$ to itself, and a nonzero
determinant $\det\big(\begin{smallmatrix}\gamma_{P_4,0}&\gamma_{P_4,2}\\ \gamma_{K_{1,3},0}&\gamma_{K_{1,3},2}\end{smallmatrix}\big)$, and the analogous one for (iii), is
sufficient.

\emph{(U3).} Let $\theta\in\aff\Simplex\setminus\{\theta^*\}$ have $\rank DF_{\le4}(\theta)|_{\one^\perp}\le1$, so $DF_T(\theta)h=0$ for all $T\in\cT_4$ and some
$h\in\one^\perp\setminus\{0\}$, which up to scaling is $h(m)=(1,m,-1-m)$ (chart 1) or $(m,1,-1-m)$ (chart 2). The edge equation
$2\theta^\top Mh=0$ and $\sum\theta=1$ define the line $\theta=\theta^*+\lambda r(m)$ with $r(m)=(Mh)\times(1,1,1)\ne0$, $\lambda\ne0$. For the other three
trees, $DF_T(\theta^*+\lambda r)h=\lambda\,w_T(\lambda,m)$ with $w_T\in\mathbb Q[\lambda,m]$ of degrees $1,2,2$ in $\lambda$. Writing
$w_{P_3}=\omega_1(m)+\lambda\omega_2(m)$, the identity $\gcd(\omega_1,\omega_2)=1$ excludes $\omega_2(m)=0$, and otherwise $m$ is a root of
$\gcd(\mathrm{Res}_\lambda(w_{P_3},w_{K_{1,3}}),\mathrm{Res}_\lambda(w_{P_3},w_{P_4}))$, whose roots are found exactly. The same argument applies to
complex $m$.

\emph{(U4).} Let $\theta\ne\theta'$ in $\Simplex$ with $F_{\le4}(\theta)=F_{\le4}(\theta')$, $u:=(\theta+\theta')/2$ and $(\theta-\theta')/2=t\,h(m)$ in one chart, $t\ne0$,
$\tau:=t^2>0$. For $T\in\cT_4$, $F_T(u+th)-F_T(u-th)=2t[T_1(u,h)+\tau T_3(u,h)]$. The edge gives $T_1=2u^\top Mh=0$ and $T_3=0$, so
$u=\theta^*+\lambda r(m)$; the wedge, star and path give $W_1+\tau W_3=0$, $S_1+\tau S_3=0$, $P_1+\tau P_3=0$, with $W_1,S_1,P_1$ divisible
by $\lambda$. If $W_3(m)\ne0$, then $\tau=-W_1/W_3$, $\lambda\ne0$, and $P_S:=(S_1W_3-W_1S_3)/\lambda$ and $P_P:=(P_1W_3-W_1P_3)/\lambda$, both
of degree two in $\lambda$, vanish; so $m$ is a real root of $R(m):=\mathrm{Res}_\lambda(P_S,P_P)$, computed with formal degree two, which
vanishes whenever the specializations have a common root. The directions with $W_3(m)=0$ are treated separately, by
gcd computations showing that they carry no double pair. For each real root $m_0$ of $R$, the common roots in $\lambda$ are
found in exact arithmetic when $m_0$ is rational, or enclosed by interval arithmetic: where the leading coefficient
$\ell_1:=P_P^{(2)}P_S^{(1)}-P_S^{(2)}P_P^{(1)}$ of the linear combination $P_P^{(2)}P_S-P_S^{(2)}P_P=\ell_1\lambda+\ell_0$ has an interval that
excludes zero, the only candidate is $\lambda_0=-\ell_0/\ell_1$, which also rules out common roots at infinity. Each candidate
is then classified by the sign of $\tau$ and the position of $\theta,\theta'$ relative to $\Simplex$.

\subsection{The channel \texorpdfstring{$M_0$}{M0}}

\emph{(N).} We have $\det M_0=167/320$, $\theta^*=(73/194,\,307/776,\,177/776)$, $c=835/3104$, $\mathsf q(x)=x^2-\frac{9539}{15520}x+\frac{3966747}{60217600}$ with discriminant
$27524569/240870400\ne0$ and nonzero constant term, and $\lambda_{1,2}=(9539\mp\sqrt{27524569})/31040\approx0.13829,0.47633$.

\emph{(U1).} In the coordinates $v$,
\[
  \begin{aligned}
    \tFthree_{P_4}&=\tfrac{403817}{1241600}v_1^3+\tfrac{170231}{776000}v_1^2v_2+\tfrac{86443}{388000}v_1v_2^2+\tfrac{67891}{388000}v_2^3,\\
    \tFthree_{K_{1,3}}&=\tfrac{60507}{310400}v_1^3-\tfrac{540297}{1552000}v_1^2v_2-\tfrac{3204}{12125}v_1v_2^2+\tfrac{21171}{388000}v_2^3 .
  \end{aligned}
\]
Both $v_1^3$ coefficients are nonzero, and
\[
  \mathrm{Res}_{v_1}\big(\tFthree_{P_4}(v_1,1),\tFthree_{K_{1,3}}(v_1,1)\big)=-\frac{7696810286293504902599896137}{59410994786467840000000000000000}\ne0,
\]
so (U1)(i) holds. In eigencoordinates the determinant for (ii) is $a+b\sqrt{27524569}\approx3.87\times10^{-4}$ and the one for (iii) is
$a-b\sqrt{27524569}\approx-2.54\times10^{-5}$, with the rationals
\[
  \begin{gathered}
  a=\frac{8330024441949916011840741596307}{46102931954299043840000000000000000},\\
  b=\frac{1810896775144710997641820311}{46102931954299043840000000000000000};
  \end{gathered}
\]
both are nonzero because $27524569$ is not a square.

\needspace{9\baselineskip}
\emph{(F0), (U2).} The tangential first derivatives of all seven tree functions vanish at the three face equitable points. The
coefficients $\psi^T_{ab}$ are exact rationals, shown in decimals:
\begin{center}\fontsize{8}{9.5}\selectfont\setlength{\tabcolsep}{5pt}
\begin{tabular}{@{}lllll@{}}
  \toprule
  Face & $f_k$ & $\det J_k$ & $(\psi^{P_4}_{11},\psi^{P_4}_{30})$; $(\psi^{K_{1,3}}_{11},\psi^{K_{1,3}}_{30})$ & $\det\big(\begin{smallmatrix}\psi^{P_4}_{11}&\psi^{P_4}_{30}\\ \psi^{K_{1,3}}_{11}&\psi^{K_{1,3}}_{30}\end{smallmatrix}\big)$ \\
  \midrule
  $\theta_1=0$ & $(0,\frac{11}{24},\frac{13}{24})$ & $-21097/24000$ & $(0.06409,0.05381)$; $(-0.25770,0.07263)$ & $5334329/288000000$ \\
  $\theta_2=0$ & $(\frac{14}{33},0,\frac{19}{33})$ & $-257573/264000$ & $(0.06860,0.22708)$; $(-0.24538,0.23646)$ & $12535117/174240000$ \\
  $\theta_3=0$ & $(\frac{23}{49},\frac{26}{49},0)$ & $-243021/392000$ & $(0.00426,0.25316)$; $(-0.01342,0.23281)$ & $84306339/19208000000$ \\
  \bottomrule
\end{tabular}
\end{center}
A nonzero last determinant implies (U2)(ii).

\emph{(U3).} In chart 1 the gcd is $m(m+1)(2044m^2-1557m-3070)/2044$: $m=0$ gives $\lambda=1535/6402$ and the point $f_2$, $m=-1$ gives
$\lambda=-885/9506$ and $f_3$, and the quadratic factor divides $\omega_1$, so its two roots give $\lambda=0$, the point $\theta^*$ itself. In
chart 2 the gcd is $m(m+1)(3070m^2+1557m-2044)/3070$, giving $f_1$, $f_3$ and $\theta^*$. So in the whole plane $\aff\Simplex$ the points of
rank at most one are exactly $\theta^*,f_1,f_2,f_3$.

\emph{(U4).} In each chart, $R$ has degree $23$, with irreducible factors over $\mathbb Q$ of degrees $1,1,2,3,6,10$. The rational roots
$m_0\in\{0,-1\}$ give $\tau=0$ exactly; the two roots of the quadratic factor give $\lambda=0$ and $\tau=0$; the three roots of the cubic
factor are the roots of $W_3$; of the remaining eight real roots, five give $\tau<0$, two give pairs with a negative coordinate, and
one gives a pair in $\Simplex$, the same pair in both charts with the roles of $\theta$ and $\theta'$ exchanged:
$\theta\approx(0.0142202,0.7148095,0.2709703)$, $\theta'\approx(0.1580727,0.0083592,0.8335681)$, with an interval bound
$\cD_4(\theta,\theta')\le1.5\times10^{-98}$, $|\Ent(\theta)-\Ent(\theta')|\approx0.170955$, and five-vertex differences of about $1.0634\times10^{-4}$
($P_5$), $5.6778\times10^{-4}$ ($S_{2,1,1}$) and $1.984722\times10^{-3}$ ($K_{1,4}$), the last with interval lower bound $1.98472\times10^{-3}$. A
double pair of $F_{\le5}$ would be a double pair of $F_{\le4}$, so (U4) holds, and $M_0\notin\cU_4$.

\emph{Independent checks.} In each chart the polynomials $P_S,P_P$ are recomputed at $20$ random rational points directly
from the tree functions, $R(m)$ at $12$ rational points as a Sylvester determinant, and the factorization and the real-root counts are
confirmed by Sturm sequences. Newton's method on the four original equations at $100$ digits converges to the pair with residual
$7\times10^{-102}$ and Jacobian singular values $3.35$, $0.480$, $0.0686$, $0.0160$. A second implementation builds the tree
polynomials by enumerating labelings, computes a Gröbner basis of the $2\times2$ minors of $DF_{\le4}$ (solutions exactly
$\theta^*,f_1,f_2,f_3$) and solves the four equations from $800$ starts. It finds only this pair, with residual $1.3\times10^{-91}$ at $90$ digits.

\subsection{The other channels}

The same certificate shows that $M_1$ and $M_2$ satisfy (N), (F0) and (U1) to (U4) with no double pair of $F_{\le4}$ in $\Simplex$, so
$M_1,M_2\in\cU_4$. Including $M_0$, $13$ of the $20$ grid channels satisfy all conditions, with $0$ to $10$ double pairs, all with
different entropies (the smallest gap among them is about $2.3\times10^{-4}$). The other $7$ have an exact coincidence of entries, and
the corresponding quantity vanishes exactly: $M_{ki}=M_{kj}$ makes one of the two determinants of (U1)(ii) and (iii) vanish,
makes $\psi^{P_4}_{11}$ and $\psi^{K_{1,3}}_{11}$ on the face $k$ vanish, so that (U2)(ii) fails there, and (U3) fails on the segment from $f_k$ to $e_k$;
$M_{ii}=M_{jj}$ gives the symmetric case. These two determinants compare a perturbation of $\theta^*$ with its reflection in one eigen-axis.

For $M_3,\dots,M_9$ the certificate is modified in three places. On each symmetric face it verifies that all tangential first
derivatives vanish, that $\psi^T_{30}=0$, that $\det J_k\ne0$ and that $(\psi^{P_4}_{11},\psi^{K_{1,3}}_{11})\ne(0,0)$. Over the seven channels,
$\min|\det J_k|=297/3200$ and $\min_k\max_T|\psi^T_{11}|=279/176000$, both on the face $\theta_2=0$ of $M_9$. For (T), with $\varphi=f_k+t\mathbf d_k$, the
four $3\times3$ minors of $K_k(\varphi)$ are polynomials in $t$. Each is divisible by $t^2$, as the factorization $K_k=D\Psi(q)B(t)$ with
$\det B(t)=-4t^2$ predicts, and the four quotients have a constant greatest common divisor, so they have no common complex
root and $K_k$ has rank three on $0<|t|\le\frac12$, including the vertices. This holds on all $15$ symmetric faces. For (U4$\sigma$), the
direction of a symmetric face is a root of $W_3$. There the wedge equation reads $\lambda(\omega_1+\lambda\omega_2)=0$, and the certificate
verifies exactly that $\lambda_0=-\omega_1/\omega_2$ gives $u=f_k$ and that $S_1,S_3,P_1,P_3$ vanish at $(\lambda_0,m)$, so every double pair in this
direction is a symmetric pair. The other directions are treated as for $M_0$. The resulting asymmetric double pairs and
separations are those of \Cref{tab:certificates}. For each channel an independent numerical search for double pairs, from
$2000$ starts of which $500$ lie on the boundary, finds the same pairs.

\subsection{The box of radius \texorpdfstring{$0.01$}{0.01}}

For $P_{0.01}:=\{M\in\Sym:\max_{ab}|M_{ab}-(M_0)_{ab}|\le0.01\}$ we use interval arithmetic on $64$ sub-boxes, splitting each of the six
entries in two; the sub-boxes have exact rational endpoints, entered by outward-rounded interval division, so they cover the
closed box. $\theta^*$ and $c$ are enclosed through the adjugate of $M$. For (N3) and (N4), Weyl's inequality bounds the motion of
each eigenvalue of $\tM$ by $\|\tM(M)-\tM(M_0)\|_F$, where $\tM(M_0)$ and its spectrum are enclosed from their exact values,
$\theta^*$ and $c$ rational, and $\lambda_{1,2}=(9539\mp\sqrt{27524569})/31040$. The enclosure of $c(M)$ is checked to lie strictly between the moved outer
eigenvalues, so that $c$ remains the middle one. For (N2), $\|\ell|_{u^\perp}\|^2=\sum_{c<d}\theta^*_c\theta^*_d\log^2(\theta^*_c/\theta^*_d)$, the variance of $\log\theta^*_Z$ under $Z\sim\theta^*$, an identity without
cancellation. The certificate gives $\min_c\theta^*_c\ge0.196$, $|\one^\top\operatorname{adj}(M)\one|\ge1.81$ and $|\det M|\ge0.488$, so that $\one^\top M^{-1}\one\ne0$. It also gives $|\lambda_1-\lambda_2|\ge0.226$, $\min_i|\lambda_i|\ge0.0827$, $|\ell(\zeta_0)|\ge0.0862$, $c\in[0.246,0.295]$
and a Weyl radius at most $0.0556$, below the half-gap of the spectrum $\{0.1383,0.2690,0.4763\}$ at $M_0$, which is
at least $0.0653$. This proves \Cref{prop:class}(e).

\subsection{Persistence}

The map $G(z;M)=F^M_{\le4}(\theta)-F^M_{\le4}(\theta')$, $z=(\theta_1,\theta_2,\theta'_1,\theta'_2)$, has at $M_0$ a zero next to $z_0$,
a rational rounding of a Newton iterate at $60$ digits with residual $1.6\times10^{-61}$; the singular values of $DG(z_0;M_0)$ are $3.347$, $0.4799$,
$0.06863$ and $0.01600$. With $X:=z_0+[-1/5000,1/5000]^4$, $P_r$ for $r=1/2000000$ and the preconditioner $Y$ an approximate inverse of
$DG(z_0;M_0)$, the Krawczyk operator \citep{krawczyk1969newton} satisfies $K(X,P_r)\subset\operatorname{int}X$, with half-widths of $K$ between $2.3\times10^{-5}$ and
$1.9\times10^{-4}$. The enclosure of $G(z_0;P_r)$ is the centered form $G(z_0;M_0)+\sum_e\partial_{M_e}G(z_0;P_r)\,[-r,r]$ over the six entries,
with the partial derivatives computed by forward differentiation in interval arithmetic, and every box is built from
rational endpoints with outward rounding. By the existence and uniqueness theorem for this operator \citep{moore2009introduction,rump2010verification}, for every
$M\in P_r$ the map $G(\cdot;M)$ has exactly one zero in $X$. Interval evaluation over $X\times P_r$ gives the following bounds: the coordinates of the two
points are at least $0.01402$ and $0.00815$; the variance $v$ is at least $0.02663$ at the first point and $0.02672$ at the second; the
entries of $M\theta$ are pairwise separated by at least $0.30995$ and $0.07025$; and $\Ent(\theta)-\Ent(\theta')\in[0.16651,0.17540]$. The boxes for
$\theta$ and $\theta'$ are disjoint, so the two points are distinct. Hence every $M\in P_r$ has an admissible double pair with entropy
gap at least $0.16651$.

\subsection{Proof of \texorpdfstring{\Cref{prop:class}}{the proposition on the condition class}}

(a) is \Cref{prop:b:open}. (b) and (c) are the certificates above. (d): the persistence certificate gives an admissible double
pair for every $M\in P_r$, so $F^M_{\le4}$ is not injective and $P_r\cap\cU_4=\emptyset$; by (a) and (b) there is a neighborhood of $M_0$
inside $\cU$, and its intersection with $\operatorname{int}(P_r)$ is an open set contained in $\cU\setminus\cU_4$ on which \Cref{cor:phase}(ii)
applies, with $\omega\ge0.16651$. (e) is the certificate of the box of radius $0.01$. \qed

\section{Numerical checks}
\label{app:numerics}

The checks below are sampled numerical diagnostics of steps of the proofs, in contrast to the certificates,
on which \Cref{prop:class,thm:symmetric} rest; \Cref{tab:diagnostics} summarizes the checks, and
everything runs on a CPU. Ranges and bounds are rounded outward.

\begin{table}[t]
  \centering
  \caption{Sampled numerical diagnostics of steps of the proofs. The ratios of the modulus search are the largest
  values the searches attained, hence lower bounds for the modulus constants; ranges and bounds are rounded
  outward. Settings and metrics are defined in \Cref{app:numerics}.}
  \label{tab:diagnostics}
  \small
  \begin{tabular}{@{}L{3.19cm}L{5.67cm}L{6.38cm}@{}}
    \toprule
    Check & population and settings & result \\
    \midrule
    Variance of the tree U-statistics & exact enumeration, $n\in\{5,6\}$, 56 cases & mean: relative error $\le2.2\times10^{-11}$; variance formula: $\le1.5\times10^{-14}$ \\
     & exact variances, 875 cases & variance over bound $\le0.124$ \\
     & weak regime, $n=10^4$, $\delta=10^{-4}$ & variance over leading term in $[1.0000,1.0013]$ \\
    Rotation path & 38 of the 40 channels formed by the twenty grid channels and twenty channels within $0.02$ of $M_0$, entrywise; $s\in\{0.04,0.02,0.01,0.005\}$ & tree orders in $[2.990,3.265]$, unicyclic orders in $[0.976,1.068]$, edge and wedge residual $\le8.3\times10^{-57}$ \\
    Rotation path & the two other grid channels, with the smallest eigenvalue gaps; $s\in\{0.02,0.01,0.005,0.0025\}$ & tree orders in $[2.967,3.386]$, unicyclic orders in $[0.941,1.078]$, edge and wedge residual $\le7.6\times10^{-58}$ \\
    Fisher information, exact, $n=5$ & $M_0$, $q=0.3$, $s=0.02$, $\Delta\in\{0.02,0.01,0.005\}$; rotation path at 5 times; straight segment at fixed $M$, $t\in\{0,0.25\}$ & fitted power of $\Delta$: $5.984$ to $5.986$; $1.995$ to $1.997$ ($5.984$ at the midpoint over these $\Delta$, where the edge derivative vanishes) \\
    Fiber path at $M_0$, 50 digits & 4007 points, 7 pieces & $\max|F_{\le4}(\theta_t;M_t)-\mathbf c|\le5.7\times10^{-48}$, $\min_{t,c}\theta_{t,c}=0.00836$, $\max\|M_t-M_0\|=0.83$ \\
    Entropy modulus search & multistart searches under $\mathcal D_5\le\eta$: $M_0$; five channels within $0.01$ of $M_0$, entrywise & largest attained $|\Delta H|^3/\mathcal D_4$: $14.38$; $13.0$ to $16.9$ \\
     & the same searches at $M_3,\dots,M_9$ & largest attained $|\Delta H|^3/\mathcal D_4$: $0.33$ to $331.6$ \\
    \bottomrule
  \end{tabular}
\end{table}

\paragraph{Tree statistics.}
The channels are $M_0$ and $M_{\rm neg}:=\frac1{10}\big(\begin{smallmatrix}6&-5&3\\-5&-4&7\\3&7&2\end{smallmatrix}\big)$, which has negative entries on the
diagonal, and the distributions are $(1/2,1/3,1/6)$, $(0,1/4,3/4)$ and $\theta^*(M_0)$. For $n\in\{5,6\}$ and the four configurations
$(M_0,(1/2,1/3,1/6),q=1/3,\Delta=1/4)$, $(M_0,(0,1/4,3/4),1/3,1/4)$, $(M_0,\theta^*(M_0),1/3,1/10)$ and $(M_{\rm neg},(1/2,1/3,1/6),1/2,1/2)$, the
mean and the variance of $\hF_T$ for the seven trees were computed by enumerating all labelings and all verdict graphs. The
largest relative error of the mean against $F_T$ is the first number in \Cref{tab:diagnostics}, and the largest relative
difference between the enumerated variance and the variance computed from the overlap patterns of the proof of
\Cref{lem:ustat} is the second. The exact variances were then computed from the overlap patterns for
$n\in\{8,12,20,50,200,10^3,10^4,10^5\}$, $\delta\in\{4,1,n^{-1/4},n^{-1/2}\}$ subject to \eqref{eq:P}, and the four configurations with the trees on
two to five vertices, $875$ cases. The table gives the largest ratio of the exact variance to the first bound of
\Cref{lem:ustat} before simplification. For $n=10^4$ and $\delta=10^{-4}$ the ratio of the variance to the leading term
$|\mathrm{Aut}(T)|(q(1-q))^{v-1}/((n)_v\Delta^{2v-2})$ lies in the range shown; for $\delta=10^{-1},10^{-2},10^{-3}$ the ranges are wider and shrink
toward one, as the attainment statement predicts.

\paragraph{Rotation path.}
For the twenty grid channels and twenty channels obtained by moving each entry of $M_0$ by at most $0.02$, the rotation path
was computed at $60$ digits on $101$ values of $t$ for the amplitudes $s=0.04\cdot2^{-k}$, $k=0,\dots,5$, following the solution $\rho(t,s)$ by continuation in
$s$ from $\rho^{(0)}(t)$ and checking it against continuation in $t$. For the two grid channels with the smallest eigenvalue gaps, $0.029$ and $0.034$, the
computation at $s=0.04$ ends before $t=1$, at $t=0.18$ where the continuation in $s$ fails and at $t=0.47$ where the check against continuation in $t$ fails,
consistent with the admissible amplitude $s_P$ of \Cref{lem:rotation} shrinking with the gap $|\lambda_1-\lambda_2|$; their orders in $s$ are fitted on $s\in\{0.02,0.01,0.005,0.0025\}$,
and those of the other $38$ channels on $s\in\{0.04,0.02,0.01,0.005\}$. The table gives the residual of the
edge and wedge functions, and the orders in $s$ fitted to $\max_t|\partial_t\hom|$ over $45$ trees on four to eight vertices and over
unicyclic graphs on at most six vertices, which \Cref{lem:rotation} predicts to be $3$ and $1$. The endpoint errors
$|M_t-M|$ at $t=1$ are below $10^{-60}$, and the ratio $|\Ent(\theta_+)-\Ent(\theta_-)|/(2s|\ell(\zeta_0)|)$ at $s=0.005$ lies in $[0.991,1.008]$ over all $40$ channels.

\paragraph{Exact Fisher information.}
For $n=5$, $M_0$, $q=0.3$ and $s=0.02$, the Fisher information of the law of the verdicts was computed exactly, summing over
all $2^{10}$ graphs and $3^5$ labelings, for $\Delta\in\{0.02,0.01,0.005\}$, along the rotation path at $t\in\{0,0.25,0.5,0.75,1\}$ and along
the straight segment from $\theta_+$ to $\theta_-$ at fixed $M_0$. Along the rotation path the information scales as $\Delta^6$, because the
edge and wedge functions are constant and the $\Delta^2$ and $\Delta^4$ terms vanish; along the segment it scales as $\Delta^2$ at
$t\in\{0,0.25\}$. At the midpoint of the segment the derivative of the edge function vanishes exactly, because the edge
function is quadratic along the segment with equal values at its ends, while the derivative of the wedge function is
not zero; over the three values of $\Delta$ the fitted power there is $5.98$, and as $\Delta\to0$ the wedge term gives
order $\Delta^4$.

\paragraph{Fiber path.}
The path of \Cref{lem:fiber-path} for $M_0$ was computed at $50$ digits on $4007$ points, with the design described after its
proof. The table gives the largest deviation of $F^{M_t}_{\le4}(\theta_t)$ from its value at the endpoints, the smallest coordinate of
$\theta_t$, attained at the endpoint $\theta^1$, and the largest entrywise distance of $M_t$ from $M_0$. The endpoint errors of $M_t$ are
below $1.1\times10^{-49}$, and the assignment of meanings to positions changes from $(1,3,2)$ to $(1,2,3)$ at the first collision and to
$(2,1,3)$ at the second. \Cref{fig:fiber} uses a coarser re-evaluation on $1081$ points that includes the two collisions and both
endpoints exactly.

\paragraph{Entropy modulus.}
For $M_0$ and five channels obtained by moving each entry of $M_0$ by at most $0.01$, the ratio $|\Ent(\theta)-\Ent(\theta')|^3/\cD_4(\theta,\theta')$
was maximized under the constraint $\cD_5(\theta,\theta')\le\eta$, with $\eta$ one quarter of the five-vertex separation of the double pair of
each channel. Sequential quadratic programming in SciPy \citep{virtanen2020scipy} was run from up to $10000$ starts for $M_0$ and $1250$ for each of the other channels, and the
best pairs were refined at $40$ or $60$ digits. For $M_3,\dots,M_9$ a search of the same kind used $3000$ starts in five classes
and was repeated with twice as many starts at $60$ digits, with $\eta$ one quarter of the separation of the asymmetric double
pairs, or $10^{-3}$ for $M_6$ and $M_9$, which have none. In both searches the local solver maximizes a surrogate with the Euclidean norm of the
differences of $F_{\le4}$ in place of the maximum norm, while the reported ratios use the maximum norm. The table reports the
largest values attained, which are lower bounds for the constants of the cubic modulus. For $M_0$ the largest ratios occur on pairs of large
amplitude nearly opposite around $\theta^*$, whose amplitude is limited by the constraint.

\end{document}